\documentclass[12pt]{iopart}

\usepackage{graphicx}
\begin{document}

\topical{Neutron Reactions in Astrophysics}
\author{R. Reifarth$^1$, C. Lederer$^{1,a}$, F. K\"appeler$^2$}
\address{$^1$ Goethe University Frankfurt, Frankfurt, Germany}
\address{$^2$ Karlsruhe Institute of Technology, Karlsruhe, Germany}
\address{$^a$ Present Affiliation: University of Edinburgh, Edinburgh, UK}

\ead{reifarth@physik.uni-frankfurt.de}

\begin{abstract}
The quest for the origin of matter in the Universe had been the subject of philosophical and theological debates over 
the history of mankind, but quantitative answers could be found only by the scientific achievements of the last century. 
A first important step on this way was the development of spectral analysis by Kirchhoff and Bunsen in the middle of the 
19$^{\rm th}$ century, which provided first insight in the chemical composition of the sun and the stars. The 
energy source of the stars and the related processes of nucleosynthesis, however, could be revealed only with the 
discoveries of nuclear physics. A final breakthrough came eventually with the compilation of elemental and isotopic 
abundances in the solar system, which are reflecting the various nucleosynthetic processes in detail. 

This review is focusing on the mass region above iron, where the formation of the elements is dominated by neutron 
capture, mainly in the slow ($s$) and rapid ($r$) processes. Following a brief historic account and a sketch of the 
relevant astrophysical models, emphasis is put on the nuclear physics input, where status and perspectives of 
experimental approaches are presented in some detail, complemented by the indispensable role of theory. 
\end{abstract}

\maketitle
\tableofcontents

\section{Neutron capture nucleosynthesis \label{sec:1}}

\subsection{Milestones and basic concepts \label{sec:1.1}}

In 1938, the quest for the energy production in stars had been solved by the work of Bethe and Critchfield \cite{BeC38}, 
von Weizs\"acker \cite{Wei38}, and Bethe \cite{Bet39a}, but the origin of the heavy elements remained a puzzle for almost 
two more decades. It was finally the discovery of the unstable element technetium in the atmosphere of red giant stars by 
Merrill in 1952 \cite{Mer52b}, which settled this issue in favor of stellar nucleosynthesis, thus questioning a primordial 
production in the Big Bang. A stellar origin of the heavy elements was strongly supported by the increasingly reliable 
compilations of the abundances in the solar system by Suess and Urey \cite{SuU56} and Cameron \cite{Cam59a},
because the pronounced features in the abundance distribution could be interpreted in terms of a series of nucleosynthesis 
scenarios associated with stellar evolution models. This key achievement is summarized in the famous fundamental papers 
published in 1957 by Burbidge, Burbidge, Fowler and Hoyle (B$^2$FH) \cite{BBF57} and by Cameron \cite{Cam57,Cam57b}.

While the elements from carbon to iron were found to be produced by charged particle reactions during the evolutionary 
phases from stellar He to Si burning, all elements heavier than iron are essentially built up by neutron reactions in the slow 
($s$) and rapid ($r$) neutron capture processes as they were termed by B$^2$FH.

\begin{center}
\begin{figure}[tb]
\includegraphics[width=15cm]{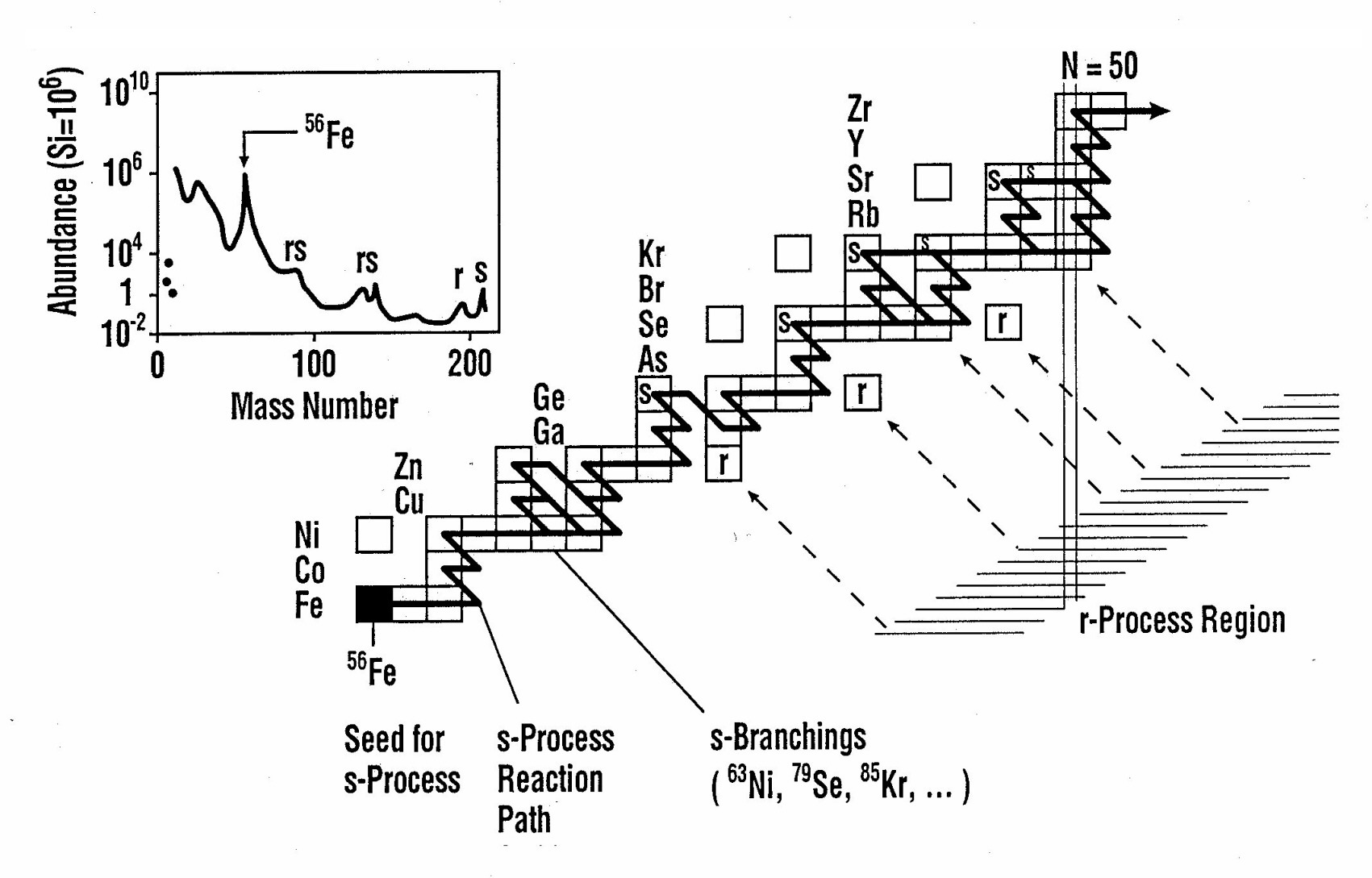}     
\caption{The formation processes of the elements between iron and the actinides. The neutron capture path of the $s$ process 
	follows the valley of stability and ends in the Pb/Bi region by $\alpha$-recycling. Due to the much higher neutron densities, 
	the $r$-process path is shifted to the far neutron-rich region, from where the reaction products decay back to stability. The 
	solar abundances are essentially composed of contributions from both processes, except for the $s$-only and $r$-only 
	isotopes, which are shielded by stable isobars against the $r$-process region or lie outside the $s$-process path, respectively. 
       An additional minor component is ascribed to the  $p$ (or $\gamma$) process to describe the rare, stable proton-rich 
	isotopes. The magic neutron number $N=50$ is shown to indicate the strong impact of nuclear structure effects, which 
	give rise to pronounced maxima in the observed abundance distribution as indicated in the inset.
	  \label{fig:1} }
\end{figure}
\end{center}

The $s$ process, which takes place during He and C burning, is characterized by comparably low neutron densities, typically 
a few times 10$^8$ cm$^{-3}$, so that neutron capture times are much slower than most $\beta$ decay times. This implies 
that the reaction path of the $s$ process follows the stability valley as sketched in Figure~\ref{fig:1} with the important 
consequence that the neutron capture cross sections averaged over the stellar spectrum are of pivotal importance for the 
resulting $s$ abundances. Although the available cross sections under stellar conditions were very scarce and rather 
uncertain, already B$^2$FH could infer that the product of cross section times the resulting $s$ abundance represents a smooth
function of mass number $A$. In the following decade, the information on cross section data was significantly improved by
dedicated measurements \cite{MaG65}, leading to a first compilation of stellar ($n, \gamma$) cross sections by Allen,
Gibbons and Macklin in 1971 \cite{AGM71}. Meanwhile, Clayton et al. \cite{CFH61} had worked out a phenomenological 
model of the $s$ process, assuming a seed abundance of $^{56}$Fe exposed to an exponential distribution of neutron 
exposures with the cross section values of the involved isotopes in the reaction path as the essential input. 

\begin{center}
\begin{figure}[tb]
\includegraphics[width=0.95\textwidth]{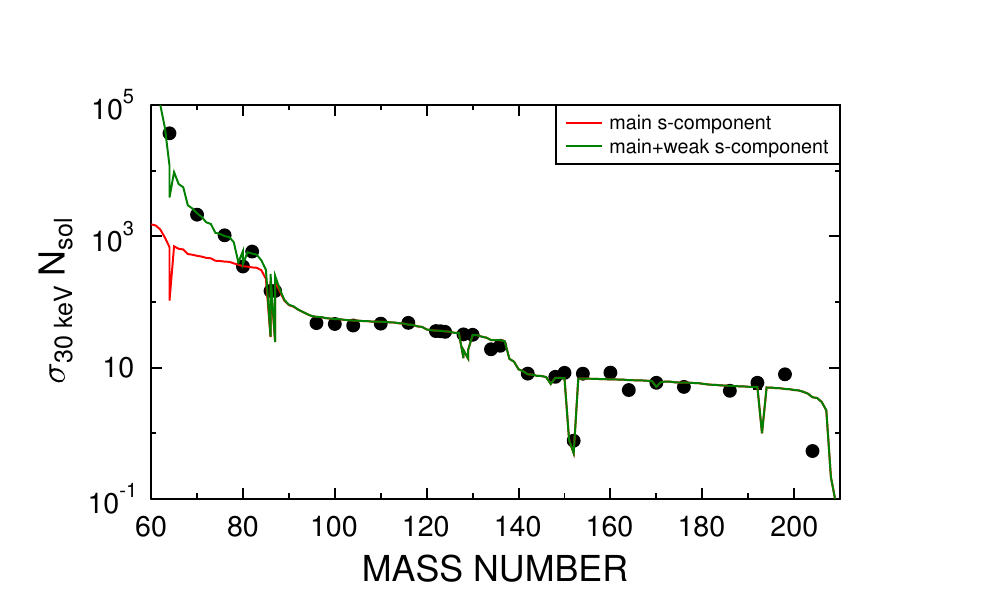}
\hfill
   \caption{The characteristic product of cross section times $s$-process 
            	abundance plotted as a function of mass number. The thick solid line 
		was obtained via the classical model for the main component, and the 
            	symbols denote the empirical products for the $s$-only nuclei.
            	Some important branchings of the neutron capture chain are 
            	indicated as well. A second, weak component had to be assumed for 
		explaining the higher $s$ abundances between Fe and $A\approx90$.
		Note that reaction flow equilibrium has only been 
	     	achieved for the main component in mass regions between magic
		neutron numbers (where $\sigma N$ values are nearly constant). 
		\label{fig:2} }
\end{figure}
\end{center}

As the cross section database was improved, this classical model turned out to be extremely useful for describing the $s$-process 
component in the solar abundance distribution. In fact, it turned out that the $s$ process itself was composed of different 
parts, i.e. the weak, main, and strong components as shown by Seeger et al. \cite{SFC65}. This $s$-process picture was 
eventually completed by the effect of important branchings in the reaction path due to the competition between neutron
capture and $\beta^-$-decay of sufficiently long-lived isotopes \cite{ClW74}. The appealing property of the classical
approach was that a fairly comprehensive picture of $s$ process could be drawn with very few free parameters and that these
parameters are directly related to the physical conditions typical for the $s$ process environment, i.e. neutron fluence, seed 
abundance, neutron density, and temperature. Moreover,
it was found that reaction flow equilibrium has been achieved in mass regions of the main component between magic
neutron numbers, where the characteristic product of cross section and $s$ abundance, $\sigma N(A)$ is nearly constant. In 
spite of its schematic nature, the classical $s$ process could be used to reproduce the solar $s$ abundances within a 
few percent as illustrated in Figure~\ref{fig:2}. 

Nevertheless, the more accurate cross section data became available, particularly around the bottle-neck isotopes with magic
neutron numbers and in $s$-process branchings, the more inherent inconsistencies of the classical model came 
to light \cite{KGB90,WVK98a}, indicating the need for a more physical prescription based on stellar evolution \cite{AKW99b}. 
This transition started with early models for stellar He burning by Weigert \cite{Wei66} and Schwarzschild and H{\"a}rm 
\cite{ScH67}, which were used by Sanders \cite{San67} to verify implicit $s$-process nucleosynthesis. The connection
to the exponential distribution of neutron exposures postulated by the classical approach was ultimately provided by Ulrich 
\cite{Ulr73} who showed that this feature follows naturally from the partial overlap of $s$-process zones in subsequent
thermal instabilities during the He shell burning phase in low-mass asymptotic giant branch (AGB) stars. Consequently, the
classical approach had been abandoned as a serious $s$-process model, but continued to serve as a convenient
approximation in the mass regions between magic neutron numbers with constant $\sigma$N$_s$ products. 
 
The second half of the solar abundances above iron is contributed by the $r$ process. In this case, the neutron densities
are extremely high, resulting in neutron capture times much shorter than average $\beta$ decay times. This implies that 
the reaction path is shifted into the neutron-rich region of the nuclide chart until the ($n, \gamma$) sequence is halted 
by inverse ($\gamma, n$) reactions by the hot photon bath. Contrary to the $s$ process, where the abundances are dependent
on the cross section values, the $r$ abundances are determined by the $\beta$-decay half lives of these waiting points close to 
the neutron drip line. 

As the consequence of the explosive supernova scenario suggested by B$^2$FH, prescriptions of the $r$-process 
abundances were severely challenged by the fact that the required nuclear physics properties for the short-lived, neutron-rich 
nuclei forming the comprehensive reaction network far from stability were essentially unknown. This information includes
$\beta$-decay rates and nuclear masses, neutron-induced and spontaneous fission rates, cross section data, and $\beta$-delayed 
neutron emission for several thousand nuclei. First attempts to reproduce the $r$-process abundances that had been 
inferred by subtraction of the $s$ abundances from the solar values \cite{AGM71} started with a simplified static 
approximation, assuming constant neutron density and temperature ($n_n\geq10^{20}$ cm$^{-3}$, $T\geq10^9$ K) 
during the explosion and neglecting neutron-induced reactions during freeze-out \cite{SFC65}. Early dynamic
$r$-process models were facing not only enormous computational problems, but had to deal with the many 
unknowns of the possible scenarios. In general, supernovae were preferred over supermassive objects and novae 
as potential $r$-process sites \cite{TAT68}, but the relevant features of such explosions, i.e. the temperature and 
density profiles, the velocity distribution during and shortly after the explosion, and the initial seed composition, were too 
uncertain to draw a plausible picture of the $r$ process by the end of the 1970ies \cite{Hil78}. 

As discussed by B$^2$FH about 35 proton-rich isotopes cannot be produced by neutron captures, because they are 
shielded from the reaction networks of the $s$ and $r$ processes as shown in Figure~\ref{fig:1}. The initial idea that these 
isotopes were produced by proton capture ($p$  process) in the hydrogen-rich envelope of massive stars during 
the supernova explosion \cite{BBF57} had to be abandoned because it led to unrealistic assumptions for densities, 
temperatures and timescales. In 1978, Woosley and Howard \cite{WoH78} suggested the shock-heated Ne/O shell 
in core-collapse supernovae as the site of the $p$ process, where temperatures are high enough for modifying a 
preexisting seed distribution by a sequence of photo-disintegration reactions. Therefore, this approach is sometimes 
also referred to as $\gamma$ process \cite{RPA90}.

\subsection{Solar abundances \label{sec:1.2}}

The abundance distribution in the solar system has served as an important source of information for the nucleosynthesis 
concepts \cite{BBF57,Cam57,Cam57b}. Following the pioneering work of Goldschmidt \cite{Gol37}, detailed abundance 
tables have been reported by Suess and Urey \cite{SuU56} and were then continuously improved by the combination of
meteoritic isotope abundances, essentially based on C1 chondrites, and of 
elemental abundances from spectroscopy of the solar photosphere. This series started with Cameron \cite{Cam82}, 
Anders and Ebihara \cite{AnE82}, Anders and Grevesse \cite{AnG89} and continued until now with compilations by 
Lodders \cite{Lod03}, Grevesse et al. \cite{GAS07},  and Lodders, Palme, Gail \cite{LPG09}.

\begin{center}
\begin{figure}[tb]
\includegraphics[width=0.95\textwidth]{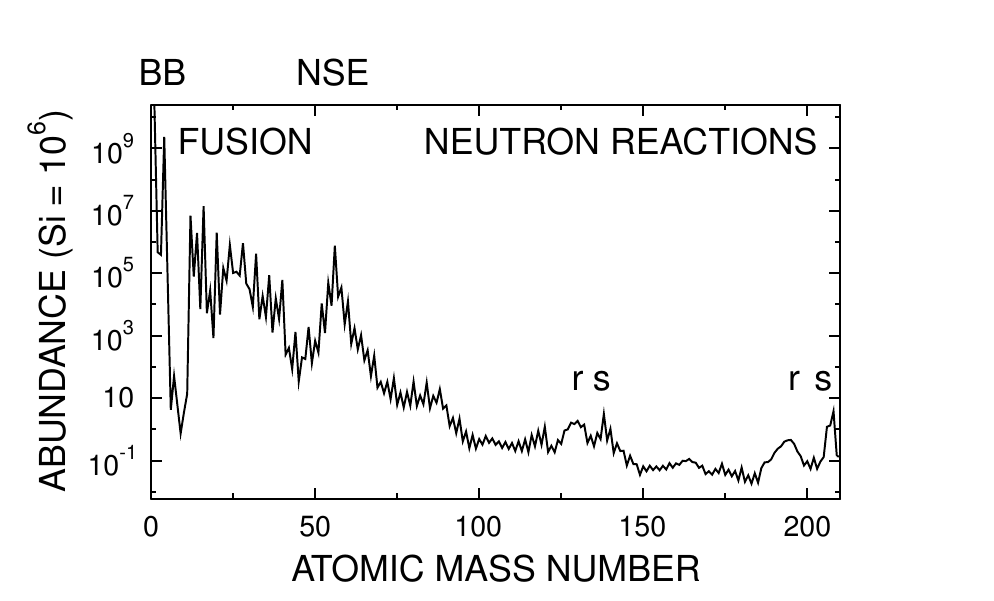}
   		\caption{The isotopic abundance distribution in the solar system with indications for the 
		main production processes (data from \cite{LPG09}). The $s$ and $r$ maxima are 
		reflecting the effect of magic neutron numbers $N=50, 82, 128$.  \label{fig:3} }
\end{figure}	 
\end{center}

The distribution plotted in Figure~\ref{fig:3} shows the solar system abundances as a function of mass number, which
clearly exhibits the influence of nuclear effects characteristic of the various nucleosynthesis sites. The distribution is 
by far dominated by the primordial H and He abundances from the Big Bang followed by the rare elements Li, Be, and 
B. Because these are difficult to produce due to the stability gaps at $A=5$ and 8, but are easily burnt in stars, the 
present abundances are essentially produced via spallation by energetic cosmic rays \cite{Ree94}.  Stellar nucleosynthesis 
of heavier nuclei starts with $^{12}$C and $^{16}$O, the products of He burning, which were partly converted to 
$^{14}$N by the CNO cycle in later stellar generations.
The light elements up to mass 50 are the result of charged-particle reactions during the advanced C, Ne, and O burning
phases. This mass region is characterized by the enhancement of the more stable $\alpha$ nuclei and by the exponentially 
decreasing abundances due to the increase of the Coulomb barrier with atomic number. The last stage dominated by 
charged-particle reactions is Si burning, where densities and temperatures are high enough to reach nuclear
statistical equilibrium. Under these extreme conditions only the most stable nuclei survive, leading to the
distinct abundance peak around $A=56$. Any further build-up of heavier elements were then provided by neutron 
capture reactions starting on these abundant isotopes, i.e. essentially on $^{56}$Fe, as a seed as discussed in the following 
section.

\section{Astrophysical models \label{sec:2}}

\subsection{The $s$ process in asymptotic giant branch stars \label{sec:2.1}}

\begin{figure}[tb]
\includegraphics[width=0.73\textwidth]{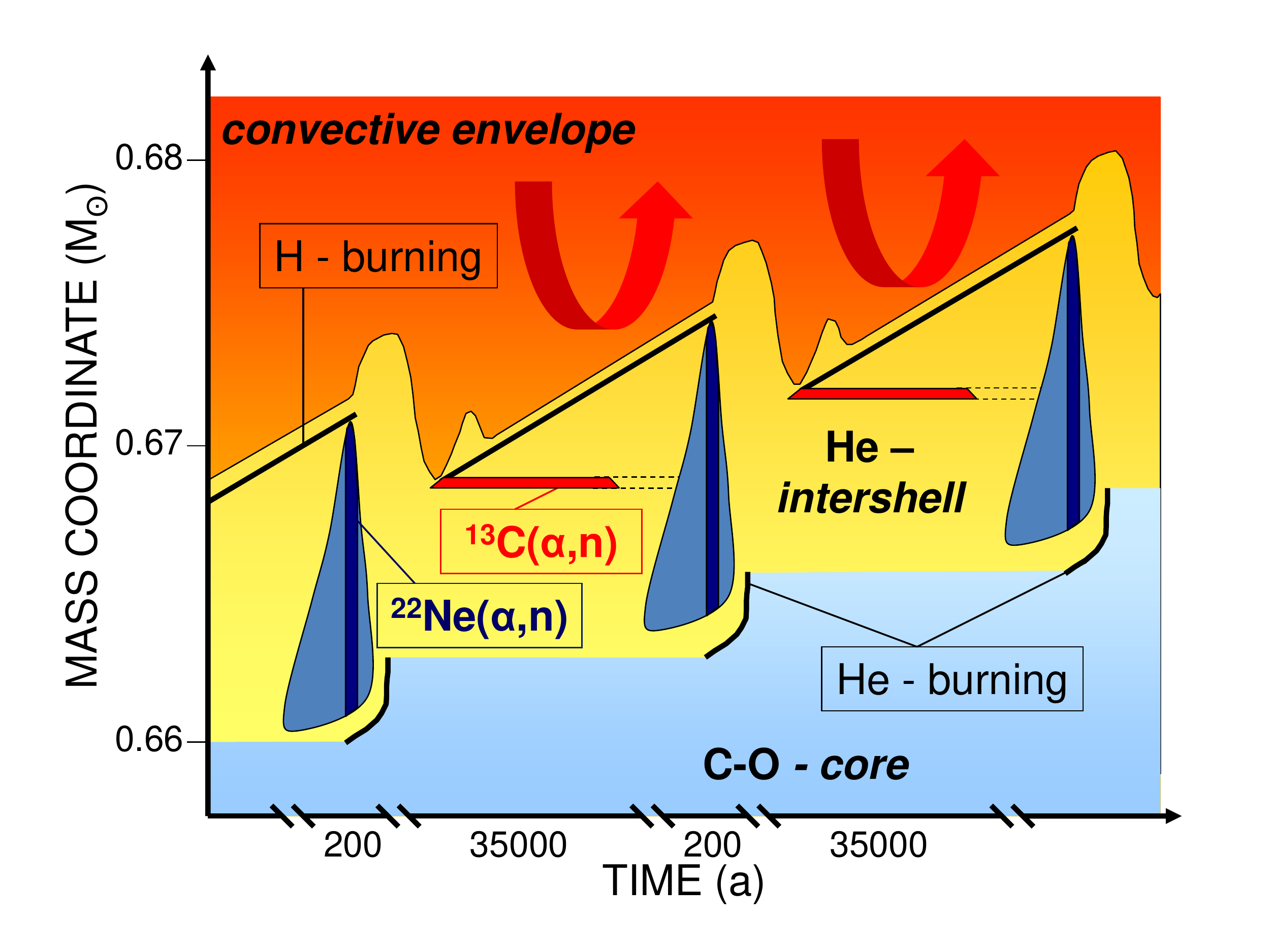}
\includegraphics[width=0.26\textwidth]{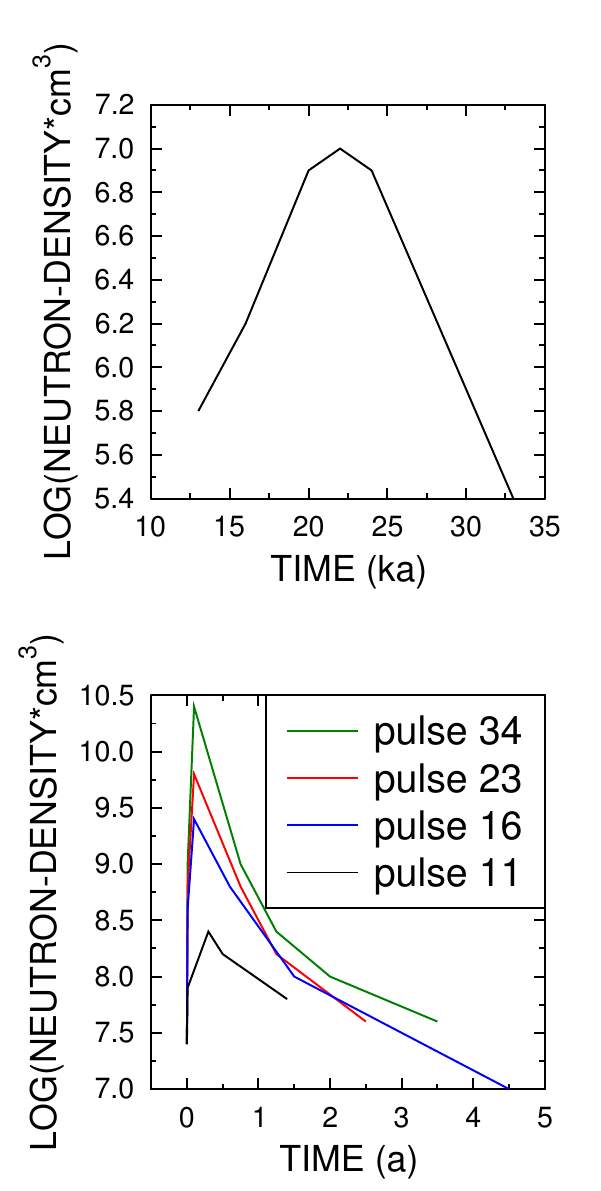}
   		\caption{Schematic structure and evolution of AGB stars, showing recurrent H and He 
		burning episodes with indications for the related $s$ process sites (left). 
   		Right: Strength and time-dependence of the neutron density in thermally pulsing 
		low-mass AGB stars contributed by ($\alpha, n$) reactions on $^{13}$C (top right) and 
		$^{22}$Ne (bottom right) \cite{BGW99}. Note that the efficiency of the latter increases with 
		pulse number (11, 16, etc) due to the higher bottom temperature in later thermal 
		pulses. \label{fig:4} }
\end{figure}

In the advanced He burning stage of low-mass stars (1 to 5 $M_\odot$, $M_\odot$ being the mass of the sun) the stellar 
structure consists of an inert stellar core of carbon and oxygen, the products of He burning, surrounded by a narrow
shell of about $10^{-2} M_\odot$ and a fully convective envelope as sketched in Figure~\ref{fig:4}. Within this 
thin layer, energy is produced in cycles, with several $10^4$ years of H burning at the bottom of the convective 
envelope alternating with He burning episodes of about 50 years. This stage of evolution represents the scenario for the 
$s$ process corresponding to the $main$ component inferred by the classical $s$ process \cite{SFC65}. 

During the H-burning period, He is produced from the bottom layers of the envelope. With the growing mass 
of the He intershell, the temperature in the intershell increases to the point, where He burning is triggered, resulting 
in a quasi-explosive thermal pulse (for details see Refs. \cite{IbR83,BGW99,Her04,SGC06,SCG08}). Due to the sudden 
energy release, the radiative energy transport during H burning is completely converted to convection in the whole 
intershell, the envelope expands, the H shell burning is temporarily extinguished, and a large amount of $^{12}$C is 
produced. The He shell burning continues radiatively for another few thousand years, before H shell burning starts 
again. In this way, thermal pulses might repeat 20 to 50 times depending on the initial stellar mass.

Significant neutron production and $s$ processing starts with the third dredge-up (TDU) after a limited number of 
thermal pulses, when the C/O core mass reaches $\sim$ 0.6 $M_\odot$ \cite{SGC06}. The term dredge-up refers to the unsual phases
in the development of stars when surface material is convectivly mixed deep into the interior of stars where the material has 
experienced nuclear reactions. As a result, freshly synthesized material is brought to the surface. If those episodes occure 
during the helium shell burning (AGB-phase), they are called third dredge-up, independent of how many pulses are actually 
particpating in the process. TDU occurs when the 
H-shell is inactive after a thermal pulse and the convective envelope penetrates into the top region of the 
He-intershell, mixing newly synthesised $^{12}$C and $s$-processed material to the surface and, at the same
time, a few protons into the top layers of the He intershell. 

After H burning resumes, these protons are captured by the abundant $^{12}$C, thus initiating the sequence 
$^{12}$C($p, \gamma$)$^{13}$N($\beta^+ \nu$)$^{13}$C in a thin region of the He-intershell, the so-called 
$^{13}$C pocket \cite{SCL97}. Neutrons are released in the pocket under radiative conditions by the $^{13}$C($\alpha, 
n$)$^{16}$O reaction at temperatures of $\sim 0.9 \times 10^8$ K \cite{GAB98}. At a neutron 
density of 10$^{6}$ to 10$^{7}$~cm$^{-3}$, about 95\% of the neutron fluence in AGB stars is reached during this 
first stage of the $s$ process because it is restricted to the thin $^{13}$C pocket and lasts for about 10,000 years. 
In other words, the pocket is strongly enriched in $s$-process elements, before it is engulfed by the subsequent 
convective thermal pulse as indicated in Figure~\ref{fig:4}. 
 
The mechanism for formation of the $^{13}$C pocket is still a matter of debate, (for a discussion see Refs.
\cite{LHW99,DeT03,HBS97,FLS96,Her00,Her04,SGC06}), but it is obvious that the H profile adopted in the pocket 
and the resulting amount of $^{13}$C determines the final $s$-abundance distribution \cite{SCG09}. 
Consequently, the pocket is often parameterized to reproduce the observed $s$-process abundance 
patterns, e.g. the solar $s$ distribution or $s$-process features found in presolar grains \cite{Zin98}. 

A second stage of the $s$ process in AGB stars occurs during the maximum extension of the thermal pulse, when 
the temperature at the base of the convective zone exceeds $2.5 \times 10^{8}$ K. Under this condition, the
$^{22}$Ne($\alpha, n$)$^{25}$Mg reaction is activated, maintaining a strong neutron burst for a 
few years. This neutron burst contributes only about 5\% to the total neutron fluence in AGB stars, but its
high neutron density of up to 10$^{10}$ cm$^{-3}$, depending on the maximum temperature at the bottom 
of the thermal pulse, is instrumental for shaping the final abundance distribution, especially in the $s$-process 
branchings. The relative strengths and dynamics of the two neutron sources characterizing the $s$ process in 
AGB stars are indicated in the right part of Figure~\ref{fig:4} \cite{GAB98}.  

\begin{figure}[tb]
\includegraphics[width=14cm]{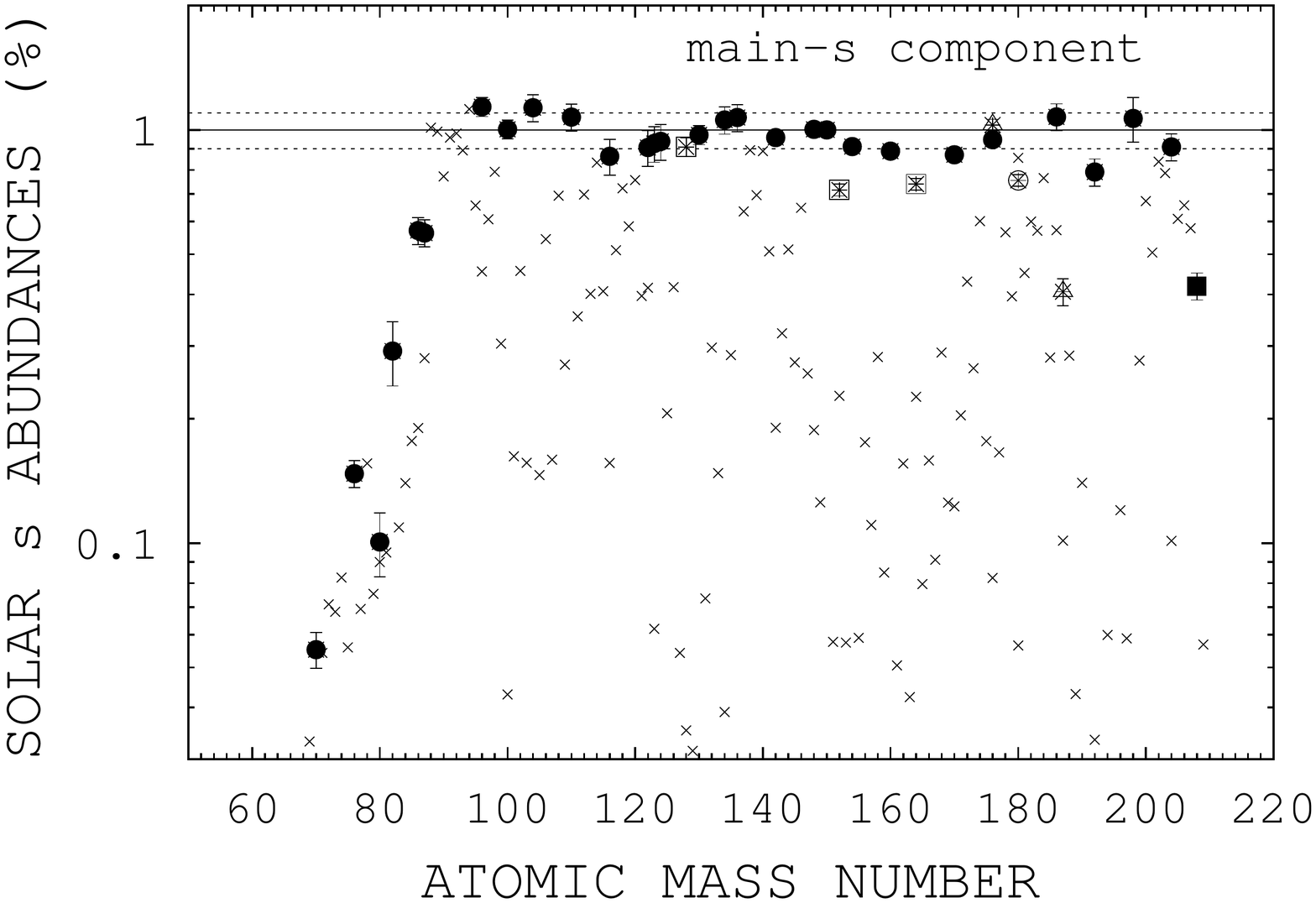}
   		\caption{The main $s$ component (in \% of the solar abundances) normalized 
		at the unbranched $s$-only isotope $^{150}$Sm versus mass number \cite{BGS10}. 
		The $s$-only isotopes between Sr and Pb/Bi (black circles) are reproduced to better than 10\% 
		except cases with sizable $p$ components or the long-lived isotopes $^{176}$Lu 
		and $^{187}$Os (open symbols) as well as the abundance of doubly magic $^{208}$Pb 
    		(black square). Isotopes denoted by crosses are only partially produced by the 
		$s$ process (for details see text). \label{fig:5}} 
\end{figure}

An important test of the current AGB model of the $s$ process is again the comparison with the solar $s$ 
abundances and with the isotope patterns of the $s$-process branchings. Based on the model described 
before excellent agreement had been demonstrated in Ref. \cite{AKW99b} between solar $s$ material
and an average of models for 1.5 and 3 $M_\odot$ AGB stars and was recently confirmed in \cite{BGS10}
as illustrated in Figure~\ref{fig:5}. For the comparison, both distributions are normalised at the unbranched
$s$-only nucleus $^{150}$Sm. The AGB model obviously describes the set of $s$-only isotopes indicated 
by full circles to better than 10\%. Different symbols were chosen for special $s$ isotopes with non-negligible 
$p$ contributions ($^{128}$Xe, $^{152}$Gd, $^{164}$Er, $^{180}$Ta) and for the long-lived isotopes 
$^{176}$Lu and $^{187}$Os. The black full square corresponds to $^{208}$Pb, which receives a contribution 
of about 50\% by the strong $s$ component produced in AGB stars of low metallicity 
\cite{TGB01b,TGA04,SGT09}.  Figure~\ref{fig:5} illustrates that low-mass AGB stars contribute the main 
component to the solar $s$ abundances from Sr to Pb/Bi, owing to the high neutron fluence in the 
$^{13}$C pocket. The high fluence has two consequences. It leads to the depletion of the $^{56}$Fe seed 
and of its progeny up to Sr at magic neutron number $N=50$, and it is establishing a reaction flow equilibrium
between magic neutron numbers characterized by $\sigma N_s\approx$ constant. Accordingly, knowledge of a 
stellar cross section yields directly the corresponding $s$ abundance of that particular isotope.

The reproduction of the main $s$ component of the solar abundance distribution represents an impressive
confirmation of the adopted $s$-process model that is characterized by the complex interplay between the 
relative strengths and dynamics of the two neutron exposures operating in low-mass AGB stars, including
the intricate abundance pattern of the various branchings along the $s$-process path. For a more complete 
account see Refs. \cite{BGW99,KGB11}. A comparison of low-mass AGB models computed with different 
evolutionary codes has been discussed by Lugaro {\it et al.} \cite{LHL03}.

The maximum temperature during thermal
pulses reaches about $3.5\times10^8$ K, leading to a substantial neutron production via the 
$^{22}$Ne($\alpha, n$)$^{25}$Mg reaction. However, the mass of the He intershell and the mixing efficiency
of $s$ processed material into the envelope are much smaller. Consequently, the predicted $s$-process 
abundance contributions are much lower than from low-mass AGB stars.

\subsection{The $s$ process in massive stars \label{sec:2.2}}

As illustrated in Figure~\ref{fig:5}, the $s$ process in low-mass AGB stars fails to reproduce the $s$ abundances 
below $A\leq90$. This gap is actually closed by the complementary weak $s$ process taking place in massive 
stars with $M\geq8M_\odot$, which explode as supernovae of type II. In massive stars, the $s$ process is 
driven by the $^{22}$Ne($\alpha, n$)$^{25}$Mg reaction, first during convective core He burning
at temperatures around 3$\times 10^8$ K and subsequently during convective shell C burning at $10^9$ K
as discussed in Refs. \cite{CSA74,LHT77,ArT85,BuG85,PAA87}. 

The available $^{22}$Ne is produced via the sequence of $^{14}$N($\alpha, n$)$^{18}$F($\beta, 
\nu$)$^{18}$O($\alpha, n$)$^{22}$Ne reactions, where $^{14}$N originates from the CNO cycle in the previous 
H-burning phase. Consequently, this first part of the weak $s$-process in massive stars is secondary-like
and decreases with metallicity. At He exhaustion in the core, not all the $^{22}$Ne is consumed 
(e.g. \cite{PAA87}) and neutron production via $^{22}$Ne($\alpha, n$)$^{25}$Mg continues during 
shell C burning, where $\alpha$ particles are provided by the reaction channel $^{12}$C($^{12}$C, 
$\alpha$)$^{20}$Ne \cite{ArT69}.  

The chemical composition of the core up to a mass of 3.5$M_\odot$ (for a star of 25$M_\odot$) 
is modified during explosive nucleosynthesis in the supernova, which destroys any previous $s$-process 
signature. However, the ejecta still contain an important mass fraction of 2.5$M_\odot$ that
preserves the original $s$-process abundances produced by the hydrostatic nucleosynthesis phases 
of the presupernova evolution. This scenario was confirmed by post-processing models and
full stellar models describing the evolution of massive stars up to the final burning phases and 
the SN explosion by the work of Raiteri {\it et al.} \cite{RBG91a,RBG91b,RGB92,RGB93} and 
a number of more recent investigations \cite{WoW95,LSC00,WHW02,TEM07,ETM09,PGH10}.

\begin{figure}[tb]
\includegraphics[width=0.9\textwidth]{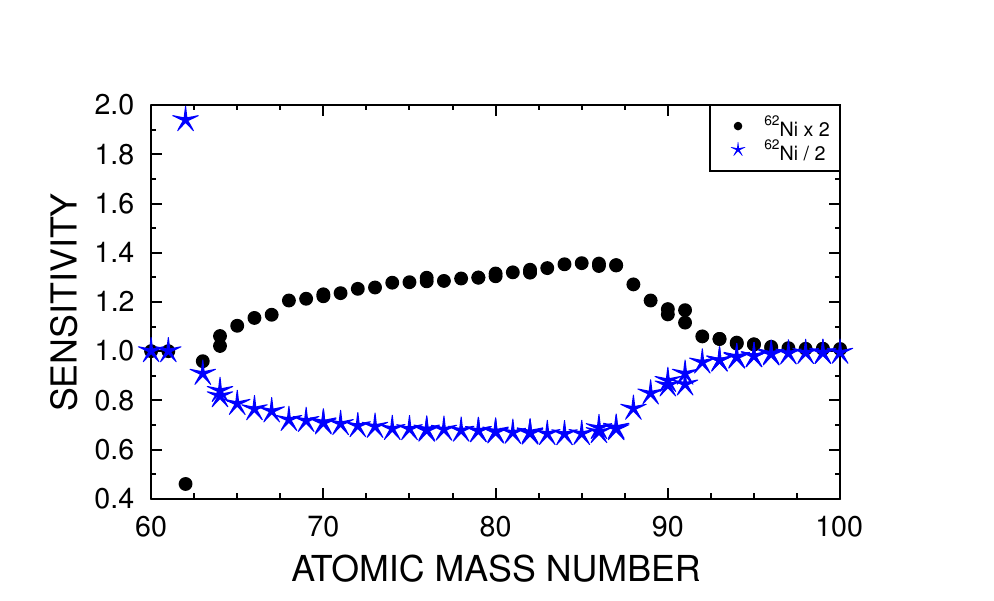}
\caption{The sensitivity to the propagation effect of cross section uncertainties is illustrated at the 
example of $^{62}$Ni. By changing the stellar neutron capture cross section of this single isotope by factors of 
two leads to drastic
changes of the entire $s$ abundances of the weak component.  \label{fig:6} }
\end{figure}

In contrast to the main $s$-process component, the neutron fluence in the weak $s$ process is too 
low for achieving reaction flow equilibrium (Figure~\ref{fig:2}). This has the important consequence that a
particular neutron capture cross section not only determines the abundance of the respective isotope (as in case of the main 
component), but affects the abundances of all heavier isotopes as well \cite{PGH10}. This propagation
effect is particularly critical for the abundant isotopes near the iron seed, where the effect was described
first at the example of the $^{62}$Ni($n, \gamma$)$^{63}$Ni cross section \cite{RaG02,RHH02,RaG05}.
An illustration is given in Figure~\ref{fig:6} by changing the neutron capture cross section of $^{62}$Ni
under stellar conditions by factors of two. 
The problem with this cross section has triggered a whole series of experimental studies on $^{62}$Ni
\cite{NPA05,TTS05,ABE08,LBC13} as well as on other key reactions of the weak $s$ process between 
Fe and Sr (see, e.g., \cite{HKU08a,HKU08b}). For a full account, see the recent update of the KADoNiS 
library \cite{DPK09}.

The cumulated uncertainties of the propagation effect are affecting all isotopes of the weak $s$ process,
up to Kr and Sr, with possible, minor contributions to the Y and Zr abundances \cite{PGH10}. 
Once the neutron-capture cross sections of the isotopes between Fe and Sr will be determinedwith 
the required accuracy of 5\%, the focus on the nuclear input for the weak s process will be on charged-particle reactions during He and C burning, which are difficult to determine (see, e.g., \cite{TEM07,BHP12}).

\subsection{Explosive scenarios: $r$ and $p$ process \label{sec:2.3}}

\begin{figure}[tb]
\includegraphics[width=0.9\textwidth]{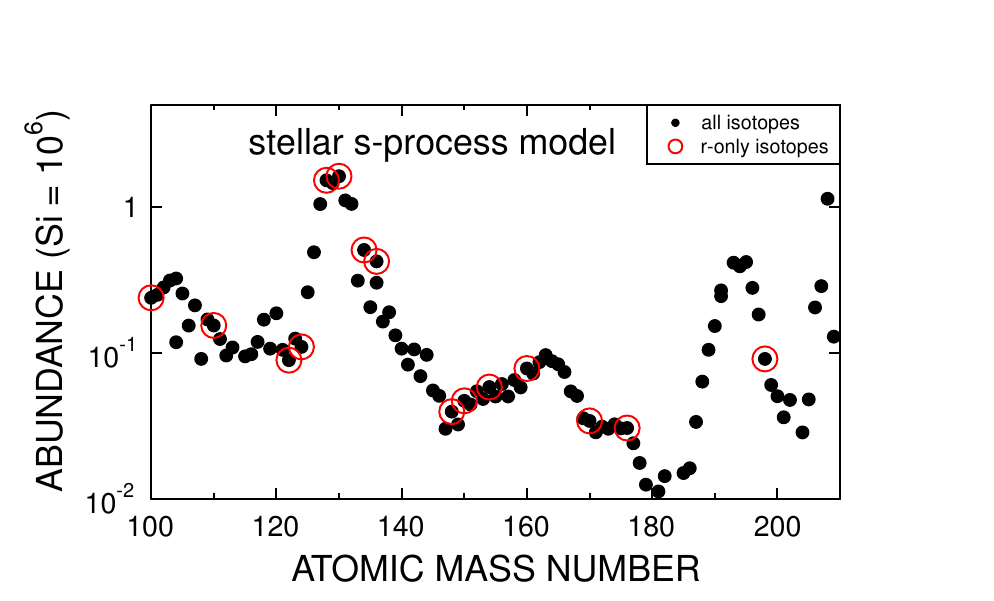}
   		\caption{The $r$-process abundances (open squares) obtained via the $r$-process 
		residual method, $N_r = N_\odot - N_s$, using the $s$-process abundances
		from Ref. \cite{KGB11}. The subset of $r$-only nuclei is marked by open circles. ($N_i$ 
		relative to Si=10$^6$). \label{fig:7}  }
\end{figure}

The second half of the heavy-element abundances in the solar distribution between Fe and the actinides, 
which are produced during  the $r$ process, can be inferred indirectly via the $r$-process residual method, 
i.e. by subtraction of the total $s$-process abundances, $N_r = N_\odot - N_s^{weak} - N_s^{main}$. 
Because the characteristic $r$-process abundance peaks fall in mass regions where the
main $s$ component reached reaction equilibrium, the classical $s$ process is still a reasonable
approximation and there is not much of a difference if the $s$-process part is taken from the classical or 
from the stellar $s$-process prescription used for the distribution shown in Figure~\ref{fig:7}.

Apart from their relevance for testing the $r$-process models sketched below, the $r$-abundance
distribution represents a key aspect for the composition of ultra-metal-poor stars 
\cite{WSG00,BPA00,CSB02,HPC02,CSB05,CoS06,SCG08}, which turned out to agree almost perfectly 
with the scaled solar $r$ component, thus indicating a unique, primary $r$ process, at least for the 
elements heavier than Ba. Significant differences in the mass region below barium require, however,
the assumption of additional $r$-process mechanisms. Accordingly, the quest for the astrophysical site(s) 
of the $r$~process remains an open problem \cite{TAK11}. 

The broad $r$-process abundance peaks at $A=80, 130$, and 195 in Figure~\ref{fig:3} are caused by 
accumulation of matter at closed neutron shells ($N=50,82,126$), far from the stability valley on the 
neutron-rich side. The positions of these peaks suggest conditions that are realized in explosive 
environments with high neutron densities of $n_n>10^{20}$~cm$^{-3}$ and temperatures of 
roughly 1~GK. The high neutron densities are driving the $r$-process path towards the 
neutron drip line until equilibrium between further neutron captures and reverse photodisintegration 
by the energetic photon bath is reached. This equilibrium is determined by the Saha equation in 
each isotopic chain and depends only on neutron density, neutron separation energy and temperature, 
completely independent of the respective ($n, \gamma$) cross sections. 

Under these conditions, the reaction path runs along a path of constant neutron separation energy, 
where matter is accumulated at the so-called waiting points defined by 
($n, \gamma)\Leftrightarrow(\gamma, n$) equilibrium. The $\beta$-decay half-lives of the waiting 
points are determining the respective $r$ abundances. Equivalent to the $s$ process, the steady 
flow condition of the $r$ process is expressed by a constant product of the beta decay rates 
$\lambda$ times the abundances $N_r$, $\lambda(Z)\times N(Z)=$const. Therefore, the only 
nuclear physics properties needed to calculate $r$~process abundances before freeze-out are 
$\beta$ decay half lives and nuclear masses (or neutron separation energies). On the basis of 
experimental $\beta$-decay rates Kratz {\it et al.} \cite{KGH86,Kra88,KBT93} verified that steady 
flow equilibrium is indeed established in the $r$~process. Recently, Wanajo showed that a  "cold 
$r$~process" scenario, operating at lower temperatures without establishing an 
($n,\gamma)\Leftrightarrow(\gamma,n$) equilibrium can also reproduce a solar-like $r$~process 
abundance pattern \cite{Wan07}. 

While the waiting point concept has been applied mostly in static $r$-process calculations, 
models for the currently favored scenarios of neutrino-driven winds from nascent neutron 
stars \cite{WoH92,CoT04,KFP07b}, collapsar scenarios for long-duration gamma ray bursts 
(GRBs) \cite{NKM13}, or neutron star mergers (e.g. \cite{FRT99}) follow the 
full comprehensive reaction network of the $r$ process in much greater detail. Accordingly, 
so far all models are jeopardized by the lack of reliable nuclear physics data far from stability 
comprising the basic information on $\beta$-decay rates \cite{KHH90} and nuclear masses 
\cite{MoN88}, but also neutrino interactions, $\beta$-delayed neutron emission, and 
$\beta$-delayed fission, not to speak of the intricacies of the respective astrophysical scenarios.

As long as neutron capture times are much shorter than $\beta$ decays, the impact of neutron 
capture cross sections is negligible for the $r$-process networks. They are, however, relevant 
for the cold $r$~process, where steady flow equilibrium is not achieved and neutron captures 
compete with $\beta$-decays \cite{Wan07}, and also when in hot $r$ process scenarios the 
reaction flow falls out of equilibrium due to exhaustion of free neutrons. 

In fact, it was shown that the onset of freeze-out is determined by the neutron capture rates 
\cite{Rau05b}. During freeze-out they affect the final $r$ process abundances, e.g. the exact 
position and width of the $r$~process peaks as well as the smoothness of the abundance 
distribution in general \cite{SuE01,ArM11}. Neutron cross sections were shown to affect also
the formation of the rare earth peak at $A\approx160$ \cite{ArM11,SEB97,MMS12}. A number
of key cross sections has been identified in sensitivity studies, e.g., in Refs. \cite{SBM09, BBH09}.  
Variation of the neutron capture rate by an order of magnitude can change the $r$ abundances 
by up to 20\%.  

As a general rule, the neutron separation energy decreases with increasing neutron number for a 
given isotopic chain. Therefore the number of levels in the product nucleus, 
which can be populated directly or indirectly, is also decreasing. In the extreme case of a nucleus 
at the neutron drip line, where the neutron separation energy is close to zero, only the ground state 
can be populated in the product nucleus and the direct radiative capture (DRC) 
is strongly favoured over the resonant (compound nucleus) capture. 
This observation can be roughly summarized by: The more $r$-like the environment, the more dominates the 
DRC over the resonant captures \cite{Gor97b,MMT83}. The crucial parameter, however, is not so much the 
separation energy, but the number of levels available in the product nucleus between ground state and neutron
separation energy. If the level density is low -as for light or neutron-magic nuclei- the DRC might dominate 
even for stable nuclei.

Presently, it is still impossible to study $r$-process related neutron capture rates in direct 
measurements, because the relevant nuclei are way too short-lived. Instead, neutron capture 
cross sections can be obtained by indirect methods. This was demonstrated close to the 
doubly-magic waiting point $^{132}$Sn by Kozub {\it et al.} \cite{KAA12}, who investigated 
the single particle properties of $^{131}$Sn by using the ($d, p$) reaction on $^{130}$Sn 
in inverse kinematics. With this information, the DRC cross section could be calculated 
on the basis of experimental data. Next generation facilities will enable a number 
of new measurements on such critical nuclei. Apart from neutron capture rates, data on neutron 
induced fission play a major role for $r$-process cycling, and are essential to investigate the 
endpoint of the $r$~process \cite{PaT04,PKR11}. More details on indirect methods can be 
found in Sec. \ref{sec:3.4}.

Another nucleosynthesis mechanism, where neutron reactions are involved, is the $p$ process,
which is responsible for the 35 nuclei on the proton rich side of the stability valley, which are 
typically 10-1000 times smaller than the $s$ and $r$~process abundances. A promising site 
for the $p$ process are the explosively burning Ne/O shells in the shock-heated envelopes of
type II supernova explosions (e.g. \cite{WoH78,RAH95}) where a pre-existing 
$s/r$-seed distribution is eroded by photon-induced reactions at temperatures of 2-3~GK, 
resulting in a large network of $(\gamma, n), (n, \gamma), (\gamma, p), (\gamma, \alpha)$ 
reactions and $\beta$-decays that extends a few mass units into the region of the unstable, 
proton-rich nuclei. A difficulty with this approach, however, are the abundances of the light 
$p$~nuclei $^{92,94}$Mo and $^{96,98}$Ru, which are exceptionally high and comparable to
the neighboring $s$ and $r$ isotopes. Therefore, separate production mechanisms have been 
proposed for these cases, such as the $\nu p$~process in neutrino driven winds of SN II ejecta 
\cite{FML06}, the $rp$~procress in hot, proton rich matter that is accreted onto the surface of 
a neutron star \cite{SAG98,SAB01}, and the $\nu$-process \cite{WHH90}.  However,  
Travaglio {\it et al.} showed recently that the $p$ nuclei including the problematic cases 
$^{92,94}$Mo and $^{96,98}$Ru could be consistently produced in Type Ia supernovae 
explosions by treating a full $p$-process network coupled to a two-dimensional supernova 
scenario \cite{TRG11}. 

Although the $p$-process nucleosynthesis is mainly driven by photon-induced reactions, 
($n, p$) reactions can push the nucleosynthesis path towards stability \cite{RGW06} and 
capture of free neutrons released in ($\gamma, n$) reactions may compete with the 
($\gamma, n$) channel \cite{ArG03} and can affect the final $p$ abundances or
can modify even the seed abundances before onset of the $p$ process \cite{RDD13}. 

Experimental possibilities for direct measurement of neutron reactions of relevance for
the $p$ process are facing similar but somewhat relaxed problems than in case of the $r$ 
process, because the $p$-process network is much closer to the stability valley and 
comprises long-lived and even stable isotopes, particularly in the region of the light $p$ 
nuclei. Examples of these cases have been identified in the sensitivity studies of Refs.
\cite{ArG03,RGW06}. For the majority of the crucial competition points in the 
reaction network, however, the specific radioactivity remains a prohibitive obstacle. 
Therefore, the use of indirect methods is called for, similar to the situation on the 
$r$-process side mentioned before. The higher $p$-process
temperatures of 2-3 GK requires that neutron-induced rates need to be known for 
$k_BT$ values of about 180-270~keV \cite{ArG03}.

\subsection{Galactic Chemical Evolution \label{sec:2.4}}

The abundance distribution of the heavy isotopes must be considered as the result of all 
previous stellar generations that were polluting the interstellar medium before the formation 
of the Solar System. Presently, it appears that only the $s$ process can be described with
sufficient confidence to determine the cosmic $s$-process abundances in the framework
of a general galactic chemical evolution (GCE) model.

Such an approach was reported by Travaglio {\it et al.} in a series of papers 
\cite{TGG99,TGB01a,TGB01b,TGA04,SGT09} using a model, in which the Galaxy is subdivided 
into three zones (halo, thick, and thin disk). With the $s$-process yields from a representative
sample of AGB stars of different mass and metallicity, and accounting for their respective 
lifetimes, it was possible to follow  the temporal enrichment of the $s$-process abundances 
in the interstellar medium \cite{TGA04,SGT09}. The resulting $s$-process distribution at 
the time when the solar system formed exhibits good agreement with the solar $s$
abundances between barium and lead, including the solar $^{208}$Pb abundance, which 
could be shown to originate mostly from low-metallicity AGB stars, thus providing a
natural explanation for the strong $s$ component. Below magic neutron number $N=82$, 
however, the abundance distribution obtained by the GCE approach turned out to 
underproduce the solar $s$-process component between the Sr-Y-Zr region and barium 
by about 20 to 30\% \cite{KGB11}. In view of this deficit, an additional type of
neutron-capture nucleosynthesis in the Galaxy, a light element primary process (LEPP) had 
been postulated \cite{TGA04}, different from the $s$ process in AGB stars and also different
from the weak $s$-process in massive stars. The yet unknown origin of the LEPP is still 
a matter of debate \cite{QiW07,FKP10,PGH10,MMB12}. The parameters of the unknown 
process (e.g., the neutron density) have been discussed in \cite{MBC07} for reproducing the
abundance patterns of HD 122563, a very metal-poor star showing a LEPP-type abundance 
pattern \cite{HAI06}. The clarification of the LEPP problem will certainly be furthered by
a general improvement of the neutron capture cross section data between Sr and Ba.

Observational constraints of the galactic chemical evolution can be provided by spectroscopic 
observation of elemental abundances in stars of different metallicities, as they are representative 
for different ages of the galaxy.  In this context, very old, ultra-metal-poor stars in the galactic halo
attracted much interest, because their abundances are characterized by only few nucleosynthesis 
events. As mentioned before, these very old stars exhibit a solar-like $r$~process distribution 
for the  elements heavier than barium \cite{SCL03,BPA00,CSB05,CoS06}, but also a clear deficit in 
the region below. 

Under certain conditions, stars may experience convective-reactive nucleosynthesis episodes. It has 
been shown with hydrodynamic simulations that neutron densities in excess of $10^{15}$~cm$^{-3}$ 
can be reached \cite{HPW11,GZY13}, if unprocessed, H-rich material is convectively mixed with a 
He-burning zone. Under such conditions, which are between the $s$ and $r$ process, the reaction 
flow occurs a few mass units away from the valley of stability. These conditions are sometimes referred 
to as the $i$ process (intermediate process). One of the most important rates, but extremely difficult 
to determine, is the neutron capture cross section of $^{135}$I. While the $^{135}$I($n, \gamma$) cross section is not 
accessible by direct measurements because of the short half-life of about 6~h, the much improved 
production rates of radioactive isotopes at FAIR \cite{Hin10}, however, offer the possibility to investigate 
the Coulomb dissociation cross section of $^{136}$I. This reaction can then in turn be used to constrain 
the $^{135}$I($n, \gamma$) rate. In general, this method is suited for ($n, \gamma$) studies on 
neutron rich isotopes, because the decreasing neutron separation energies are favoring the DRC 
channel compared to the more complex compound mechanism as mentioned before (see also Sec. 
\ref{sec:3.4}). The neutron separation energy of $^{136}$I is only 3.8~MeV, since  $^{136}$I is neutron magic.

\section{Neutron reactions \label{sec:3}}

\subsection{Definitions and current database \label{sec:3.1}}

In stellar interiors, the plasma is considered to be in thermodynamic equilibrium. Hence, particles are quickly 
thermalized and their velocities $v_x$ follow a Maxwell-Boltzmann distribution, which only depends on the 
mass $m_x$ of the particle and on temperature $T$:
\begin{equation}
\phi(v_x)dv_x=4\pi v_x^2(\frac{m_x}{2\pi k_BT})^{3/2}\exp(-\frac{m_xv_x^2}{2k_BT})dv_x.
\end{equation}
Here, $\phi(v_x)dv_x$ denotes the probability to find a particle with a velocity between $v_x$ and $v_x+dv_x$,  
and $k_B$ is the Boltzmann constant. For interacting nuclei $X$ and $Y$ the reaction rate per particle 
pair can then be written as:
\begin{equation}
<\sigma v>=\int_{0}^{\infty}{\phi(v)\sigma(v)vdv}=4\pi(\frac{\mu}{2\pi k_BT})^{3/2}\int_{0}^{\infty}\sigma(v)v^3\exp(-\frac{\mu v^2}{k_BT})dv,
\end{equation}
where $v$ is the relative velocity between $X$ and $Y$ and $\mu=m_xm_y/(m_x+m_y)$ the reduced 
mass. 
Maxwellian average cross sections (MACS) are defined as the reaction rate scaled by the most 
probable velocity $v_T=\sqrt{k_BT/\mu}$ of the Maxwell-Boltzmann distribution:
\begin{equation}
 <\sigma>=\frac{<\sigma v>}{v_{T}}=\frac{2}{\sqrt{\pi}}\frac{1}{(k_BT)^{2}}\int_{0}^{\infty}\sigma(E)E\exp(-\frac{E}{k_BT})dE
\label{macsequation}
\end{equation}

According to the temperatures at the various $s$-process sites, MACS values need to be known for thermal
energies $k_BT= 8$ to 90~keV. Therefore, neutron capture cross sections should be known up to several hundred 
keV, keeping in mind that the energy range above 400~keV contributes only about 1\% to the thermal spectrum
at $k_BT=90$~keV. The effect on the MACS is even smaller because the cross sections are typically decreasing with 
increasing neutron energies. The situation is similar for $r$-process nucleosynthesis, where temperatures are typically $\leq1$~GK, 
corresponding to $k_BT$ values of at most 90~keV. The $p$ process operates at higher temperatures, with $k_BT$ 
values up to 270~keV, requiring neutron cross sections up to at least 1~MeV.

In a first compilation by Macklin and Gibbons \cite{MaG65} MACS data were calculated from $k_BT=5-90$~keV 
using the neutron capture information up to 1965. In the following decades, new and better experimental 
techniques became available, resulting in a wealth of new and more accurate measurements. Thanks to higher 
neutron fluxes and better detectors, measurements on many new isotopes were performed with significantly
improved uncertainties, reaching a few percent in favorable cases. These data were collected in further MACS
compilations by K\"appeler {\it et al.} \cite{KBW89} and Bao {\it et al.} \cite{BBK00}. The latter became available
online as the "Kadonis Astrophysical Database of Nucleosynthesis in Stars" (KADoNiS) and has been continuously
updated to the present version v0.3 \cite{DPK09}. Recently, charged particle reaction rates for the astrophysical 
p~process have been added as well \cite{SDP10}. 

Neutron cross sections for $s$-process studies are determined by two complementary methods. The time-of-flight 
(TOF) and the activation technique require both the production of a neutron beam. Because neutrons are 
unstable, it is not (yet) possible to perform experiments in inverse kinematics. Therefore, experiments are 
essentially limited to direct measurements of ($n, \gamma$) cross section on reasonably long-lived samples,
but the potential for neutron capture studies via the inverse ($\gamma , n$) channel will be briefly touched 
as well. 
 
\subsection{Time-of-flight experiments  \label{sec:3.2}}

\begin{figure}[htb]
\begin{center}
\includegraphics[width=10.0cm,angle=00]{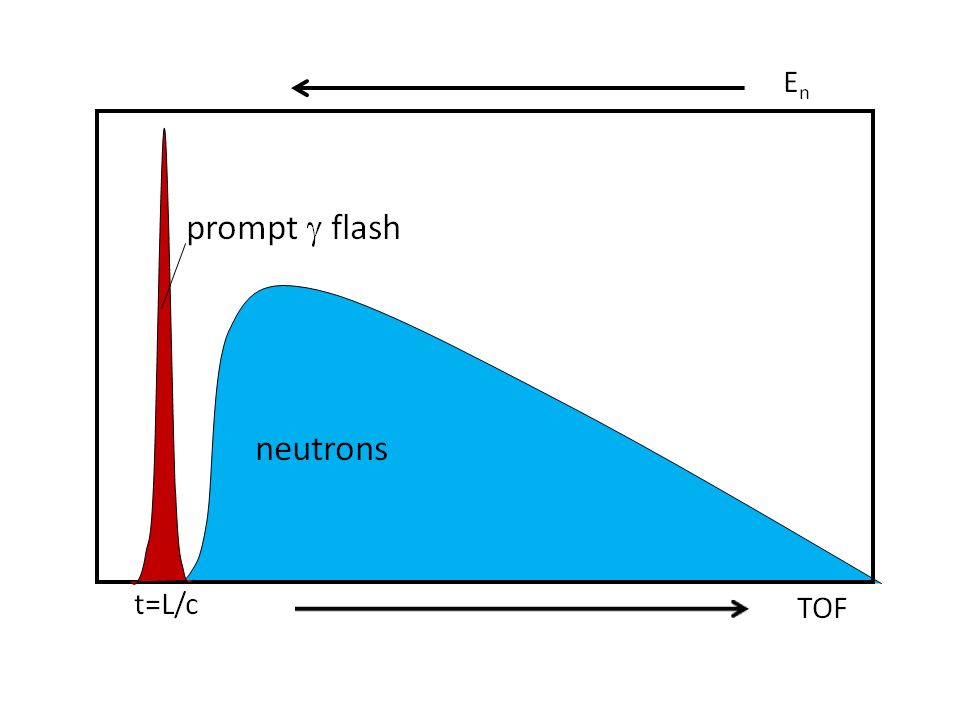}
\caption{\label{fig:8} Schematic time-of-flight spectrum. The sharp peak at $t=L/c$ is caused 
by prompt photons produced by the impact of a particle pulse on the target. Neutrons reach the 
measurement station at later times and give rise to a broad distribution depending on their initial 
energies.}
\end{center}
\end{figure}

The TOF method enables cross section measurements as a function of neutron energy. 
Neutrons are produced quasi-simultaneously by a pulsed particle beam, thus allowing
one to determine the neutron flight time $t$ from the production target to the sample 
where the reaction takes place. For a flight path $L$, the neutron energy is 
\begin{equation}
 E_{n}=m_nc^{2}(\gamma-1)
\label{tofformularel}
\end{equation}
where $m_n$ is the neutron mass and $c$ the speed of light. The relativistic correction 
$\gamma=\left(\sqrt{1-(L/t)^2/c^2}\right)^{-1}$ can usually be neglected in the neutron 
energy range of interest in nucleosynthesis studies and Eq. \ref{tofformularel}
reduces to 
\begin{equation}
 E_{n}=\frac{1}{2}m_n\left(\frac{L}{t}\right)^2.
\label{tofformula}
\end{equation}
The TOF method requires that the neutrons are produced at well defined times. This is achieved by
irradiation of an appropriate neutron production target with a fast-pulsed beam from particle 
accelerators. The TOF spectrum measured at a certain distance from the target is sketched
in Figure~\ref{fig:8}. The essential features are a sharp peak at $t=L/c$, the so-called 
$\gamma$-flash due to prompt photons produced by the impact of a particle pulse on the 
target, followed by a broad distribution of events when the neutrons arrive at the sample position,
corresponding to the initial neutron energy spectrum.  

Neutron TOF facilities are mainly characterized by two features, the energy resolution $\Delta E_n$ 
and the flux $\phi$. The neutron energy resolution is determined by the uncertainties of the flight path 
$L$ and of the neutron flight time $t$:

\begin{equation}
\frac{\Delta E_n}{E_n}=2\sqrt{\frac{\Delta t^{2}}{t^{2}}+\frac{\Delta L^{2}}{L^{2}}} 
\end{equation}

The neutron energy resolution can be improved by increasing the flight path, but only at the expense of
the neutron flux, which scales with $1/L^2$. The ideal combination of energy resolution and neutron 
flux is, therefore, always an appropriate compromise. The energy resolution is affected by the Doppler 
broadening due to the thermal motion of the nuclei in the sample, by the pulse width of the particle 
beam used for neutron neutron production, by the uncertainty of the flight path including the size of 
the production target, and by the energy resolution of the detector system. 

The by far most important type of neutron reactions in astrophysics are neutron captures, which constitute
the main production mechanism for the formation of the heavier chemical elements from iron to the actinides.
The corresponding cross sections usually show a strong resonant structure, caused by the existence of excited 
levels in the compound nucleus. The excitation function for a reaction can accordingly be divided into three regions, 
the resonance region, where the experimental setup allows to identify individual resonances, the unresolved 
resonance region, where the average level spacing is still larger than the natural resonance widths, and the 
continuum region, where resonances start to overlap. The border between the first two regions is determined
by the average level spacing and by the TOF resolution of the experiment. 

Beside the resonant component related to the formation of a compound nucleus, there is also a small direct 
component to the cross section, which corresponds to events where the neutron is directly captured into the
final state with cocomitant emission of high-energy $\gamma$-rays. Because of its smooth dependence on 
neutron energy the direct component is not easily distinguishable from experimental background and, 
therefore, hard to identify in the resonance region. Direct capture is important for light as well as for the 
very neutron-rich, heavier nuclei involved in the $r$-process network, because
in these cases the compound contributions are very small due to the low level densities in these nuclei. 

In TOF measurements, capture cross sections are determined via the prompt $\gamma$-ray cascade emitted 
in the decay of the compound nucleus. The most common detection principles for measuring neutron capture
cross sections are the use of total absorption calorimeters or total energy detection systems. 

The  total energy technique is based on a device with a $\gamma$-ray detection efficiency,
$\varepsilon_\gamma$, proportional to $\gamma$ energy, $E_\gamma$.  Provided that the overall 
efficiency is low and that no more than one $\gamma$ is detected per event, the efficiency for detecting 
a capture event becomes - on average - independent of cascade multiplicity and de-excitation pattern, 
but depends only on the excitation energy of the compound nucleus, i.e. the neutron separation energy. 
A detector with an intrinsic proportionality of $E_\gamma$ and $\varepsilon_\gamma$ was first 
developed by Moxon and Rae \cite{MoR63} by combining a $\gamma$-to-electron converter
with a thin plastic scintillator. Because of this conversion, Moxon-Rae detectors are essentially insensitive 
to low-energy $\gamma$ rays and were, therefore, used in TOF measurements on radioactive samples 
\cite{WiK78,WiK79a}. The efficiency of Moxon-Rae detectors for capture events is typically less than a few 
percent.

Higher efficiencies of about 20\% can be obtained by an extension of the Moxon-Rae principle originally 
proposed by Maier-Leibnitz \cite{Rau63,MaG67b}. In these total energy detection systems the proportionality 
$E_\gamma$ - $\varepsilon_\gamma$ is extrinsically realized by an $a~posteriori$ treatment of the recorded 
pulse-height. This Pulse Height Weighting technique is commonly used with liquid scintillation detectors 
about one liter in volume, small enough for the condition to detect only one $\gamma$ per cascade. Present
applications at neutron facilities n\_TOF (Cern, Switzerland) and at GELINA (IRMM, Belgium) are using deuterated 
benzene (C$_6$D$_6$) as scintillator
because of its low sensitivity to scattered neutrons. Initially, the background due to scattered neutrons had 
been underestimated, resulting in overestimated cross sections, particularly in cases with large 
scattering-to-capture ratios as pointed out by Koehler {\it et al.} \cite{KWG00} and Guber {\it et al.} 
\cite{GLS05,GLS05b}. With an optimized design, an extremely low neutron sensitivity of 
$\varepsilon_n/\varepsilon_\gamma\approx3\times10^{-5}$  has been obtained at n\_TOF \cite{PHK03},
which is especially important for light and intermediate-mass nuclei, where elastic scattering usually 
dominates the capture channel. 

A total absorption calorimeter consists of a set of detectors arranged in 4$\pi$ geometry, thus covering the 
maximum solid angle. Because the efficiency for a single $\gamma$-ray of the capture cascade is usually 
close to 100\% in such arrays, capture events are characterized by signals corresponding essentially to the 
Q-value of the reaction. Provided good resolution in $\gamma$ energy, gating on the Q-value represents,
therefore, a possibility of significant background suppression. 

Total absorption calorimeters exist at several TOF facilities. Most are using BaF$_2$ as scintillator material, 
which combines excellent timing properties, fairly good energy resolution, and low sensitivity to neutrons 
scattered in the sample. In fact, neutron scattering dominates the background in calorimeter-type detectors, 
because the keV-cross sections for scattering are typically 10 to 100 times larger than for neutron capture.
In measurements at moderated neutron sources this background is usually reduced by an absorber surrounding 
the sample. Such a detector has been realized first at the Karlsruhe Van de Graaff accelerator \cite{WGK90a}.
This design, which consists of 42 crystals, is also used at the n\_TOF facility at CERN \cite{GAA09}, while a
geometry with 160 crystals has been adopted for the DANCE detector at Los Alamos \cite{HRF01,RBA04}. In specific 
neutron spectra, e.g. in measurements with the Karlsruhe array, where the maximum neutron energy was 
about 200~keV, scattered neutrons can be partially discriminated via TOF between sample and scintillators, 
because the capture $\gamma$ rays reach the detector before the scattered neutrons \cite{WGK90a}.
There are also $4\pi$ arrays made of NaI crystals \cite{MAA85, BDS94}, but in the astrophysically important 
keV region these detectors are suffering from the background induced by scattered neutrons, which are easily 
captured in the iodine of the scintillator.

At white neutron sources, the highest neutron energy for which the neutron capture cross section can be 
determined is limited by the recovery time of the detectors after the $\gamma$~flash. While the accessible 
neutron energy range is practically not restricted for C$_6$D$_6$ detectors, BaF$_2$ arrays are more 
sensitive, depending on the respective neutron source. At n\_TOF, for example, the BaF$_2$ calorimeter 
can be used only up to few tens of keV, whereas there are no strong limitations at Los Alamos. Independent 
of the detection system used, measurements at higher neutron energies are increasingly difficult because 
the ($n, \gamma$) cross section decreases with neutron energy, while at the same time competing reaction 
channels, e.g. inelastic neutron scattering, are becoming stronger. Nevertheless, present techniques are 
covering the entire energy range of astrophysical relevance up to about 500 keV  with sufficient accuracy.

Apart from neutron capture, there are also other neutron-induced reactions of relevance for stellar 
nucleosynthesis. While neutron-induced fission of the very short-lived isotopes involved in the $r$ 
process is presently inaccessible to experiments, the impact of ($n, p$) and ($n, \alpha$) reactions 
can dominate over neutron capture, especially in the mass region below Fe. Examples are the neutron 
poison reaction $^{14}$N($n, p$)$^{14}$C \cite{BBR88,KoB89} and the recycling reaction
 $^{33}$S($n, \alpha$)$^{30}$Si impeding the $s$ production of $^{36}$S
\cite{SJL95,WWB87}.

Ionization chambers are popular for cross section measurements of reactions with charged particles ($cp$) 
in the exit channel. These detectors have the advantage that they are rather insensitive to neutrons and 
can be operated directly inside the neutron beam, thus covering a large solid angle. Different designs have 
been used for astrophysical cross section measurements at various TOF facilities, for example Frisch-gridded 
chambers in ($n, cp$) measurements at Karlsruhe and GELINA, (e.g. \cite{SJL95} and \cite{DWW07}, 
respectively) or a compensated chamber at ORELA \cite{GKA00}. Among the current gas detectors the 
Microbulk Micromegas type \cite{AAB10} is a fairly new development, which is based on the MicroMegas 
principle \cite{GRR96} and has the advantage of fast timing and the capability for spatial resolution 
\cite{ACK11}. The very low material budget of this type allows one to share the same neutron beam line with 
another experiment by operating the detector in parasitic mode, in parallel to other experiments. Its low 
sensitivity to the $\gamma$~flash and its comparatively high gain make the detector particularly suited for 
measurements of ($n, cp$) cross sections, where the ejectile energies are typically a few MeV only. 

Fast ionization chambers are frequently used in fission cross sections measurements, where background
problems are relaxed because of the comparably high Q-values (e.g. Ref. \cite{CCK08}). More detailed 
information on the fission process can be obtained with parallel plate avalanche counters \cite{TBD05}, 
for example on the fission fragment angular distribution \cite{TLA14}. Further advantages of these 
detectors are excellent time and spatial resolution and their practically negligible response to the 
$\gamma$ flash.

Compared to gas detectors, silicon detectors are superior in energy resolution and better suited 
for particle identification in TOF experiments. They are, however, sensitive to neutron damage, 
and are, therefore, usually operated outside the neutron beam. Silicon detectors have been 
successfully used in measurements of ($n, p$) and ($n, \alpha$) cross sections (see, e.g. 
\cite{KoB89,KoG91}) as well as for monitoring the neutron flux \cite{MMA04}.  Recently, a chemical 
vapor deposited (CVD) diamond detector has been developed at n\_TOF for measurements
of ($n, cp$) cross sections \cite{WGG13}. Diamond detectors are radiation hard and can be 
operated almost background-free even directly in the neutron beam. However, their energy 
resolution does not reach that of silicon detectors.

\subsection{The activation method\label{sec:3.3}}

To obtain stellar cross sections from an activation experiment, the neutron spectrum should 
ideally correspond to the perfectly thermal spectrum at the respective $s$-process site. By
serendipity, the $^7$Li($p, n$)$^7$Be reaction, which represents the most prolific neutron source 
at low energy accelerators fulfills this requirement almost perfectly \cite{RaK88,LKM12,FFK12}. 
Adjusting the proton energy at $E_p=1912$~keV, 30 keV above the reaction threshold, yields 
a neutron spectrum with an energy dependence close to 
\begin{equation}
\Phi_n(E)=E\exp(-\frac{E}{k_BT})
\end{equation}
as required in the MACS definition of Eq. \ref{macsequation} for the case of $k_BT=25$~keV
almost perfectly mimicking the situation during He shell flashes in AGB stars (Sec. \ref{sec:2.1}).

This quasi-stellar neutron source is ideally suited for the determination MACS data for a number 
reasons:

(i) The experimental setup is rather simple. Because the proton energy is just above the reaction 
threshold, all neutrons are emitted in a forward cone with an opening angle of 120 degrees. 
Because the accelerator can be operated in direct mode and the sample can be placed directly 
at the neutron production target, the neutron flux in activation
measurements is orders of magnitude higher than can be achieved in TOF experiments. 
The higher flux results in an enormous gain in sensitivity, thus allowing for measurements of very 
small cross sections and/or on very small samples. (ii) Samples can be used in natural isotopic 
composition or even as compounds, because the reaction products are detected off-line in a 
low-background environment via their characteristic decay modes. In favorable cases this allows 
one to measure the cross sections of several isotopes at the same time or to determine the partial 
cross sections for feeding isomeric states, an aspect that is important for the reaction flow in some 
of the branchings. (iii) Backgrounds are strongly reduced. For example, the effect of scattered 
neutrons and the ambient $\gamma$ background, which is immanent to TOF experiments, are 
completely eliminated. Sample-related effects due to multiple scattering and self-shielding can be 
avoided by using very small samples. In general, the comparably short measurement times allows 
one to study systematic uncertainties experimentally by repeated activations under modified conditions. 

However, the method is limited to cases where neutron capture leads to an unstable isotope with a 
half-life longer than about a second. Moreover, the quasi-stellar spectrum is not exactly thermal and the
respective corrections (which are usually very small though)  require the knowledge of the energy dependence 
of the cross sections. A more severe problem results from the fact that quasi-stellar spectra can only be 
produced for very few temperatures. Apart from the $^7$Li($p, n$)$^7$Be reaction for $k_BT=25$~keV, 
other possibilities are provided by ($p, n$) reactions on $^{18}$O for $k_BT=5$~keV 
\cite{HDJ05} and $^3$H for $k_BT=52$~keV \cite{KNA87}, both with significantly reduced intensities. This 
means that the MACS determined via activation may have to be inter- or extrapolated to cover the full 
range of $s$-process temperatures. Unless the energy-dependence of the cross section is well 
defined experimentally, this procedure may cause sizable uncertainties.

In view of its attractive features, the activation technique has been extensively used at the 
Karlsruhe Van de Graaff. Across the nuclear chart, MACS measurements have been
carried out on 120 isotopes according to the KADoNiS compilation \cite{DPK09}. Typically, the 
induced activity can be measured via the $\gamma$ emission of the product nucleus, which implies 
favorable signal-to-background ratios and the unambiguous identification of the reaction products. 
The excellent selectivity achieved in this way can often be used to study more than one reaction 
in a single irradiation. A prominent example is Xe, where five cross sections can be measured 
simultaneously \cite{Bee91}. If suited compounds are used for the sample material, cross sections 
of all constituents can be determined as in case of RbBr, where MACS data for all four stable 
isotopes could be obtained \cite{HKU08b}.

The technique allows very good accuracy as demonstrated  by the 1.4\% uncertainty
claimed for the MACS of $^{197}$Au \cite{RaK88}. Examples for the unique sensitivity are
measurements of very small cross sections at the $\mu$b-level, e.g. on $^{15}$N with a 
MACS-value of only 6 $\mu$b for $k_BT = 25$ keV. Similarly small cross sections were obtained 
for $^{14}$C \cite{RHF08} and $^{18}$O \cite{MSG96}. Both are particular, because the short 
half-lives of the reaction products required the use of a fast cyclic activation scheme \cite{BRW94}.

The excellent sensitivity of the activation method was mandatory for measurements on
very small samples of the unstable branch-point isotopes $^{135}$Cs \cite{PDA04}, $^{147}$Pm 
\cite{RAH03}, $^{155}$Eu \cite{JaK95b}, $^{163}$Ho \cite{JaK96a}, $^{171}$Tm \cite{RHH03},
and $^{182}$Hf \cite{VDH07}. Among these cases, the 28 ng of $^{147}$Pm represent the
smallest sample ever used in an ($n, \gamma$) cross section measurement in the keV region. 
In dealing with MACS measurements on unstable isotopes, the possibility for using sample sizes 
of a few $\mu$g or even less is particularly crucial, because these materials are hard to produce.
Another important aspect in using small radioactive samples is that it allows one to keep the radiation 
hazards and the related backgrounds at a manageable level. Here, $^{60}$Fe deserves special 
mentioning as the measurement of this MACS is complicated by the small ($n,\gamma$) cross section
as well as by the difficulty in obtaining a suitable sample \cite{URS09,SND10}.

While fast cyclic activation is called for in case of short-lived reaction products, long-lived product 
nuclei are more difficult to deal with, in particular if the induced activities are near or below
the detection limit. If decay counting of the reaction product becomes too difficult, the reaction 
products can be directly counted using Accelerator Mass Spectrometry (AMS), an ultra-sensitive 
technique to measure isotopic ratios as small as $10^{-16}$. AMS was first been applied for 
cross section measurements by Paul et al. for the $^{25}$Mg($p, n$)$^{26}$Al reaction 
\cite{PHK80}. A prominent example for an ($n, \gamma$) reaction is the case of $^{62}$Ni,
where the product nucleus $^{63}$Ni has not only a half-life of 101 years, but decays without 
any $\gamma$ emission and a very low $\beta$-endpoint energy of 70 keV \cite{NPA05,DFK10}. 
Meanwhile, AMS has been applied successfully for MACS measurements on $^9$Be and $^{13}$C 
\cite{WCD08}, $^{14}$N \cite{WBD12}, $^{40}$Ca \cite{DHK09}, $^{54}$Fe \cite{CDW06}, 
$^{58}$Ni \cite{RDF07}, and $^{78}$Se \cite{DHK06a}. 

\subsection{Indirect measurements \label{sec:3.4}}

Depending on the reaction type the corresponding indirect approaches are very different. Within this 
review, only the techniques important for neutron studies, i.e. inverse reactions and surrogate reactions, 
will be discussed. 

The Coulomb dissociation (CD) method can be used to determine
the desired cross section  of the reaction A($n, \gamma$)B via the inverse reaction B($\gamma, n$)A
by applying the detailed balance theorem.  It has been shown that this method can be successfully
applied, if the structure of the involved nuclei is not too complicated, as in the case of the reaction 
$^{14}$C($n, \gamma$)$^{15}$C \cite{RHF08}. If the reaction product B is short-lived, the 
CD method can be applied at radioactive beam facilities \cite{NFA09,DAB03}. 

Limitations of the CD method are (i) the applicability of this method to heavier nuclei close to the 
valley of stability due to the high level density in the compound nucleus, and (ii) because the 
resolution of current facilities of $\geq100$~keV is not sufficient to constrain the astrophysically 
relevant cross section.

Restriction (i) is alleviated in $r$-process studies, because the level density is rapidly decreasing
as the Q-values drops towards the neutron drip line. This implies that fewer levels are important,
and the part of the capture cross section, which can be constrained via the inverse reaction, increases. 
Restriction (ii) motivated the 
development of improved experimental approaches such as NeuLAND@FAIR, which aims for an 
energy resolution of better than 50~keV \cite{BAA11}.

If the product is stable or very long-lived, also real photons can be used to study B($\gamma, n$)A
reactions \cite{SMV03,Wel08}. In principle the same restrictions apply as for the Coulomb dissociation method.

Surrogate reactions have been successfully used for obtaining neutron-induced fission cross sections
\cite{EAB05}. This approach is using a charged particle reaction for producing the same compound
system as in the neutron reaction of interest. In this way, a short-lived target isotope can be replaced by 
a stable or longer-lived target. For neutron capture reactions, however, the method is challenged 
because the compound nucleus that is produced in the surrogate reaction is characterized by a
spin-parity distribution that can be very different from the spin-parity distribution of the compound 
nucleus occurring in the direct ($n, \gamma$) reaction \cite{FDE07,EBD12}.
 
\section{Facilities and achievements \label{sec:4}}

\subsection{White neutron sources \label{sec:4.1}}

Commonly, the term "white" refers to accelerator-based neutron sources, where an initially hard 
neutron spectrum region is softened by a moderator for producing a continuous neutron spectrum 
covering the energy range from the high production energies in the MeV to GeV region down to 
thermal values.

The highest neutron yields are obtained in spallation reactions of high-energy beams. As spallation 
neutrons are very energetic, a moderator near the neutron production target is needed to shift the 
spectrum to the lower neutron energies of astrophysical interest. TOF facilities at spallation 
neutron sources are presently operating at CERN \cite{AAA01,CCV12, GTB13}, at Los Alamos National 
Laboratory \cite{LBR90} and at J-PARC \cite{JPC04}. 

At n\_TOF, intense 20 GeV pulses of $7\times10^{12}$ protons are producing about $2\times10^{15}$ 
neutrons per pulse in a massive lead target, corresponding to 300 neutrons per incident proton. The 
target is cooled with water, which acts also as moderator. The resulting neutron spectrum ranges from 
thermal energies of 25~meV to  a few GeV.  The flight path of 185~m and the proton pulse width of 
7~ns, offer a very good energy resolution. During the approximately 10 years of successful operation, 
a large number of $(n, \gamma$) reactions of stable nuclei on the $s$~process path has been studied 
with C$_6$D$_6$ scintillation detectors (Sec. \ref{sec:3.2}), i.e. isotopes of Fe and Ni 
\cite{Led13}, Zr \cite{TMF11a,TMF11b,TMF13}, La \cite{TAA07}, Au \cite{LCD11}, Pb \cite{DAA07b}, 
and Bi \cite{DAA06a} as well as the potential neutron poison isotopes of magnesium \cite{MKB12}. 
Particular projects are the capture cross sections of $^{186,187,188}$Os for determining the age of 
the universe via the Re/Os clock \cite{MFM10,MHK10,FMM10} and the cross sections of the unstable 
branch point isotopes $^{63}$Ni ($t_{1/2}=101.5$ yr) \cite{LMA13} and $^{151}$Sm ($t_{1/2}=93$ yr) 
\cite{AAA04c,MAA06}. In addition, the ($n, \alpha$) cross section of the long-lived isotope $^{59}$Ni
($t_{1/2}=76$ kyr) has recently been determined with a special CVD diamond mosaic-detector 
\cite{WGG13}. 

At the Los Alamos Neutron Science Center (LANSCE), a proton beam with an energy of 800~MeV 
reacts with a W target. The total neutron yield at the target station is smaller than at n\_TOF, but 
the time-integrated neutron flux at the sample position is on average about 20 times higher than 
at n\_TOF due to a shorter flight path of 
20~m and a higher repetition rate of 20 Hz (vs. 0.4 Hz at n\_TOF). On the other hand, the proton 
pulse width of 120~ns is limiting the energy resolution in the keV region, and as a consequence, 
only average cross sections can be determined above about 10~keV. Measurements of astrophysical 
interest started with ($n, p$) cross sections \cite{KoB89,KGO93,KKV97} and have been later extended
to neutron capture data for $^{62,63}$Ni \cite{ABE08,WBC12} and $^{147,151}$Sm \cite{KRU12,RAA06}.

At the Material and Life science experimental facility at J-PARC in Japan, an intense, pulsed neutron
beam is produced via spallation of 3~GeV protons on a mercury target. Besides the focus on material 
and life sciences, one beam line at a distance of 22~m from the target is dedicated to neutron cross 
section measurements, using the Accurate Neutron-Nucleus Reaction Measurement Instrument (ANNRI). 
ANNRI consists of 2 detection setups, a 4$\pi$ Ge detector array and a NaI(Tl) spectrometer.  Neutron 
fluxes at the current beam power of 120 kW are  about a factor then higher than at the 20~m station at 
LANSCE. The final beam power of 1 MW will boost the flux by another factor of 10. Due to the long 
pulse width of 1 $\mu$s the neutron energy resolution is deteriorating already at several hundred eV. 
Current efforts concentrate on neutron capture measurements on minor actinides and long-lived fission 
products, e.g. $^{244,246}$Cm \cite{KFG11}.

Electron accelerators have been used for neutron production via bremsstrahlung-induced $(\gamma, f)$ 
and $(\gamma, n)$ reactions by irradiation of high-$Z$ targets with pulsed electron beams. A prominent 
example was the Oak Ridge Electron Linear Accelerator (ORELA) \cite{orela}, where important experimental 
techniques have been developed \cite{MaG67b,MHW79} and where most of the initial MACS data for 
$s$-process studies have been determined in numerous measurements by Macklin and collaborators 
(see, e.g., \cite{MHW75,MHW77,MaY87b,Mac88}, including first studies on unstable isotopes 
\cite{Mac82b,Mac83,Mac85a,Mac85b}. After an update of the experimental setup, astrophysics at ORELA
was resumed by Koehler {\it et al.} with ($n, \gamma$) and ($n, \alpha$) measurements on isotopes of 
Sr, Mo, Ba, Sm, and Pt \cite{KSW97,KSG98,KWG01a,RKK03,KGR04,KoG13}. 

When ORELA was closed in 2012, GELINA at Geel/Belgium \cite{BeS78,BDG09} remained the only 
electron-driven, moderated facility for neutron studies in the energy regime of the $s$ process. The 
accelerator delivers a 140~MeV electron beam with a pulse width of 1~ns onto a rotating U target, 
and the water-moderated neutron spectra from 25~meV up to 20~MeV can be accessed at 10 flight 
paths from 10 to 400~m.  The astrophysics activities are essentially carried by visiting groups 
and are devoted to $(n, \gamma)$ cross sections of isotopes on the $s$~process path, e.g., Si 
\cite{GKD03}, Kr \cite{MBB05}, Nd \cite{GSK97a}, and Ba/Pb \cite{BCM97}.  As a complement 
to the ($n, \gamma$) studies, ($n ,\alpha$) and ($n, p$) cross sections were measured for 
$^{17}$O \cite{WWG02},  $^{26}$Al \cite{DWW07} and for $^{36}$Cl \cite{DWG07}, respectively.  

At the ELBE accelerator at Dresden-Rossendorf a 40 MeV electron beam on an unmoderated liquid 
metal target is used to operate a very compact neutron source characterized by a remarkably high 
TOF resolution \cite{BBE13}. Applications  in astrophysics are hampered, however, by the hard 
neutron spectrum that is concentrated in the energy range from 0.2 to 10 MeV with an intensity 
maximum around 1 MeV, too high for most astrophysical $(n, \gamma$) measurements. 

About a factor of $10^4$ higher neutron fluxes than at conventional TOF facilities can be produced 
with lead slowing down spectrometers (LSDS), however, at the expense of neutron energy resolution, 
which is $>30$\% (FWHM). An LSDS consists of a pulsed neutron source surrounded by high purity 
lead blocks, typcally 1 m$^3$ in volume. Neutrons get trapped in the lead for several hundreds of 
microseconds and are slowed down by inelastic and elastic scattering. Losses due to neutron capture 
on Pb are very small, because the capture cross sections are orders of magnitude smaller than 
for scattering. At LSDS facilities cross section measurements can be performed on minute samples 
of a few tens of ng, and are, therefore, well suited for measurements on short-lived isotopes and/or 
samples of high specific activity. The first LSDS was built in the 1950ies \cite{Ber55}, and the concept 
has been revived, e.g. at CERN \cite{AAA02}, Los Alamos \cite{RHO05}, and KURRI \cite{YKK93}. 
Neutron cross section measurements with LSDS spectrometers have been mainly used for fission and 
capture cross section measurements for reactor applications (see e.g. \cite{MSB85,KLY02}), but have 
potential for measurements on $s$~process branching points as well. 

\subsection{Low-energy accelerators \label{sec:4.2}}

Neutron production via ($p, n$) reactions at low-energy electrostatic accelerators was the initial
source of information on stellar cross sections \cite{MaG65}. When electron linacs came into 
routine operation in the late 1960ies most experimental activities were attracted by the higher 
flux and better energy resolution of the new neutron sources.  About 15 years later, however, the 
specific advantages of low-energy machines led to a revival, particularly for applications in
astrophysics.

Fast pulsed electrostatic accelerators are important because they are complementary 
to white neutron sources due to a combination of unique features, i.e. the possibility
to tailor the neutron spectrum, low backgrounds, and competitive flux at sufficient TOF 
resolution. Neutron spectra can be restricted to the immediate region 
of interest for $s$-process applications by the proper choice of the proton energy and
the thickness of the neutron production target. For the example of the $^7$Li($p, n$)$^7$Be 
reaction, which has been used in most cases because it represents the most prolific ($p, n$) 
source, neutrons in the energy range from a few to 220 keV are produced by bombarding 
layers of metallic Li or LiF typically 10 to 30 $\mu$m in thickness. With proton beams 
30 and 100 keV above the reaction threshold at 1881 keV, continuous neutron spectra 
with maximum energies of 100 and 225 keV are obtained, where the first choice offers a 
significantly better signal to background ratio at lower neutron energies. The neutron 
production target consists only of a thin Li layer on a comparably thin backing without 
any moderator, thus allowing for very short flight paths down to a few centimeters. In this 
way, neutron intensities at the position of the sample are becoming comparable to those of 
white neutron sources in spite of the lower neutron production yields and higher repetition 
rates \cite{RHH04}. And, not to forget, low-energy accelerators bear the option for activation measurements 
in quasi-stellar neutron spectra (Sec. \ref{sec:3.3}).

Thanks to these options, particularly the Karlsruhe Van de Graaff has been extensively 
used for astrophysics studies. In the optimized TOF mode, typical beam parameters were an
average proton current of 2 $\mu$A at a repetition rate of 250 kHz, and a pulse width of 0.7 ns,
perfectly matching the 0.5 ns time resolution of the 4$\pi$ BaF$_2$ array that was originally 
developed for ($n, \gamma$) studies at this accelerator \cite{WGK90a}. With flight paths around
80 cm, a flux similar to that of  electron linacs could be obtained at a TOF resolution of 1.2 
ns/m, fully adequate for the determination of MACS data in the temperature range of the 
various $s$-process scenarios.

The Karlsruhe activities are covering the mass range of the entire $s$-process path from carbon
to bismuth. Particularly fruitful was the setup with the 4$\pi$ BaF$_2$ detector, which has been 
extensively used for determining an accurate and comprehensive set of MACS data in the mass 
range of the main component from Nb  to Ta. In this program, emphasis was put on the $s$-only
 isotopes of  Sn \cite{WVT96a}, Te \cite{WVK92}, Xe \cite{RKV04}, Ba \cite{VWG94b}, 
Sm \cite{WGV93}, Gd \cite{WVK95b}, and Lu \cite{WVK06a} to characterize important 
$s$-process branchings as well as to establish a grid for normalization of the $N_s\sigma$ 
distribution via  the MACS data of $^{110}$Cd \cite{WVK02}, $^{142}$Nd \cite{WVK98a}, and 
$^{150}$Sm  \cite{WGV93}. The full potential of the 4$\pi$ BaF$_2$ array has been
demonstrated in the MACS measurement on Nature's rarest stable isotope  $^{180}$Ta$^{\rm m}$
\cite{WVA01,WVa04}. Although the total world supply of that isotope could be obtained, the sample
consisted of only 150 mg Ta with a modest enrichment in $^{180}$Ta$^{\rm m}$ of 5.5\%.
In this case, the option of using the Q-value difference was crucial for discriminating events
of the dominant impurity $^{181}$Ta.    

An important feature of low-energy accelerators is the possibility of tailoring the neutron spectra by 
the choice of neutron source reaction, proton energy, and thickness of the neutron production target
\cite{RHK09}. Special examples are the kinematic collimation of the neutron beam providing the 
possibility to reduce neutron flight paths to a few centimeters with an enormous increase of the flux 
at the sample position (e.g. \cite{WiK79a, WiK81}). This option allowed also to produce a narrow 
energy beam for a measurement of the ($n, n'$) cross section \cite{MHK10} for complementing the 
nuclear physics input for dating the Universe by means of the Re/Os chronometry \cite{FMM10}.

The flexibility of  low-energy accelerators has also been used for experiments with unique sensitivity to
the DRC channel at TIT Tokyo \cite{NIM91}, which provided the MACS data for the very abundant 
isotopes $^{12}$C \cite{ONI94}, $^{13}$C \cite{SOK97}, and $^{16}$O \cite{INM95}, which are
important neutron poisons for the $s$ process in spite of their minute ($n, \gamma$) cross sections.

To complete this section, it is worth mentioning the potential of reactor filtered neutron beams, which
are produced by means of sharp interference dips on the low-energy side of broad scattering resonances.
A first astrophysically motivated measurement that was reported using the $24\pm2$ keV spectrum
obtained with an iron filter \cite{BPS79} already showed that this method has the advantage of
very clean, well-defined spectra covering a relatively narrow energy range. A unique set of neutron 
filters has been installed at the Kyiv reactor, providing more than 10 neutron lines in the energy range 
from thermal to several hundred keV with intensities of 10$^6$ - 10$^8$~cm$^{-2}$s$^{-1}$
\cite{GKK08, Gri11}. Apart from activation measurements for normalization of uncertain differential
($n, \gamma$) cross sections, filtered neutron beams are important for measurements of elastic and
inelastic scattering cross sections, which are needed for the quantitative assessment of the influence
of thermally populated excited states on the MACS. The uncertainty of the respective correction,
the so-called stellar enhancement factor (Sec. \ref{sec:4.3}), can be significantly reduced provided 
that all reaction channels are constrained by experimental data as demonstrated for the important case
of $^{187}$Os \cite{FMM10}. In that study, a quasi-monoenergetic neutron beam was produced
via the $^7$Li($p, n$) reaction \cite{MHK10}, but with much lower intensity and less resolution than 
at Kyiv, for example.

\subsection{Role of Theory \label{sec:4.3}}

Typical reaction networks for explosive nucleosynthesis contain about 2000 mostly unstable isotopes
connected by more than 20000 reactions. These numbers underline the importance of theoretical predictions for 
the astrophysical rates of all these reactions. Whenever possible, the theoretical predictions are supported, guided, 
and constrained by experimental data, but for the foreseeable future, by far the largest part of this information 
will depend on models suited for global predictions of nuclear properties. The current half-life limit for direct 
measurements of neutron capture rates is a few years. With the new facilities on the horizon (Sec. \ref{sec:5.1}), 
this limit will be reduced to about 100 days \cite{CoR07}, still not sufficiently low to contribute to our 
understanding of the $r$ process, even of the freeze-out phase \cite{SuE01}. The same holds true for the 
$i$ process \cite{HPW11,GZY13} with the neutron capture cross section of $^{135}$I ($t_{1/2}=6$~h) as 
an important example (Sec. \ref{sec:2.4}). The continuous attempts for constructing such global approaches 
are reflected in huge databases, which are often available online, e.g. by works of Rauscher {\it et al.} 
\cite{RaT00,Rau12,Rau12a} and Goriely \cite{Gor05}, or by the data obtained with the Talys code \cite{KHD05}.

Even if experimental cross section data are available, very often these data are not fully commensurable with 
the energy regime of the stellar situation. In such cases, cross sections 
obtained with theoretical models can be normalized to the available experimental data. If the uncertainty of
the energy dependence of the calculated cross section is small, this method provides a safe 
extrapolation to the relevant stellar energies from the measured energies. 
This approach was applied to a number of isotopes, where 
the cross section has been determined only by activation in a quasi-stellar spectrum corresponding to 
$k_BT=25$~keV (for examples see \cite{BBK00}).

Cross section data from laboratory experiments are subject to corrections for temperature effects,
because low-lying excited nuclear states are thermally populated in the hot environment of the interior of 
stars. As their cross sections generally exceed the respective ground state values, the corresponding correction 
is often denoted as stellar enhancement factor (SEF). The SEF corrections depend critically on the 
excitation energies of the first few nuclear states and are important if the lowest excited states occur 
within an energy window corresponding to three to four times the thermal energy of the stellar 
environment. Under the conditions of the $s$-process component in AGB stars, this affects states 
below about 100 keV, but at the higher temperatures in massive stars this window is extended to 
300 keV or higher. The problem of the SEF corrections and their uncertainties has been recently
addressed in detail by Rauscher {\it et al.} \cite{RaT00,Rau12}. 

On a laboratory scale, the excited states decay quasi instantly via $\gamma$-emission and are, 
therefore, out of reach for direct measurements. The only chance in the foreseeable future might 
be to use the NIF facility at LLNL \cite{FBB13}, where laser-heated fusion plasmas can be produced
with temperatures above 10 keV, but presently the determination of SEF corrections remains 
subject to statistical model calculations. 

The input for such calculations, however, can be derived 
from experimental data on the ground state cross section complemented by the statistical properties 
of neutron resonances and by the cross sections for the competing elastic and inelastic scattering 
channels as demonstrated in case of $^{187}$Os, where the first excited state occurs already at 
9.8 keV \cite{FMM10}. A similar approach has also been used for the MACS calculation of the branch 
point nucleus $^{192}$Ir by adjusting the parameters used in the Talys code on the basis of 
experimental cross section data for the Ir and Pt isotopes \cite{KoG13}. Also indirect methods 
can be useful to constrain the relevant model parameters. For example, time-reversed experiments 
can provide the ratio for populating excited states and ground state, thus contributing to the calculation
of excited state cross sections. 

\subsection{Stellar decay rates \label{sec:4.4}}

In addition to the potential temperature-dependence of the neutron capture rates \cite{BBK00}, analyses 
of $s$-process branchings are facing severe theoretical problems in determining the weak interaction 
properties. Although all rates for $\beta$-decay and electron capture (EC) of relevance in the $s$ process 
are known under terrestrial conditions, contributions of thermally populated excited states as well 
as atomic effects in the strongly ionized stellar plasma can dramatically modify the laboratory values 
\cite{TaY87}. The calculated $\beta$ rates in stellar environments are subject to nuclear uncertainties, 
which remain difficult to estimate, because the related uncertainties depend strongly on the 
experimentally unknown decay properties of excited states. 

Under stellar conditions, ground state and low-lying excited states are in thermal equilibrium. The 
combined $\beta$-decay rates of the involved states determine, therefore, the $\beta$-decay 
probability of each species. So far, almost all stellar decay rates have been inferred in a semi-empirical 
way by Takahashi and Yokoi \cite{TaY87} on the basis of decay systematics derived from analogue 
states in neighboring nuclei. In order to illustrate the effect of the remaining nuclear uncertainties, 
these calculations have been iterated by modifying the unknown transition rates by an error value of 
log $ft =\pm0.5$ \cite{Gor99}. For typical $s$-process conditions (temperature $3\times 10^8$~K and 
electron density $N_e = 10^{27}$~cm$^{-3}$), the final rates differed by a maximum factor of 
three, but the individual variation depends strongly on the excitation energies and decay patterns of
the individual excited states.

In case of isomeric excited states with sufficiently long half-lives one may attempt to determine the 
weak-decay rate experimentally. An example is the decay of $^{79}$Se, where the 96-keV isomer 
($t_{1/2} = 3.9$ min) could be sufficiently populated for measuring its $\beta$-decay half life of 
about 5 d \cite{KlK88}, leading to a drastic reduction of the terrestrial half life of $3\times10^5$ yr to
a few years under $s$-process conditions. Apart from isomeric states, $\beta$ decay properties 
of excited states have not been measured in the laboratory yet. In addition, stellar decays 
are affected by modes, which are energetically not possible on earth, i.e. EC from the continuum 
or bound $\beta$ decay. These effects, which may alter the decay rates by orders of magnitude 
\cite{TaY87}, represent a permanent experimental challenge. 

If the $\beta$-decay Q-value is small, the correspondingly long terrestrial half-lives of such nuclei 
can be dramatically shortened in a hot stellar plasma, where the degree of ionization is high or even complete. 
This opens the possibility for bound-state $\beta$ decay, where the electrons are emitted directly into empty 
atomic orbits. The mechanism of bound-state $\beta$-decay has been successfully verified by following 
the decay of fully stripped ions in the Experimental Storage Ring at GSI/Darmstadt. Half-lives for bound-state 
$\beta$ decay have been measured for $^{163}$Dy \cite{JBB92}, which is stable in its neutral state, and for
$^{187}$Re \cite{BFF96}, which is important for the Re/Os cosmic chronometer \cite{FMM10}. Additional
examples are $^{205}$Hg and $^{207}$Tl \cite{FIK08}. 

For the theoretical calculations of stellar rates, Gamow-Teller strength distributions B(GT) for low lying states 
are needed \cite{FFN80,LaM00,LaM03}. Charge-exchange reactions, e.g. ($p, n$), provide access to the 
strength of these transitions via the proportionality between the cross sections at low momentum
transfer $q$ close to 0$^{\circ}$ and B(GT),
\begin{equation}
\frac{d\sigma^{CE}}{d\Omega}(q=0)=\hat{\sigma}_{GT}(q=0)B(GT),
\end{equation}
where $\hat{\sigma}_{GT}(q=0)$ is the unit cross section for GT transitions at q=0 \cite{TGC87}. As shown in 
Ref. \cite{Fre05} this kind of information can also be obtained via ($d$, $^2$He) reactions, which can be studied 
with high-resolution magnetic spectrometers. In order to access GT distributions for unstable nuclei, experiments 
could be carried out in inverse kinematics with radioactive ion beams.  

\section{Current developments and new opportunities \label{sec:5}}

Current efforts in laboratory developments are focused on facilities with higher fluxes and on detectors with high 
efficiencies and low inherent backgrounds. These features are crucial for studies of unstable isotopes, where 
often sample sizes are limited due to restricted availability or a high specific activity, but are likewise important
for accurate measurements of small cross sections. 

\subsection{New facilities \label{sec:5.1}}

Currently, some facilities are under construction, which will yield neutron fluxes up to two orders of magnitude 
higher than currently available. This large increase is either based on the installation of a shorter flight path, 
like at n\_TOF EAR-2, or on using an innovative new accelerator design to run small scale accelerators with 
unprecedented beam intensities (for example the projects FRANZ and SARAF). 

\begin{figure}[htb]
\begin{center}
\includegraphics[width=12.0cm,angle=00]{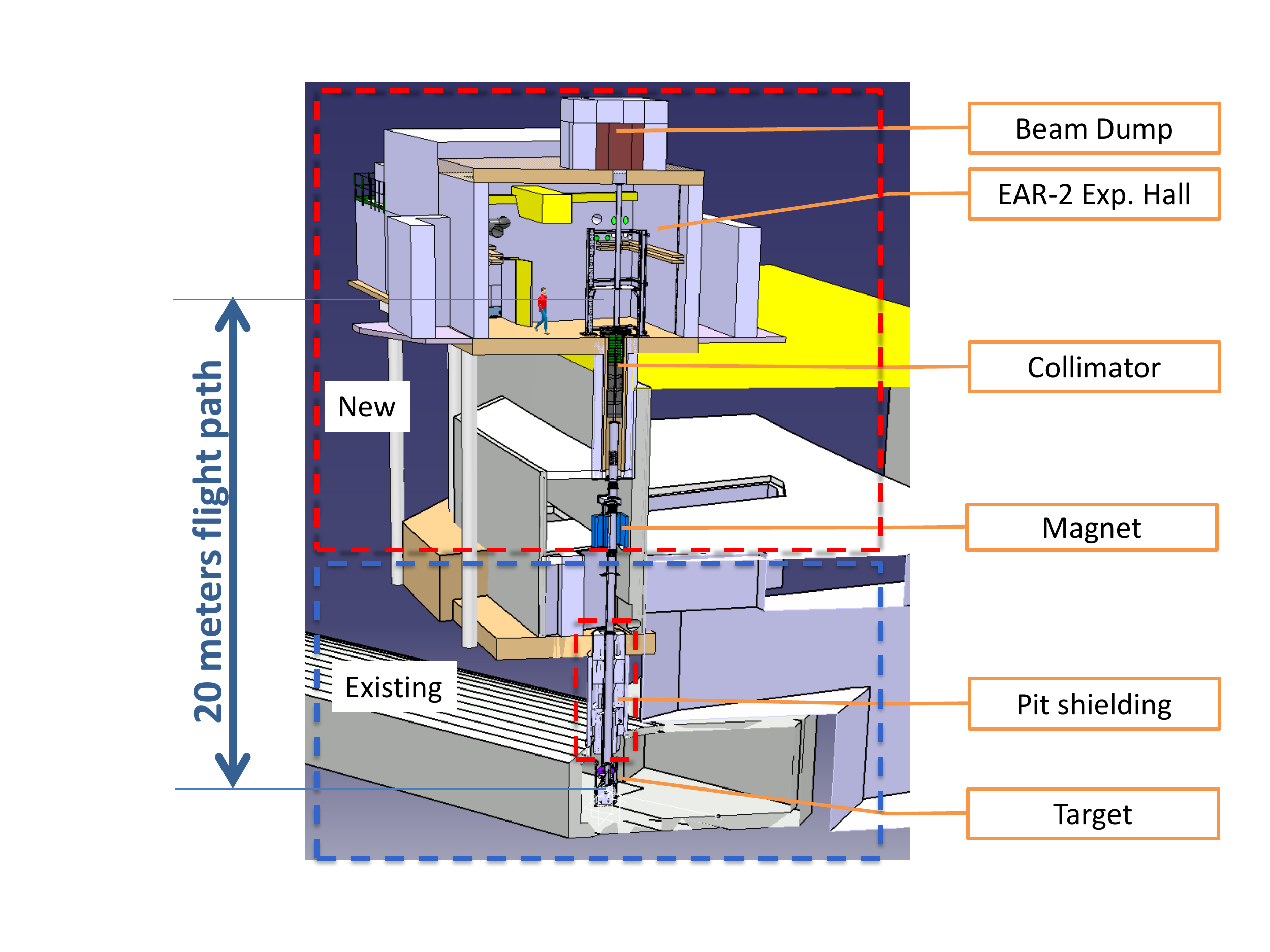}
\caption{\label{fig:9} Sketch of the second experimental area (EAR-2) currently under construction at 
n\_TOF/CERN. EAR-2 will be placed at 90 degrees with respect to the CERN-PS proton beam. The shorter 
flight path of 20~m results in a flux which is 20-30 times higher than at the 185~m measurement station 
EAR-1. (\copyright~n\_TOF@CERN)}
\end{center}
\end{figure}

The EAR-2 project \cite{CCV12,GTB13} is an upgrade of the n\_TOF facility at CERN \cite{AAA01,Chi13} by 
complementing the existing experimental area at 185~m (EAR-1) by a new experimental area at a shorter flight 
path of 20~m. Figure \ref{fig:9} shows a sketch of the EAR-2 project, which is expected to start operation 
the summer of 2014.

The new flight path will be installed vertically at an angle of 90 degrees with respect to the 185~m flight path 
serving the existing experimental area (Sec. \ref{sec:4.1}). Charged particles are removed from the neutron 
beam by a permanent magnet, and the neutron beam is shaped by collimators for optimized conditions 
with respect to capture and fission experiments. Compared to EAR-1, the main advantages of EAR-2 are 
higher average and instantaneous neutron fluxes (by factors of 25 and 250, respectively). In addition, 
the $\gamma$~flash is reduced, because $\gamma$~rays and ultra-relativistic 
particles produced in the target are predominantly emitted in the direction of the proton beam. The 
performance of the new flight path is presently optimized by detailed simulations and will be studied
in detail during an extended commissioning campaign.

\begin{figure}[htb]
\begin{center}
\includegraphics[width=15.0cm,angle=00]{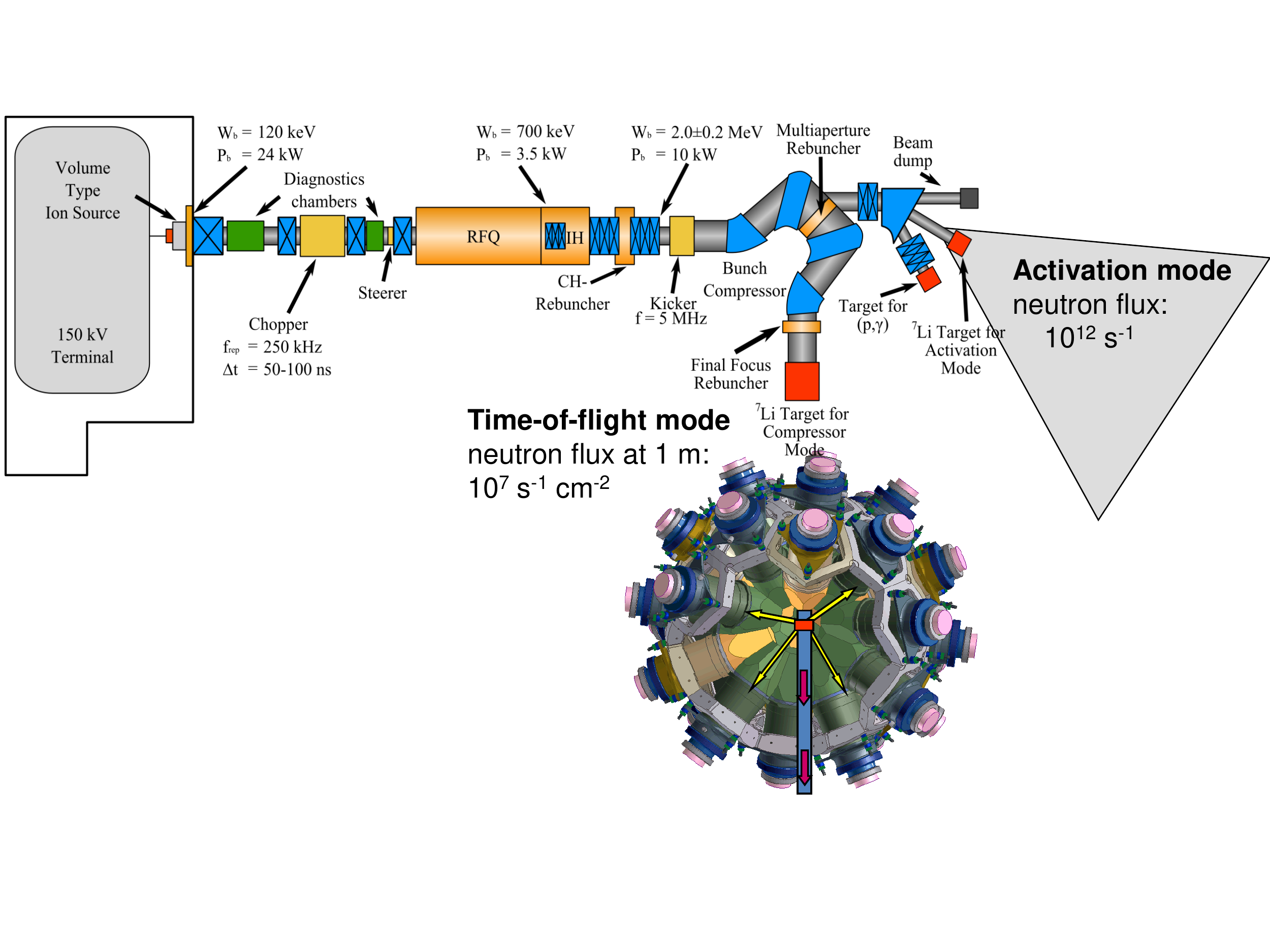}
\caption{\label{fig:10} Sketch of the neutron facility FRANZ, currently under construction at 
		Goethe University 
Frankfurt/Germany}
\end{center}
\end{figure}

A different approach that will be well suited for both, time-of-flight and activation measurements, is 
pursued at Goethe University Frankfurt, Germany \cite{RCH09,RCM07,CMR06,MCM06}. The Frankfurt Neutron Source 
of the Stern Gerlach Zentrum (FRANZ), which is currently under construction, is based on a high intensity 
proton accelerator using the $^{7}$Li($p,n$) reaction for neutron production in the energy range up to 
500~keV. The layout of the FRANZ facility is shown in Figure \ref{fig:10}. 

The proton beam of 100-250~mA DC, produced in a volume type ion source, is first accelerated to 
$2.0$ MeV in a radiofrequency quadrupole section (RFQ) coupled to an interdigital H-type (IH) structure. 
The final proton beam with an energy between 1.8 and 2.2~MeV and an energy resolution of 20~keV
is obtained by a drift tube cavity downstream of the IH part.  Nominal proton currents will be limited 
to 20~mA DC, resulting in a 1000 times higher neutron flux than what was available for TOF 
measurements at Karlsruhe. For activation measurements, the total neutron yield will be  $10^{12}$ 
neutrons per second. For TOF measurements, the proton beam is compressed to bunches of 1~ns, 
using a chopper system at the entrance of the RFQ with a repetition rate up to 250 kHz and a 
Mobley-type bunch compressor. In this configuration, the total neutron yield will be $2\times10^{11}$ 
neutrons per second. As the experimental conditions are very similar, most of the equipment 
was transfered after the shutdown of the Karlsruhe Van de Graaff, including the $4\pi$ BaF$_2$ array
mentioned in Sec. \ref{sec:3.2}. 

Because the amount of sample material can be reduced by the gain in flux, TOF measurements at FRANZ
appear to be feasible already with samples of 10$^{14}$ atoms. This number represents a break-through 
with respect to the production of unstable samples, because beam intensities of the order of 10$^{10}$ to 
10$^{12}$ s$^{-1}$ are expected at future Rare Isotope Facilities such as FRIB~\cite{FRIB12}, 
RIKEN~\cite{Tan98}, or FAIR \cite{Fai07}.

 A low-energy approach is also being realized at the Soreq Applied Research Accelerator Facility SARAF 
(Soreq NRC,  Israel) \cite{NMB06,FPA09}, where deuteron and proton beams of 2 mA and energies up 
to 40 MeV are obtained with a superconducting linac. Recently, first neutrons were successfully produced 
via the $^{7}$Li($p,n$) reaction by means of an innovative windowless Liquid-Lithium 
Target (LiLiT) \cite{FPA09}. For technical reasons astrophysics experiments at this facility will be 
concentrated on the activation method.

A similar concept has been adopted for planning the  LEgnaro NeutrOn Source LENOS. Proton beams 
of initially 5~MeV energy and 50~mA current will be degraded to achieve an energy distribution, which can 
be used to produce different neutron spectra of quasi-Maxwellian shape via the the $^{7}$Li($p,n$) reaction 
\cite{MMP09}. There is also potential to measure cross sections of radioactive species by combining the 
neutron beam with radioactive beams provided by the future SPES project \cite{MPM12}.

Intense neutron sources with low-energy particle beams are also under development for medical
applications. Projects for cancer treatment by boron neutron capture therapy (BNCT), e.g. at Birmingham
\cite{CGM04}, are using the $^7$Li($p, n$) reaction and are also working in the energy regime of 
astrophysical interest.

\begin{table}[!htb]
\caption{Comparison of integrated neutron flux between 10 and 100 keV and energy resolution of 
	operating and future TOF facilities.  \label{tab:1} }
\begin{tabular}{l c c c}
Facility          &  $\phi_n$ ($10^{4}$ s$^{-1}$cm$^{-2}$) & $\phi_{n}$ (pulse$^{-1}$cm$^{-2}$) & $\Delta E/E$ (in units of $10^{-3}$) \\ 
\hline
n\_TOF EAR-1 (185~m) 	  	& 0.4 	 	& $1\times10^4$ 	&  1-2   		\\ 
LANSCE (20~m)        		& 13  	 	& $7\times10^3$	&  8 - 26   	\\ 
GELINA(30~m)         		& 1.4    		&   18     			& 1.3  	   	\\ 
FZK (1~m)                		& 1  			&      0.01          	& 3 - 10    	\\ 
						&			&				&			\\
n\_TOF EAR-2 (20~m)  		& 8  			& $2\times10^5$	& 10-20     	\\ 
FRANZ (1~m)          		& 600  		&    6          		& 3 - 10		\\ 
\end{tabular}
\end{table}

The performance of some existing and future TOF facilities is compared in Table \ref{tab:1}
in terms of neutron flux and resolution. The flux values (in units of cm$^{-2}$s$^{-1}$ 
and cm$^{-2}$pulse$^{-1}$) represent integrals over the neutron energy range from 10 to 
100~keV, which are most important for astrophysical applications. Among existing facilities,
LANSCE is offering the highest neutron flux per cm$^{2}$s, however, with limited 
neutron energy resolution due to the long pulse width of the primary proton beam. Comparable 
fluxes are obtained at GELINA (30~m station), n\_TOF (EAR-1 at 185~m), and previously at 
Karlsruhe (80~cm). In terms of neutron flux per pulse, n\_TOF has the highest intensity, 
resulting in very good signal to background conditions, which are of advantage for measurements
on radioactive samples. At GELINA, shorter flight paths of 10~m are also available,
thus offering higher neutron flux, but at the expense of neutron energy resolution. 

\begin{figure}[htb]
\begin{center}
\includegraphics[width=15.0cm,angle=00]{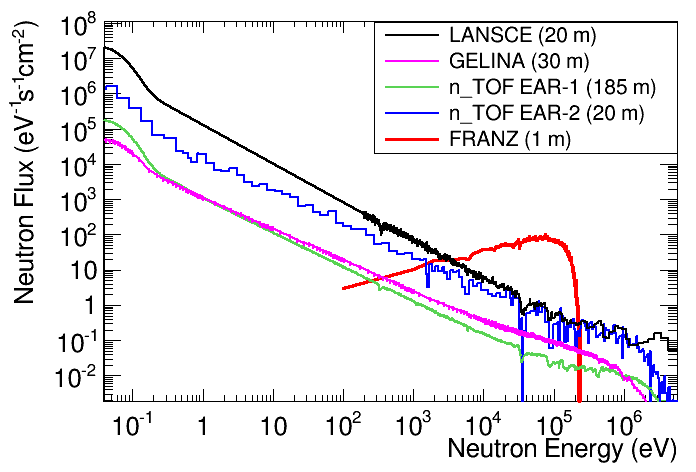}
\caption{\label{fig:11} Comparison of the neutron flux at existing and future facilities as listed in 
Table \ref{tab:1}.}
\end{center}
\end{figure}

The neutron flux at the future FRANZ facility was calculated to be factors of $60-600$ higher than 
at existing installations, using the code PINO \cite{RHK09}. Monte Carlo simulations of the neutron flux 
at the future n\_TOF EAR-2 suggest that the neutron flux per time unit will increase by a factor of 
about 25 with respect to the existing n\_TOF EAR-1, making the flux comparable to LANSCE. The 
instantaneous neutron flux, however, is expected to increase by a factor of 250, a key advantage
of the CERN facility. The neutron flux over the full energy range covered by these TOF facilities is 
compared in Figure~\ref{fig:11}.

Note that activations at existing low-energy accelerators and particularly at the future FRANZ facility can be 
performed in quasi-stellar neutron fields that are still orders of magnitude more intense than achievable 
even in the most advanced TOF installations.

\subsection{Detectors \label{sec:5.2}}

In general, neutron capture events are detected via the prompt $\gamma$-ray cascades. The best
signature of these cascades is the total energy that is defined by the Q-value of the reaction. Depending 
on the level structure of the product nucleus, the energy distribution of the emitted $\gamma$-rays 
can be very broad, which means the possibility of background discrimination based on the energy of
single $\gamma$-rays is not restricted by the energy resolution of the detector. Therefore, most 
approaches sacrifice energy resolution for detection efficiency or neutron sensitivity. 

The sensitivity to neutrons is extremely important, since the neutron scattering cross section in the 
keV-regime can be several orders of magnitude higher than the neutron capture cross section under 
investigation. Accordingly, the detectors are exposed to sample-scattered neutrons, which may cause 
significant backgrounds. The most successful technique for suppressing such backgrounds is based 
on liquid C$_6$D$_6$ scintillation detectors, which can be optimized for extremely low neutron 
sensitivity \cite{PHK03}. Advanced detectors of this type are used on a regular basis, in particular 
if isotopes with very small ($n, \gamma$) cross sections are under investigation \cite{MKB12}. 

Important progress in this field may come from a new concept based on composite detectors 
of low efficiency. Using the directional information obtained via the Compton effect the idea
aims for a strong reduction of the ambient background from neutron interactions in the detectors and 
room walls \cite{DGT13}. 

An important and broadly used alternative approach is the calorimetrical detection of all 
emitted $\gamma$-rays. This technique, where capture events are discriminated against backgrounds 
via the specific Q-value of the investigated sample, requires 4$\pi$ detectors of high-Z materials 
with sufficient resolution in $\gamma$ energy and detection efficiencies near 100\%. The first array 
suited for the keV region was the Karlsruhe 4$\pi$ BaF$_2$ detector \cite{WGK90a} (Sec. \ref{sec:3.2}).
BaF$_2$, which is among the scintillators with the lowest response to neutrons and exhibits also 
excellent timing properties (Table~\ref{tab:2}), became also the material of choice \cite{HRF01} for 
other 4$\pi$ calorimeters \cite{RBA04,GAA09}. 

Meanwhile, more scintillators can be produced in sufficiently large volumes. Some attractive 
alternatives are compared with established detector materials in Table \ref{tab:2} with respect to 
density or absorption length, wavelength of the scintillation light, signal intensity given by the number 
of photons produced per $\gamma$-ray energy, and decay time. The entries in the last column 
correspond to a quality factor for the sensitivity to scattered neutrons.

Among the new scintillators, Ce doped LaBr$_3$ represents an interesting alternative, because of 
the superior energy resolution indicated by the large number of photons per keV $\gamma$ energy. 
Whether the background reduction obtained with a narrow Q-value window compensates the 
effect of the rather high neutron sensitivity may well depend on the specific experimental situation.
From the examples listed in Table~\ref{tab:2}, high-efficiency scintillators such as BaF$_2$ represent 
a balanced compromise between a modest energy resolution and low neutron sensitivity. In that
respect LaCl$_3$ might offer a promising unexplored solution. Extreme cases are C$_6$D$_6$ detectors, 
which combine an extremely low neutron sensitivity with poor resolution in $\gamma$-energy, whereas 
the excellent energy resolution of Ge-detectors is to be paid with a comparably high sensitivity to
scattered neutrons. The latter choice, which has been adopted for the ANNRI array at J-PARC/Japan 
\cite{NFG11}, provides additional information on the decay patterns of the capture cascades,
however. 

Another important aspect is the decay time of the scintillation light. Most time-of-flight facilities 
suffer from an intense $\gamma$-flash caused by the impact of the particle beam on the neutron 
production target that results in a high energy deposition in the detectors at very short flight times 
(see Figure~\ref{fig:8}). Particularly at short flight paths, the dead time caused by the $\gamma$ flash 
impedes or even prevents the registration of neutron-induced events at later times. The longer the 
decay time of the detector material and the shorter the flight path, the lower the maximum energy 
for which capture events can be observed. As soon as this energy limit falls below 10~keV, the setup 
is of limited value for astrophysical purposes. 

\begin{table}
\caption{Characteristics of some scintillator materials. }
    \label{tab:2}
    \begin{tabular}{lcccccc}
  Scintillator           		& Density 	& Abs.  		& Wave- 	& Photons	& Decay	 	& Quality$^b$		\\
                         			&            	&  length$^a$ & length	& per keV    	& time		& s$_\gamma$/s$_n$ \\
                         			& (g/cm$^3$) & cm  	& (nm)    		&     		& (ns)   		&       			\\
     \hline
C$_6$D$_6$             			& 0.954  & 14        	& 425		& $\approx$10& 2.8    		&  80000 		\\ 
C$_6$F$_6$             			& 1.61   	& 10.1  	& 430		& 10       		& 3.3   		&  610 			\\
 BaF$_2$$^c$	              			& 4.88  	& 3.5    	& 220; 310   	& 1.8; 10		& 0.6; 630 	&  300 			\\
 Bi$_4$Ge$_3$O$_{12}$$^c$	& 7.13  	& 2.1  	& 480    		& 0.7; 7.5 	& 60; 300 	&  520 			\\
 NaI(Tl)                    			& 3.67  	& 4.6  	& 415   		& 38     		& 250   		&  23			\\
 CsI(Na)                    			& 4.51  	& 3.8	& 420   		& 41     		& 630     		&  22 			\\
							&		&		&			&			&			&				\\
CeF$_3$$^c$               			& 6.16  	& 2.7 	& 300; 340   	& 0.2; 4.3 	& 3; 27   		&  740 			\\
LaCl$_3$(Ce)                	 		& 3.85  	& 4.3  		& 350   		& 49     		& 28  		&  430 			\\
LaBr$_3$(Ce)                			& 5.08   	& 3.4		& 380   		& 63     		& 16 	    	&  27			\\
Lu$_{1.8}$Y$_2$SiO$_5$(Ce)	& 7.1     	& 2.2 		& 420   		& 32     		& 41    		 &  29 			\\
\end{tabular}
$^a$Thickness of detector material needed to reduce the intensity of 1~MeV $\gamma$-rays by $1/e$. \\
$^b$Ratio of total photon absorption cross section for 1~MeV $\gamma$-rays and Maxwellian averaged neutron 
capture cross section of the scintillator material at 30~keV.\\
$^c$Material with two components in scintillation light.
  \end{table}

The prompt detection of charged particles requires very thin samples and is, therefore, challenging in 
terms of the observed count rate. The detector is usually more massive than the sample, thus
requiring the possibility of discriminating reactions in the sample and in the detector. One approach is to 
place the detectors outside the neutron beam sacrificing solid angle coverage. Fast silicon detectors of 
different sizes and granularities are suited for this purpose, i.e. for cross section measurements of 
($n, \alpha$) reactions \cite{WLK12} and for neutron beam monitors using the $^{6}$Li($n, \alpha$) 
reaction \cite{MMA04}. Alternatively, efficiencies of nearly 100\% are obtained with highly transparent 
detectors, which can be be placed directly inside the neutron beam. Most common are gas detectors, e.g. 
ionization chambers, which have been operated in the single Frisch-gridded version \cite{DWW07,SKK93} 
or with reduced background in the compensated mode \cite{GKA00}. 

More recently, MicroMegas detectors
\cite{GRR96,ACK11} were successfully used at CERN. Advantages of this detector are its very low 
background, fast timing, spatial resolution, and a sufficiently low noise level for application in ($n, \alpha$) 
measurements. In fission studies, parallel plate avalanche counters (PPAC) turned out to be extremely 
insensitive to neutrons and $\gamma$-rays. This feature is crucial in the neutron energy range above 
100 MeV, where other detectors are overwhelmed by the $\gamma$ flash \cite{PTA10,TLA14}. Diamond 
detectors became another option for in-beam detection of charged particles with good time and energy 
resolution. Because they are only available in rather small sizes so far, a composite diamond detector 
was recently used for an ($n, \alpha$) cross-section measurement by the n\_TOF collaboration at CERN
\cite{WGG13}.

In activation measurements the induced activities of the product nuclei are measured off-line (Sec. \ref{sec:3.3}), 
using techniques according to the half-life and the decay mode of the reaction product. In most cases, HPGe 
detectors are applied, ideally with segmented configurations, e.g. Clover type detectors. In this way, a high 
efficiency can be combined with a reduced load per detector, an important aspect if the sample itself is radioactive.
The power of this approach has been demonstrated at the example of the $^{147}$Pm($n, \gamma$) reaction 
\cite{RAH03}. This measurement could be performed on only $10^{14}$ atoms by using the high 
efficiency of a two-Clover setup in close geometry and the coincidence option for detecting $\gamma$ 
cascades in the decay of $^{148}$Pm.
  
Special examples are reactions, in which the product nuclei decay either without $\gamma$ emission or are
long-lived isotopes so that the induced activities fall below the detection limit. The latter cases can be 
studied via  accelerator mass spectrometry, still maintaining the excellent sensitivity of the activation technique 
as discussed in Sec. \ref{sec:3.3}. Examples from the first category, however, suffer from significantly reduced
sensitivity, because the required detection of the decay electrons implies the use of very thin samples to avoid 
excessive absorption losses \cite{RSK00}.

\subsection{Advanced concepts \label{sec:5.3}}

While existing techniques are sufficient to study most neutron reactions on stable isotopes with the required 
accuracy, neutron capture data for radioactive nuclei, which are of key importance in advance nucleosynthesis 
models (Sec. \ref{sec:2}), are presenting a number of experimental challenges. An inherent difficulty in
measurements on radioactive samples results from the radiation emitted by the sample itself, thus imposing 
stringent limits on the sample size that can be handled in such experiments. Other major problems are the
production of sufficiently clean and isotopically pure samples and the coordination of production and 
experiment in order to minimize the ingrowth of the daughter nuclei.

An obvious solution for dealing with small samples is to increase the neutron flux \cite{CoR07}, either by
increasing the neutron production rate or by decreasing the distance from the neutron production target. 
As spallation sources are operating with a fixed primary beam, a higher flux can be obtained by reducing
the flight path, e.g. at the n\_TOF facility, but at the expense of a correspondingly reduced resolution in 
neutron energy. Actual developments of new low-energy accelerators, e.g. the FRANZ project, offer the 
possibility for increasing the flux by both options, higher intensity of the primary beam as well as a reduction 
in flight path. For example, if FRANZ would be operated with a flight path of 10~cm instead of 1~m
with a nominal time resolution of 1~ns, the neutron-energy resolution is still comparable to the one at
DANCE (Table \ref{tab:1}) but the neutron flux at the sample position would be about  
$10^9$~cm$^{-2}$s$^{-1}$, orders of magnitude higher than in any other TOF facility.  

\begin{figure}
  \includegraphics[width=.7\textwidth]{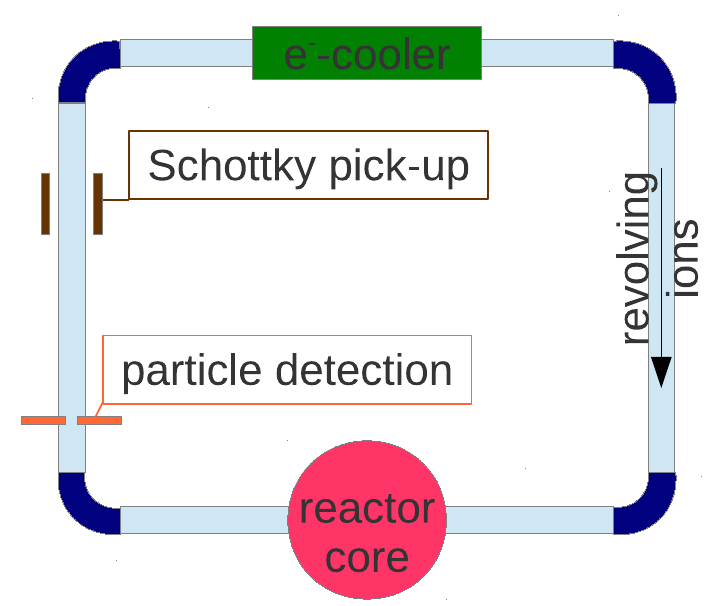}
  \caption{Proposal for ($n, \gamma$) measurements on short-lived isotopes 
\cite{ReL14}. The main components are an ion storage ring with beam lines and 
focusing elements (light blue), dipole magnets (dark blue) and electron cooler (green). 
The ring is coupled to a thermal reactor with an interaction region near the core
core (red). Capture events are identified by particle detectors outside the nominal orbit
(orange) and/or by Schottky pick-up electrodes (brown).
  \label{fig:12}}
\end{figure}

A completely different approach is to investigate neutron-induced reactions in inverse kinematics 
\cite{ReL14}. This requires a beam of radioactive ions cycling in a storage ring with 100~AkeV or 
less and a neutron target. Radioactive ions close to stability can be produced with high intensities 
using ISOL-techniques and storage rings for low beam energies, which require extremely high 
vacuum, are under construction, e.g. the CRYRING at GSI/FAIR \cite{HHS13} or the CSR at 
MPK/Heidelberg \cite{HBB11}. The neutron target could be either a reactor coupled with the 
storage ring to obtain an interaction zone near the core \cite{ReL14} or a bottle of ultra-cold 
neutrons. The scheme of such a setup is sketched in Fig.~\ref{fig:12}.

\section{Status and requests \label{sec:6}}

\subsection{Compilations of stellar ($n, \gamma$) cross sections \label{sec:6.1}}

In 1971, the first comprehensive collection of recommended MACS values for 
$k_BT$ =  30 keV by Allen {\it et al.} \cite{AGM71} listed 130 experimental cross 
sections with typical uncertainties between 10 and 25\%. To provide a full set of 
data for the early pioneering studies of the $s$-process, the experimental results were 
complemented by 109 semi-empirical values estimated from cross section trends 
derived from neighboring nuclei. 

The next compilation published in 1987 \cite{BaK87} was prepared again for MACS 
values at a single thermal energy of $k_BT = 30$ keV, according to the needs of the classical 
approach for a scenario of constant temperature and neutron density \cite{SFC65}.
A major achievement, however, was the significant improvement of the accuracy, which 
was reaching the 1 - 2\% level for a number of important $s$-process isotopes.

When the classical approach had been challenged by refined stellar models with a range
of $s$-process conditions, from He shell burning in thermally pulsing low mass AGB stars 
\cite{GBP88,HoI88} to shell C burning in massive stars \cite{RBG91a,RBG91b} and with
($\alpha, n$) reactions on $^{13}$C and $^{22}$Ne as the main neutron sources, the
compilations had to be extended to provide MACS data for thermal energies between 5
and 100 keV \cite{BVW92}.

The following MACS compilation of Bao {\it et al.} \cite{BBK00} was already comprising a network of 364 
($n, \gamma$) reactions, including also relevant partial cross sections, and provided detailed 
information on previous MACS results, which were eventually condensed into recommended 
values for thermal energies from 5 to 100 keV. Where experimental information was missing, 
calculations with the Hauser-Feshbach statistical model code {\sc NON-SMOKER} \cite{RaT00}
were empirically corrected for known systematic deficiencies in the nuclear input of the 
calculation. Stellar enhancement factors (SEF) were included as well. 

For easy access, this compilation was published in electronic form via the \textsc{KADoNiS} 
project (http://www.kadonis.org) \cite{DHK05}. The present update (\textsc{KADoNiS} v0.3 
\cite{DPK09}), includes 38 improved and 14 new cross sections compared to \cite{BBK00}, in 
total, data for 356 isotopes, including 77 radioactive nuclei on or close to the $s$-process path. 
This version counted experimental data for 13 of these radioactive nuclei, i.e. for $^{14}$C, 
$^{60}$Fe, $^{93}$Zr, $^{99}$Tc, $^{107}$Pd, $^{129}$I, $^{135}$Cs, $^{147}$Pm, 
$^{151}$Sm, $^{154}$Eu, $^{163}$Ho, $^{182}$Hf, and $^{185}$W, while empirically 
corrected Hauser-Feshbach rates with typical uncertainties of 25 to 30\% were quoted otherwise. 
Presently, a new update is in preparation and should be available in 2014.

The present version of \textsc{KADoNiS} consists of two parts: the $s$-process 
library and a collection of available experimental $p$-process reactions. The 
$s$-process library will be complemented in the near future by ($n, p$) and 
($n, \alpha$) cross sections measured at $k_BT$ = 30 keV, as it was already included 
in \cite{BaK87}. The $p$-process database will be a collection of all available 
charged-particle reactions measured within or close to the Gamow window of the 
$p$ process ($T_9$= 2-3 GK).

\begin{figure}[tb]
\includegraphics[width=14cm]{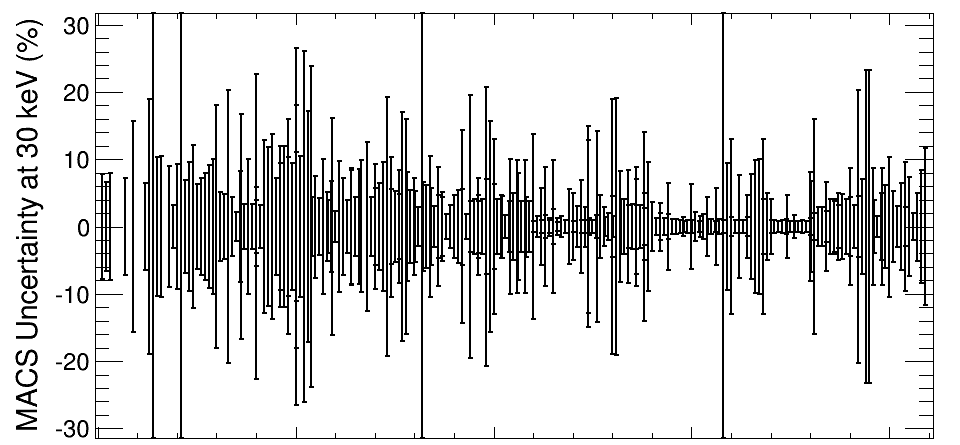}
\includegraphics[width=14cm]{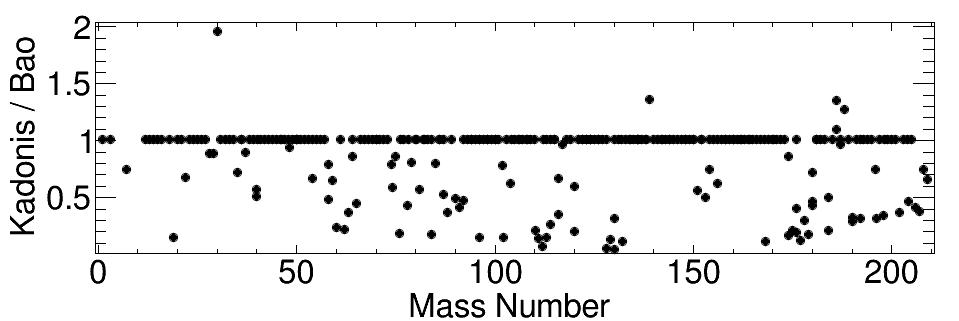}
  	\caption{Top: MACS uncertainties versus mass number. Bottom: Current recommended uncertainties relative to the 
			uncertainties in 2000 (Ref.~\cite{BBK00}).\label{fig:13}  }
\end{figure}	

\subsection{Further requirements \label{sec:6.2}}

The recommended uncertainties of the ($n, \gamma$) cross sections for $s$-process nucleosynthesis calculations 
is summarized in Figure~\ref{fig:13} as a function of 
mass number (top) together with the improvements of the last decade represented by the ratio of the uncertainties (bottom). 
Though the 
necessary accuracy for stable isotopes of 1 to 5\% has been locally achieved, further 
improvements are clearly required, predominantly below $A$ = 120 and in the mass region 
above 180. It is worthewhile noting that some uncertainties actually increased. Those are cases, where theoretical estimates
revealed larger uncertainties than before. No experimental data are available for those isotopes.

Further efforts in this field are the more important as Figure~\ref{fig:13} reflects 
only the situation for a thermal energy of 30 keV. In most cases, however, 
the experimental data had to be extrapolated to determine the MACS values at
lower and higher temperatures. Accordingly, this implies significantly larger 
uncertainties.  

The lack of accurate data is particularly crucial for the weak $s$-process 
in massive stars, which is responsible for most of the $s$ abundances between 
Cu and Sr. Since the neutron exposure of the weak $s$ process is not
sufficient for achieving flow equilibrium, cross section uncertainties may
affect the abundances of a sequence of heavier isotopes (see Figure~\ref{fig:6}).

A further extension of \textsc{KADoNiS} is planned to include more radioactive 
isotopes, which are relevant for $s$-process nucleosynthesis at higher neutron 
densities (up to 10$^{11}$ cm$^{-3})$ \cite{CGS06}. Because these isotopes are 
more than one atomic mass unit away from the canonical $s$-process path on the 
neutron-rich side of the stability valley, their stellar ($n, \gamma$) values have to 
be extrapolated from known cross sections by means of the statistical Hauser-Feshbach 
model. According to the \textsc{KADoNiS} homepage already 73 additional 
unstable isotopes have been identified. The desired accuracy for radioactive 
branch-point isotopes is 10 to 20\%.

From the above, it is obvious that experimental work should focus on two aspects,
(i) on a large set of stable isotopes, where data are incomplete or too uncertain, and 
(ii) on the largely unexplored field of radioactive isotopes, be it the branch points in the 
$s$ path or the neutron-rich nuclei at the border of stability that are reached under 
extreme $s$-process conditions at neutron densities of 10$^{12}$ to 10$^{15}$ 
cm$^{-3}$.

\section{Summary \label{sec:7}}

Neutron reactions are of pivotal importance for our understanding of how the heavy element 
abundances are formed during the late stellar phases. Maxwellian averaged cross sections 
are particularly important to constrain $s$-process models related to the H/He shell burning 
in AGB stars and also for the He and C shell burning phases in massive stars. The fact that 
the respective $s$-process abundance distributions can be deduced in quantitative detail is 
crucial for defining the abundances 
produced by explosive nucleosynthesis and for deriving a reliable picture of Galactic chemical 
evolution. 

Continued improvements of laboratory neutron sources and of measurement techniques were
instrumental for establishing a comprehensive collection of neutron-induced reaction rates
in the astrophysically relevant energy range from a few up to about 300~keV. Apart from very 
few exceptions, experimental data are available for all stable isotopes between Fe and Pb, 
although not always with sufficient accuracy and in the entire energy range of interest. Such
deficits are particularly found for the very small cross sections of the abundant light elements, 
which represent potential neutron poisons, and for neutron magic nuclei, which are the bottle 
necks in the $s$-process reaction flow. For the unstable species, which are needed under special 
$s$-process conditions characterized by high neutron densities and for explosive nucleosynthesis, 
experimental data are still very scarce and must so far be complemented by theory. The main 
role of theory, however, refers to corrections concerning the stellar environment, i.e. with 
respect to the effect of thermally populated excited states to the enhancement of weak 
interaction rates in the stellar plasma. 

The main challenges for the future will be related to the further improvement of the laboratory 
neutron sources and to the development of advanced experimental methods. Progress in these
fields is mandatory for tackling the yet unsatisfactory situation with neutron reactions of unstable 
isotopes. It appears that promising developments in both areas are presently under way with the 
potential for innovative solutions. As a consequence, future experiments can be performed with much 
higher sensitivities, i.e. by using very small amounts of sample material. This is crucial for 
dealing with unstable isotopes, because the sample activity can be reduced and the stringent 
problem of sample preparation can be solved by using the intense radioactive beam facilities. 
Within the next decade, these options will provide ample opportunities to extend neutron reaction
studies into the region of unstable isotopes.

\begin{center}
{\bf Acknowledgements}
\end{center}
The authors would like to thank C. Guerrero, F. Gunsing, V. Vlachoudis, as well as  J. Heyse and P. Schillebeeckxs for providing neutron fluxes and further information about n\_TOF and GELINA. Thanks are also due to R. Gallino and S. Bisterzo for their permission to use Figure~\ref{fig:5}. This work was supported by the Helmholtz Young Investigator project VH-NG-327 and the BMBF project 05P12RFFN6 and the Helmholtz International Center for FAIR.

\section*{References}
\bibliographystyle{iopart-num}
\bibliography{refbib}

\newcommand{\noopsort}[1]{} \newcommand{\printfirst}[2]{#1}
  \newcommand{\singleletter}[1]{#1} \newcommand{\swithchargs}[2]{#2#1}
\providecommand{\newblock}{}
\begin{thebibliography}{100}
\expandafter\ifx\csname url\endcsname\relax
  \def\url#1{{\tt #1}}\fi
\expandafter\ifx\csname urlprefix\endcsname\relax\def\urlprefix{URL }\fi
\providecommand{\eprint}[2][]{\url{#2}}

\bibitem{BeC38}
Bethe H and Critchfield C 1938 {\em Phys. Rev.\/} {\bf 54} 248

\bibitem{Wei38}
von Weizs{\"a}cker C 1938 {\em Physik. Zeitschrift\/} {\bf 39} 639

\bibitem{Bet39a}
Bethe H 1939 {\em Phys. Rev.\/} {\bf 55} 103

\bibitem{Mer52b}
Merrill S 1952 {\em Ap. J.\/} {\bf 116} 21

\bibitem{SuU56}
Suess H and Urey H 1956 {\em Rev. Mod. Phys.\/} {\bf 28} 53--74

\bibitem{Cam59a}
Cameron A 1959 {\em Ap. J.\/} {\bf 129} 676

\bibitem{BBF57}
Burbidge E, Burbidge G, Fowler W and Hoyle F 1957 {\em Rev. Mod. Phys.\/} {\bf
  29} 547

\bibitem{Cam57}
Cameron A 1957 Stellar evolution, nuclear astrophysics, and nucleogenesis,
  chalk river report crl--41 Tech. rep. A.E.C.L. Chalk River, Canada

\bibitem{Cam57b}
Cameron A 1957 {\em Pub. Astron. Soc. Pacific\/} {\bf 9} 201

\bibitem{MaG65}
Macklin R and Gibbons J 1965 {\em Rev. Mod. Phys.\/} {\bf 37} 166

\bibitem{AGM71}
Allen B, Gibbons J and Macklin R 1971 {\em Adv. Nucl. Phys.\/} {\bf 4} 205

\bibitem{CFH61}
Clayton D, Fowler W, Hull T and Zimmerman B 1961 {\em Ann. Phys.\/} {\bf 12}
  331

\bibitem{SFC65}
Seeger P, Fowler W and Clayton D 1965 {\em Ap. J. Suppl.\/} {\bf 11} 121

\bibitem{ClW74}
Clayton D and Ward R 1974 {\em Ap. J.\/} {\bf 193} 397

\bibitem{KGB90}
K{\"a}ppeler F, Gallino R, Busso M, Picchio G and Raiteri C 1990 {\em Ap. J.\/}
  {\bf 354} 630 -- 643

\bibitem{WVK98a}
Wisshak K, Voss F, K{\"a}ppeler F, Kazakov L and Reffo G 1998 {\em Phys. Rev.
  C\/} {\bf 57} 391 -- 408

\bibitem{AKW99b}
Arlandini C, K{\"a}ppeler F, Wisshak K, Gallino R, Lugaro M, Busso M and
  Straniero O 1999 {\em Ap. J.\/} {\bf 525} 886--900

\bibitem{Wei66}
Weigert A 1966 {\em Z. Astrophys.\/} {\bf 64} 395

\bibitem{ScH67}
Schwarzschild M and H{\"a}rm R 1967 {\em Ap. J.\/} {\bf 150} 961

\bibitem{San67}
Sanders R 1967 {\em Ap. J.\/} {\bf 150} 971

\bibitem{Ulr73}
Ulrich R 1973 {\em Explosive Nucleosynthesis\/} ed Schramm D and Arnett W
  (Austin: University of Texas) p 139

\bibitem{TAT68}
Truran J, Arnett W, Tsuruta S and Cameron A~W 1968 {\em Astrophys. Space
  Sci.\/} {\bf 1} 129

\bibitem{Hil78}
Hillebrandt W 1978 {\em Space Sci. Rev.\/} {\bf 21} 639

\bibitem{WoH78}
Woosley S and Howard W 1978 {\em Ap. J. Suppl.\/} {\bf 36} 285

\bibitem{RPA90}
Rayet M, Prantzos N and Arnould M 1990 {\em Astron. Astrophys.\/} {\bf 227} 271

\bibitem{Gol37}
Goldschmidt V 1937 {\em Norske Vidensk. Akad. Skr.\/} {\bf Mat.--Naturv. Kl.
  IV} 1

\bibitem{Cam82}
Cameron A 1982 {\em Essays in Nuclear Astrophysics\/} ed Barnes C, Schramm D
  and Clayton D (Cambridge: Cambridge Univ. Press) pp 23--43

\bibitem{AnE82}
Anders E and Ebihara M 1982 {\em Geochim. Cosmochim. Acta\/} {\bf 46} 2363

\bibitem{AnG89}
Anders E and Grevesse N 1989 {\em Geochim. Cosmochim. Acta\/} {\bf 53} 197

\bibitem{Lod03}
Lodders K 2003 {\em Ap. J.\/} {\bf 591} 1220

\bibitem{GAS07}
Grevesse N, Asplund M and Sauval A 2007 {\em Space Sci. Rev.\/} {\bf 130}
  105--114

\bibitem{LPG09}
Lodders K, Palme H and Gail H~P 2009 {\em Landolt--B{\"o}rnstein, New Series,
  Vol. VI/4B, Chap. 4.4\/} ed Tr{\"u}mper J (Berlin: Springer) pp 560--630
  arXiv:0901.1149 [astro-ph.EP]

\bibitem{Ree94}
Reeves H 1994 {\em Rev. Mod. Phys.\/} {\bf 66} 193

\bibitem{BGW99}
Busso M, Gallino R and Wasserburg G 1999 {\em Ann. Rev. Astron. Astrophys.\/}
  {\bf 37} 239

\bibitem{IbR83}
Iben I~J and Renzini A 1983 {\em Ann. Rev. Astron. Astrophys.\/} {\bf 21} 271

\bibitem{Her04}
Herwig F 2004 {\em Ap. J. Suppl.\/} {\bf 155} 651

\bibitem{SGC06}
Straniero O, Gallino R and Cristallo S 2006 {\em Nucl. Phys. A\/} {\bf 777} 311

\bibitem{SCG08}
Sneden C, Cowan J and Gallino R 2008 {\em Ann. Rev. Astron. Astrophys.\/} {\bf
  46} 241

\bibitem{SCL97}
Straniero O, Chieffi A, Limongi M, Busso M, Gallino R and Arlandini C 1997 {\em
  Ap. J.\/} {\bf 478} 332

\bibitem{GAB98}
Gallino R, Arlandini C, Busso M, Lugaro M, Travaglio C, Straniero O, Chieffi A
  and Limongi M 1998 {\em Ap. J.\/} {\bf 497} 388--403

\bibitem{LHW99}
Langer N, Heger A, Wellstein S and Herwig F 1999 {\em Astron. Astrophys.\/}
  {\bf 346} L37

\bibitem{DeT03}
Denissenkov P and Tout C 2003 {\em Mon. Not. Royal Astron. Soc.\/} {\bf 340}
  722

\bibitem{HBS97}
Herwig F, Bl{\"o}cker T, Sch{\"o}nberner D and El~Eid M 1997 {\em Astron.
  Astrophys.\/} {\bf 324} L81 -- L84

\bibitem{FLS96}
Freytag B, Ludwig H and Steffen M 1996 {\em Astron. Astrophys.\/} {\bf 313} 497

\bibitem{Her00}
Herwig F 2000 {\em Astron. Astrophys.\/} {\bf 360} 952

\bibitem{SCG09}
Straniero O, Cristallo S and Gallino R 2009 {\em Publ. Astron. Soc.
  Australia\/} {\bf 26} 133

\bibitem{Zin98}
Zinner E 1998 {\em Ann. Rev. Earth Planet. Sci.\/} {\bf 26} 147--188

\bibitem{BGS10}
Bisterzo S, Gallino R, Straniero O, Cristallo S, K{\"a}ppeler F and Aoki W 2010
  {\em Mon. Not. Royal Astron. Soc.\/} {\bf 404} 1529

\bibitem{TGB01b}
Travaglio C, Gallino R, Busso M and Gratton R 2001 {\em Ap. J.\/} {\bf 549}
  346--352

\bibitem{TGA04}
Travaglio C, Gallino R, Arnone E, Cowan J, Jordan F and Sneden C 2004 {\em Ap.
  J.\/} {\bf 601} 864

\bibitem{SGT09}
Serminato A, Gallino R, Travaglio C, Bisterzo S and Straniero O 2009 {\em Publ.
  Astron. Soc. Australia\/} {\bf 26} 153

\bibitem{KGB11}
K{\"a}ppeler F, Gallino R, Bisterzo S and Aoki W 2011 {\em Rev. Mod. Phys.\/}
  {\bf 83} 157

\bibitem{LHL03}
Lugaro M, Herwig F, Lattanzio J, Gallino R and Straniero O 2003 {\em Ap.J.\/}
  {\bf 586} 1305

\bibitem{CSA74}
Couch R, Schmiedekamp A and Arnett W 1974 {\em Ap. J.\/} {\bf 190} 95

\bibitem{LHT77}
Lamb S, Howard W, Truran J and Iben I 1977 {\em Ap. J.\/} {\bf 217} 213

\bibitem{ArT85}
Arnett W and Thielemann F~K 1985 {\em Ap. J.\/} {\bf 295} 589

\bibitem{BuG85}
Busso M and Gallino R 1985 {\em Astron. Astrophys.\/} {\bf 151} 205

\bibitem{PAA87}
Prantzos N, Arnould M and Arcoragi J 1987 {\em Ap. J.\/} {\bf 315} 209

\bibitem{ArT69}
Arnett W and Truran J 1969 {\em Ap. J.\/} {\bf 157} 339

\bibitem{RBG91a}
Raiteri C, Busso M, Gallino R, Picchio G and Pulone L 1991 {\em Ap. J.\/} {\bf
  367} 228--238

\bibitem{RBG91b}
Raiteri C, Busso M, Gallino R and Picchio G 1991 {\em Ap. J.\/} {\bf 371}
  665--672

\bibitem{RGB92}
Raiteri C, Gallino R and Busso M 1992 {\em Ap. J.\/} {\bf 387} 263--275

\bibitem{RGB93}
Raiteri C, Gallino R, Busso M, Neuberger D and K{\"a}ppeler F 1993 {\em Ap.
  J.\/} {\bf 419} 207--223

\bibitem{WoW95}
Woosley S and Weaver T 1995 {\em Ap. J. Suppl.\/} {\bf 101} 181 -- 235

\bibitem{LSC00}
Limongi M, Straniero O and Chieffi A 2000 {\em Ap. J. Suppl.\/} {\bf 129} 625

\bibitem{WHW02}
Woosley S, Heger A and Weaver T 2002 {\em Rev. Mod. Phys.\/} {\bf 74} 1015

\bibitem{TEM07}
The L, El~Eid M and Meyer B 2007 {\em Ap. J.\/} {\bf 655} 1058

\bibitem{ETM09}
El~Eid M, The L~S and Meyer B 2009 {\em Space Sci. Rev.\/} {\bf 147} 1

\bibitem{PGH10}
Pignatari M, Gallino R, Heil M, Wiescher M, K{\"a}ppeler F, Herwig F and
  Bisterzo S 2010 {\em Ap. J.\/} {\bf 710} 1557

\bibitem{RaG02}
Rauscher T and Guber K 2002 {\em Phys. Rev. C\/} {\bf 66} 028802

\bibitem{RHH02}
Rauscher T, Heger A, Hoffman R and Woosley S 2002 {\em Ap. J.\/} {\bf 576} 323

\bibitem{RaG05}
Rauscher T and Guber K 2005 {\em Phys. Rev. C\/} {\bf 71} 059903

\bibitem{NPA05}
Nassar H, Paul M, Ahmad I, Berkovits D, Bettan M, Collon P, Dababneh S,
  Ghelberg S, Greene J, Heger A, Heil M, Henderson D, Jiang C, K{\"a}ppeler F,
  Koivisto H, O�Brien S, Pardo R, Patronis N, Pennington T, Plag R, Rehm K,
  Reifarth R, Scott R, Sinha S, Tang X and Vondrasek R 2005 {\em Phys. Rev.
  Lett.\/} {\bf 94} 092504

\bibitem{TTS05}
Tomyo A, Temma Y, Segawa M, Nagai Y, Makii H, Ohsaki T and Igashira M 2005 {\em
  Ap. J.\/} {\bf 623} L153

\bibitem{ABE08}
Alpizar-Vicente A, Bredeweg T, Esch E~I, Greife U, Haight R, Hatarik R,
  O'Donnell J, Reifarth R, Rundberg R, Ullmann J, Vieira D and Wouters J 2008
  {\em Phys. Rev. C\/} {\bf 77} 015806

\bibitem{LBC13}
{Lederer} C, {Massimi} C, {Berthoumieux} E, {Colonna} N, {Dressler} R,
  {Guerrero} C, {Gunsing} F, {K{\"a}ppeler} F, {Kivel} N, {Pignatari} M,
  {Reifarth} R, {Schumann} D, {Wallner} A, {Altstadt} S, {Andriamonje} S,
  {Andrzejewski} J, {Audouin} L, {Barbagallo} M, {B{\'e}cares} V, {Be{\v
  c}v{\'a}{\v r}} F, {Belloni} F, {Berthier} B, {Billowes} J, {Boccone} V,
  {Bosnar} D, {Brugger} M, {Calviani} M, {Calvi{\~n}o} F, {Cano-Ott} D,
  {Carrapi{\c c}o} C, {Cerutti} F, {Chiaveri} E, {Chin} M, {Cort{\'e}s} G,
  {Cort{\'e}s-Giraldo} M~A, {Dillmann} I, {Domingo-Pardo} C, {Duran} I,
  {Dzysiuk} N, {Eleftheriadis} C, {Fern{\'a}ndez-Ord{\'o}{\~n}ez} M, {Ferrari}
  A, {Fraval} K, {Ganesan} S, {Garc{\'{\i}}a} A~R, {Giubrone} G,
  {G{\'o}mez-Hornillos} M~B, {Gon{\c c}alves} I~F, {Gonz{\'a}lez-Romero} E,
  {Gramegna} F, {Griesmayer} E, {Gurusamy} P, {Harrisopulos} S, {Heil} M,
  {Ioannides} K, {Jenkins} D~G, {Jericha} E, {Kadi} Y, {Karadimos} D,
  {Korschinek} G, {Krti{\v c}ka} M, {Kroll} J, {Langer} C, {Lebbos} E, {Leeb}
  H, {Leong} L~S, {Losito} R, {Lozano} M, {Manousos} A, {Marganiec} J,
  {Marrone} S, {Martinez} T, {Mastinu} P~F, {Mastromarco} M, {Meaze} M,
  {Mendoza} E, {Mengoni} A, {Milazzo} P~M, {Mingrone} F, {Mirea} M,
  {Mondalaers} W, {Paradela} C, {Pavlik} A, {Perkowski} J, {Plag} R, {Plompen}
  A, {Praena} J, {Quesada} J~M, {Rauscher} T, {Riego} A, {Roman} F, {Rubbia} C,
  {Sarmento} R, {Schillebeeckx} P, {Schmidt} S, {Tagliente} G, {Tain} J~L,
  {Tarr{\'{\i}}o} D, {Tassan-Got} L, {Tsinganis} A, {Tlustos} L, {Valenta} S,
  {Vannini} G, {Variale} V, {Vaz} P, {Ventura} A, {Vermeulen} M~J, {Versaci} R,
  {Vlachoudis} V, {Vlastou} R, {Ware} T, {Weigand} M, {Wei{\ss}} C, {Wright}
  T~J, {{\v Z}ugec} P and {n TOF Collaboration} 2014 {\em Phys. Rev. C\/} {\bf
  89} 025810

\bibitem{HKU08a}
Heil M, K{\"a}ppeler F, Uberseder E, Gallino R and Pignatari M 2008 {\em Phys.
  Rev. C\/} {\bf 77} 015808

\bibitem{HKU08b}
Heil M, K{\"a}ppeler F, Uberseder E, Gallino R, Bisterzo S and Pignatari M 2008
  {\em Phys. Rev. C\/} {\bf 78} 025802

\bibitem{DPK09}
Dillmann I, Plag R, K{\"a}ppeler F and Rauscher T 2009 {\em EFNUDAT Fast
  Neutrons - scientific workshop on neutron measurements, theory \&
  applications\/} ed Hambsch F~J (Geel: JRC-IRMM) pp 55 -- 58
  http://www.kadonis.org

\bibitem{BHP12}
Bennett M, Hirschi R, Pignatari M, Diehl S, Fryer C, Herwig F, Hungerford A,
  Nomoto K, Rockefeller G, Timmes F and Wiescher M 2012 {\em Mon. Not. Royal
  Astron. Soc.\/} {\bf 420} 3047--3070

\bibitem{WSG00}
Westin J, Sneden C, Gustafsson B and Cowan J 2000 {\em Ap. J.\/} {\bf 530} 783

\bibitem{BPA00}
Burris D, Pilachowski C, Armandroff T, Sneden C, Cowan J and Roe H 2000 {\em
  Ap. J.\/} {\bf 544} 302

\bibitem{CSB02}
Cowan J, Sneden C, Burles S, Ivans I, Beers T, Truran J, Lawler J, Primas F,
  Fuller G, Pfeiffer B and Kratz K~L 2002 {\em Ap. J.\/} {\bf 572} 861

\bibitem{HPC02}
{Hill} V, {Plez} B, {Cayrel} R, {Beers} T~C, {Nordstr{\"o}m} B, {Andersen} J,
  {Spite} M, {Spite} F, {Barbuy} B, {Bonifacio} P, {Depagne} E, {Fran{\c c}ois}
  P and {Primas} F 2002 {\em Astron. Astrophys.\/} {\bf 387} 560

\bibitem{CSB05}
Cowan J, Sneden C, Beers T~C, Lawler J~E, Simmerer J, Truran J~W, Primas F,
  Collier J and Burles S 2005 {\em Ap. J.\/} {\bf 627} 238

\bibitem{CoS06}
Cowan J and Sneden C 2006 {\em Nature\/} {\bf 440} 1151--1156

\bibitem{TAK11}
Thielemann F~K, Arcones A, Käppeli R, Liebend{\"o}rfer M, Rauscher T, Winteler
  C, Fr{\"o}hlich C, Dillmann I, Fischer T, Martinez-Pinedo G, Langanke K,
  Farouqi K, Kratz K~L, Panov I and Korneev I 2011 {\em Progress in Particle
  and Nuclear Physics\/} {\bf 66} 346--353

\bibitem{KGH86}
Kratz K~L, Gabelmann H, Hillebrandt W, Pfeiffer B, Schl{\"o}sser K and
  Thielemann F~K 1986 {\em Z. Physik\/} {\bf 325} 489--490

\bibitem{Kra88}
Kratz K~L 1988 {\em Rev. Mod. Astron.\/} {\bf 1} 184

\bibitem{KBT93}
Kratz K~L, Bitouzet J~P, Thielemann F~K, M{\"o}ller P and Pfeiffer B 1993 {\em
  Ap. J.\/} {\bf 403} 216--238

\bibitem{Wan07}
Wanajo S 2007 {\em Ap. J.\/} {\bf 666} L77

\bibitem{WoH92}
Woosley S and Hoffman R 1992 {\em Ap. J.\/} {\bf 395} 202

\bibitem{CoT04}
Cowan J and Thielemann F~K 2004 {\em Physics Today\/} {\bf 57} 45

\bibitem{KFP07b}
Kratz K~L, Farouqi K and Pfeiffer B 2007 {\em Prog. Particle Nucl. Phys.\/}
  {\bf 59} 147--155

\bibitem{NKM13}
Nakamura K, Kajino T, Mathews G, Sato S and Harikae S 2013 {\em Int. J. Mod.
  Phys. E\/} {\bf 22} 1330022

\bibitem{FRT99}
Freigburghaus C, Rosswog S and Thielemann F~K 1999 {\em Ap. J.\/} {\bf 525}
  L121

\bibitem{KHH90}
Kratz K~L, Harms V, Hillebrandt W, , Pfeiffer B, Thielemann F~K and W{\"o}hr A
  1990 {\em Z. Phys. A.\/} {\bf 336} 357

\bibitem{MoN88}
M{\"o}ller P and Nix R 1988 {\em Atomic Data Nucl. Data Tables\/} {\bf 39} 213

\bibitem{Rau05b}
Rauscher T 2005 {\em Nucl. Phys. A\/} {\bf 758} 655c

\bibitem{SuE01}
Surman R and Engel J 2001 {\em Phys. Rev. C\/} {\bf 64} 035801

\bibitem{ArM11}
Arcones A and Martinez-Pinedo G 2011 {\em Phys. Rev. C\/} {\bf 83} 045809

\bibitem{SEB97}
Surman R, Engel J, Bennett J and Meyer B 1997 {\em Phys. Rev. Lett.\/} {\bf 79}
  1809 -- 1812

\bibitem{MMS12}
Mumpower M, McLaughlin G and Surman R 2012 {\em Phys. Rev. C\/} {\bf 85} 045801

\bibitem{SBM09}
Surman R, Beun J, McLaughlin G and Hix W 2009 {\em Phys. Rev. C\/} {\bf 79}
  045809

\bibitem{BBH09}
Beun J, Blackmon J, Hix W, McLaughlin G, Smith M and Surman R 2009 {\em J.
  Phys. G\/} {\bf 36} 025201

\bibitem{Gor97b}
{Goriely} S 1997 {\em Astron. Astrophys.\/} {\bf 325} 414--424

\bibitem{MMT83}
Mathews G, Mengoni A, Thielemann F~K and Fowler W 1983 {\em Ap. J.\/} {\bf 270}
  740

\bibitem{KAA12}
Kozub R, Arbanas G, Adekola A, Bardayan D, Blackmon J, Chae K, Chipps K,
  Cizewski J, Erikson L, Hatarik R, Hix W, Jones K, Krolas W, Liang J, Ma Z,
  Matei C, Moazen B, Nesaraja C, Pain S, Shapira D, Shriner J, Smith M and Swan
  T 2012 {\em Phys. Rev. Lett.\/} {\bf 109} 172501

\bibitem{PaT04}
Panov I and Thielemann F 2004 {\em Astron. Lett.\/} {\bf 30} 647

\bibitem{PKR11}
Panov I, Korneev I, Rauscher T and Thielemann F 2011 {\em Bull. Russ. Acad.
  Sci. Phys.\/} {\bf 75} 484

\bibitem{RAH95}
Rayet M, Arnould M, Hashimoto M, Prantzos N and Nomoto K 1995 {\em Astron.
  Astrophys.\/} {\bf 298} 517 -- 527

\bibitem{FML06}
Fr{\"o}hlich C, Mart\'inez-Pinedo G, Liebend{\"o}rfer M, Thielemann F~K, Bravo
  E, Hix W, Langanke K and Zinner N 2006 {\em Phys. Rev. Lett.\/} {\bf 96}
  142502

\bibitem{SAG98}
Schatz H, Aprahamian A, G{\"o}rres J, Wiescher M, Rauscher T, Rembges J,
  Thielemann F~K, Pfeiffer B, M{\"o}ller P, Kratz K~L, Herndl H, Brown B and
  Rebel H 1998 {\em Physics Reports\/} {\bf 294} 167

\bibitem{SAB01}
Schatz H, Aprahamian A, V B, Bildsten L, Cumming A, Ouellette M, Rauscher T,
  Thielemann F~K and Wiescher M 2001 {\em Phys. Rev. Letters\/} {\bf 86} 3471
  -- 3474

\bibitem{WHH90}
Woosley S, Hartmann D, Hoffman R and Haxton W 1990 {\em Ap. J.\/} {\bf 356} 272

\bibitem{TRG11}
Travaglio C, R{\"o}pke F, Gallino R and Hillebrandt W 2011 {\em Ap. J.\/} {\bf
  739} 93

\bibitem{RGW06}
Rapp W, G{\"o}rres J, Wiescher M, Schatz H and K{\"a}ppeler F 2006 {\em Ap.
  J.\/} {\bf 653} 474

\bibitem{ArG03}
Arnould M and Goriely S 2003 {\em Phys. Rep.\/} {\bf 384} 1

\bibitem{RDD13}
Rauscher T, Dauphas N, Dillmann I, Fr{\"o}hlich C, F{\"u}l{\"o}p Z and
  Gy{\"u}rky G 2013 {\em Rep. Prog. Phys.\/} {\bf 76} 066201

\bibitem{TGG99}
Travaglio C, Galli D, Gallino R, Busso M, Ferrini F and Straniero O 1999 {\em
  Ap. J.\/} {\bf 521} 691

\bibitem{TGB01a}
Travaglio C, Galli D and Burkert A 2001 {\em Ap. J.\/} {\bf 547} 217

\bibitem{QiW07}
Qian Y~Z and Wasserburg G 2007 {\em Phys. Rep.\/} {\bf 442} 237

\bibitem{FKP10}
Farouqi K, Kratz K~L, Pfeiffer B, Rauscher T, Thielemann F~K and Truran J 2010
  {\em Ap. J.\/} {\bf 712} 1359--1377

\bibitem{MMB12}
Maiorca E, Magrini L, Busso M, Randich S, Palmerini S and Trippella O 2012 {\em
  Astrophys. J.\/} {\bf 747} 53

\bibitem{MBC07}
Montes F, Beers T, Cowan J, Elliot T, Farouqi K, Gallino R, Heil M, Kratz K~L,
  Pfeiffer B, Pignatari M and Schatz H 2007 {\em Ap. J.\/} {\bf 671} 1685

\bibitem{HAI06}
Honda S, Aoki W, Ishimaru Y, Wanajo S and Ryan S 2006 {\em Ap. J.\/} {\bf 643}
  1180

\bibitem{SCL03}
Sneden C, Cowan J, Lawler J, Ivans I, Burles S, Beers T, Primas F, Hill V,
  Truran J, Fuller G, Pfeiffer B and Kratz K~L 2003 {\em Ap. J.\/} {\bf 591}
  936--953

\bibitem{HPW11}
{Herwig} F, {Pignatari} M, {Woodward} P~R, {Porter} D~H, {Rockefeller} G,
  {Fryer} C~L, {Bennett} M and {Hirschi} R 2011 {\em Ap. J.\/} {\bf 727} 89
  (\textit{Preprint} \eprint{1002.2241})

\bibitem{GZY13}
{Garc{\'{\i}}a-Hern{\'a}ndez} D~A, {Zamora} O, {Yag{\"u}e} A, {Uttenthaler} S,
  {Karakas} A~I, {Lugaro} M, {Ventura} P and {Lambert} D~L 2013 {\em Astron.
  Astrophys.\/} {\bf 555} L3 (\textit{Preprint} \eprint{1306.2134})

\bibitem{Hin10}
Hinkelmann M and {for the FAIR Joint Core Team} 2010 Facility for antiproton
  and ion research Tech. rep. FAIR Newsletter No. 15
  https://www-alt.gsi.de/documents/DOC-2010-Apr-60-1.pdf

\bibitem{KBW89}
K{\"a}ppeler F, Beer H and Wisshak K 1989 {\em Rep. Prog. Phys.\/} {\bf 52}
  945--1013

\bibitem{BBK00}
Bao Z, Beer H, K{\"a}ppeler F, Voss F, Wisshak K and Rauscher T 2000 {\em
  Atomic Data Nucl. Data Tables\/} {\bf 76} 70--154

\bibitem{SDP10}
Sz{\"u}cs T, Dillmann I, Plag R and F{\"u}l{\"o}p Z 2010 {\em NIC-XI\/} ed
  Blaum K, Christlieb N and Martinez-Pinedo G (Trieste: Proceedings of Science)
  p 247

\bibitem{MoR63}
Moxon M and Rae E 1963 {\em Nucl. Instr. Meth.\/} {\bf 24} 445

\bibitem{WiK78}
Wisshak K and K{\"a}ppeler F 1978 {\em Nucl. Sci. Eng.\/} {\bf 66} 363 -- 377

\bibitem{WiK79a}
Wisshak K and K{\"a}ppeler F 1979 {\em Nucl. Sci. Eng.\/} {\bf 69} 39 -- 46

\bibitem{Rau63}
Rau F 1963 {\em Nukleonik\/} {\bf 5} 191 -- 197

\bibitem{MaG67b}
Macklin R and Gibbons J 1967 {\em Phys. Rev.\/} {\bf 159} 1007--1012 based on
  H. Maier-Leibnitz, priv. comm. and Rau, F., Nukleonik, 5 (1963) 191

\bibitem{KWG00}
Koehler P, Winters R, Guber K, Rauscher T, Harvey J, Raman S, Spencer R,
  Blackmon J, Larson D, Bardayan D and Lewis T 2000 {\em Phys. Rev. C\/} {\bf
  62} 055803

\bibitem{GLS05}
Guber K, Leal L, Sayer R, Koehler P, Valentine T, Derrien H and Harvey J 2005
  {\em Nucl. Instr. Meth. B\/} {\bf 241} 218--222

\bibitem{GLS05b}
Guber K, Leal L, Sayer R, Koehler P, Valentine T, Derrien H and Harvey J 2005
  {\em Nuclear Data for Science and Technology\/} ed Haight R, Chadwick M,
  Kawano T and Talou P (New York: AIP) pp 1706--1711 aIP Conference Series 769

\bibitem{PHK03}
Plag R, Heil M, K{\"a}ppeler F, Pavlopoulos P, Reifarth R and Wisshak K 2003
  {\em Nucl. Instr. Meth. A\/} {\bf 496} 425 -- 436

\bibitem{WGK90a}
Wisshak K, Guber K, K{\"a}ppeler F, Krisch J, M{\"u}ller H, Rupp G and Voss F
  1990 {\em Nucl. Instr. Meth. A\/} {\bf 292} 595 -- 618

\bibitem{GAA09}
{Guerrero} C, {Abbondanno} U, {Aerts} G, {{\'A}lvarez} H, {{\'A}lvarez-Velarde}
  F, {Andriamonje} S, {Andrzejewski} J, {Assimakopoulos} P, {Audouin} L,
  {Badurek} G, {Baumann} P, {Be{\v c}v{\'a}{\v r}} F, {Berthoumieux} E,
  {Calvi{\~n}o} F, {Calviani} M, {Cano-Ott} D, {Capote} R, {Carrapi{\c c}o} C,
  {Cennini} P, {Chepel} V, {Chiaveri} E, {Colonna} N, {Cortes} G, {Couture} A,
  {Cox} J, {Dahlfors} M, {David} S, {Dillmann} I, {Domingo-Pardo} C, {Dridi} W,
  {Duran} I, {Eleftheriadis} C, {Ferrant} L, {Ferrari} A, {Ferreira-Marques} R,
  {Fujii} K, {Furman} W, {Goncalves} I, {Gonz{\'a}lez-Romero} E, {Gramegna} F,
  {Gunsing} F, {Haas} B, {Haight} R, {Heil} M, {Herrera-Martinez} A, {Igashira}
  M, {Jericha} E, {K{\"a}ppeler} F, {Kadi} Y, {Karadimos} D, {Kerveno} M,
  {Koehler} P, {Kossionides} E, {Krti{\v c}ka} M, {Lampoudis} C, {Leeb} H,
  {Lindote} A, {Lopes} I, {Lozano} M, {Lukic} S, {Marganiec} J, {Marrone} S,
  {Mart{\'{\i}}nez} T, {Massimi} C, {Mastinu} P, {Mendoza} E, {Mengoni} A,
  {Milazzo} P~M, {Moreau} C, {Mosconi} M, {Neves} F, {Oberhummer} H, {O'Brien}
  S, {Pancin} J, {Papachristodoulou} C, {Papadopoulos} C, {Paradela} C,
  {Patronis} N, {Pavlik} A, {Pavlopoulos} P, {Perrot} L, {Pigni} M~T, {Plag} R,
  {Plompen} A, {Plukis} A, {Poch} A, {Praena} J, {Pretel} C, {Quesada} J,
  {Rauscher} T, {Reifarth} R, {Rubbia} C, {Rudolf} G, {Rullhusen} P, {Salgado}
  J, {Santos} C, {Sarchiapone} L, {Savvidis} I, {Stephan} C, {Tagliente} G,
  {Tain} J~L, {Tassan-Got} L, {Tavora} L, {Terlizzi} R, {Vannini} G, {Vaz} P,
  {Ventura} A, {Villamarin} D, {Vicente} M~C, {Vlachoudis} V, {Vlastou} R,
  {Voss} F, {Walter} S, {Wiescher} M and {Wisshak} K 2009 {\em Nucl. Instr.
  Meth. A\/} {\bf 608} 424

\bibitem{HRF01}
Heil M, Reifarth R, Fowler M, Haight R, K{\"a}ppeler F, Rundberg R, Seabury E,
  Ullmann J, Wilhelmy J, Wisshak K and Voss F 2001 {\em Nucl. Instr. Meth. A\/}
  {\bf 459} 229 -- 246

\bibitem{RBA04}
{Reifarth} R, {Bredeweg} T~A, {Alpizar-Vicente} A, {Browne} J~C, {Esch} E~I,
  {Greife} U, {Haight} R~C, {Hatarik} R, {Kronenberg} A, {O'Donnell} J~M,
  {Rundberg} R~S, {Ullmann} J~L, {Vieira} D~J, {Wilhelmy} J~B and {Wouters} J~M
  2004 {\em Nucl. Instr. Meth. A\/} {\bf 531} 530--543

\bibitem{MAA85}
Muradyan G, Adamchuk A, Adamchuk V, Shchepkin Y and Voskanyan M 1985 {\em Nucl.
  Sci . Eng.\/} {\bf 90} 60

\bibitem{BDS94}
Block R, Danon Y, Slovacek R, Werner C and Youk G 1994 {\em Nuclear Data for
  Science and Technology\/} ed Dickens J (La Grange Park, Illinois: American
  Nuclear Society) p~81

\bibitem{BBR88}
Brehm K, Becker H, Rolfs C, Trautvetter H, K{\"a}ppeler F and Ratynski W 1988
  {\em Z. Phys. A\/} {\bf 330} 167 -- 172

\bibitem{KoB89}
Koehler P and O'Brien H 1989 {\em Phys. Rev. C\/} {\bf 39} 1655

\bibitem{SJL95}
Schatz H, Jaag S, Linker G, Steininger R, K{\"a}ppeler F, Koehler P, Graff S
  and Wiescher M 1995 {\em Phys. Rev. C\/} {\bf 51} 379 -- 391

\bibitem{WWB87}
Wagemans C, Weigmann H and Barthelemy R 1987 {\em Nucl. Phys. A\/} {\bf 469}
  497

\bibitem{DWW07}
De~Smet L, Wagemans C, Wagemans J, Heyse J and Van~Gils J 2007 {\em Phys. Rev.
  C\/} {\bf 76} 045804

\bibitem{GKA00}
Gledenov Y, Koehler P, Andrzejewski K, Guber K and Rauscher T 2000 {\em Phys.
  Rev. C\/} {\bf 62} 42801

\bibitem{AAB10}
Andriamonje S, Attie D, Berthoumieux E, Calviani M, Colas P, Dafni T,
  Fanourakis G, Ferrer-Ribas E, Galan J, Geralis T, Giganon A, Giomataris I,
  Gris A, Guerrero~Sanchez C, Gunsing F, Iguaz F, Irastorza I, De~Oliveira R,
  Papaevangelou T, Ruz J, Savvidis I, A T and Tomás A 2010 {\em J.
  Instrumentation\/} {\bf 5} P02001

\bibitem{GRR96}
Giomataris Y, Rebourgeard P, Robert J and Charpark G 1996 {\em Nucl. Instr.
  Meth. Phys. Res. A\/} {\bf 376} 29

\bibitem{ACK11}
Andriamonje S, Calviani M, Kadi Y, Losito R, Vlachoudis V, Berthoumieux E,
  Gunsing F, Giomataris Y, Papaevangelou T, Guerrero C, Colonna N, Weiss C and
  {the n\_TOF Collaboration} 2011 {\em J. Korean Phys. Soc.\/} {\bf 59} 1601

\bibitem{CCK08}
Calviani M, Cennini P, Karadimos D, Ketlerov V, Konovalov V, Furman W,
  Goverdowski A, Vlachoudis V, Zanini L and the~n\_TOF Collaboration 2008 {\em
  Nucl. Instr. Meth. A\/} {\bf 594} 220– 227

\bibitem{TBD05}
Tassan-Got L, Berthier B, Duran I, Ferrant L, Isaev S, de~la Naour C, Paradela
  C, Stephan C, Trubert D and the~n\_TOF Collaboration 2005 {\em International
  Conference on Nuclear Data for Science and Technology\/} ed Haight R,
  Chadwick M, Kawano T and Talou P (New York: American Insitute of Physics) p
  1529 {AIP} Conference Series 769

\bibitem{TLA14}
{Tarr{\'{\i}}o} D, {Leong} L~S, {Audouin} L, {Duran} I, {Paradela} C,
  {Tassan-Got} L, {Le Naour} C, {Bacri} C~O, {Petitbon} V, {Mottier} J,
  {Caama{\~n}o} M, {Altstadt} S, {Andrzejewski} J, {Barbagallo} M,
  {B{\'e}cares} V, {Be{\v c}v{\'a}{\v r}} F, {Belloni} F, {Berthoumieux} E,
  {Billowes} J, {Boccone} V, {Bosnar} D, {Brugger} M, {Calviani} M,
  {Calvi{\~n}o} F, {Cano-Ott} D, {Carrapi{\c c}o} C, {Cerutti} F, {Chiaveri} E,
  {Chin} M, {Colonna} N, {Cort{\'e}s} G, {Cort{\'e}s-Giraldo} M~A, {Diakaki} M,
  {Domingo-Pardo} C, {Dzysiuk} N, {Eleftheriadis} C, {Ferrari} A, {Fraval} K,
  {Ganesan} S, {Garc{\'{\i}}a} A~R, {Giubrone} G, {G{\'o}mez-Hornillos} M~B,
  {Gon{\c c}alves} I~F, {Gonz{\'a}lez-Romero} E, {Griesmayer} E, {Guerrero} C,
  {Gunsing} F, {Gurusamy} P, {Jenkins} D~G, {Jericha} E, {Kadi} Y,
  {K{\"a}ppeler} F, {Karadimos} D, {Koehler} P, {Kokkoris} M, {Krti{\v c}ka} M,
  {Kroll} J, {Langer} C, {Lederer} C, {Leeb} H, {Losito} R, {Manousos} A,
  {Marganiec} J, {Mart{\'{\i}}nez} T, {Massimi} C, {Mastinu} P~F, {Mastromarco}
  M, {Meaze} M, {Mendoza} E, {Mengoni} A, {Milazzo} P~M, {Mingrone} F, {Mirea}
  M, {Mondalaers} W, {Pavlik} A, {Perkowski} J, {Plompen} A, {Praena} J,
  {Quesada} J~M, {Rauscher} T, {Reifarth} R, {Riego} A, {Roman} F, {Rubbia} C,
  {Sarmento} R, {Schillebeeckx} P, {Schmidt} S, {Tagliente} G, {Tain} J~L,
  {Tsinganis} A, {Valenta} S, {Vannini} G, {Variale} V, {Vaz} P, {Ventura} A,
  {Versaci} R, {Vermeulen} M~J, {Vlachoudis} V, {Vlastou} R, {Wallner} A,
  {Ware} T, {Weigand} M, {Wei{\ss}} C, {Wright} T~J and {{\v Z}ugec} P 2014
  {\em Nuclear Instruments and Methods in Physics Research A\/} {\bf 743}
  79--85

\bibitem{KoG91}
Koehler P and Graff S 1991 {\em Phys. Rev. C\/} {\bf 44} 2788

\bibitem{MMA04}
{Marrone} S, {Mastinu} P~F, {Abbondanno} U, {Baccomi} R, {Marchi} E~B,
  {Bustreo} N, {Colonna} N, {Gramegna} F, {Loriggiola} M, {Marigo} S, {Milazzo}
  P~M, {Moreau} C, {Sacchetti} M, {Tagliente} G, {Terlizzi} R, {Vannini} G,
  {Aerts} G, {Berthomieux} E, {Cano-Ott} D, {Cennini} P, {Domingo-Pardo} C,
  {Ferrant} L, {Gonzalez-Romero} E, {Gunsing} F, {Heil} M, {Kaeppeler} F,
  {Papaevangelou} T, {Paradela} C, {Pavlopoulos} P, {Perrot} L, {Plag} R,
  {Tain} J~L, {Wendler} H and {n TOF Collaboration} 2004 {\em Nucl. Instr.
  Meth. Phys. Res. A\/} {\bf 517} 389 -- 398

\bibitem{WGG13}
Wei{\ss} C, Griesmayer E, Guerrero C, Altstadt S, Andrzejewski J, Audouin L,
  Badurek G, Barbagallo M, B{\'e}cares V, Be{\u{c}}v{\'a}{\u{r}} F, Belloni F,
  Berthoumieux E, Billowes J, Boccone V, Bosnar D, Brugger M, Calviani M,
  Calvi{\~n}o F, Cano-Ott D, Carrapiço C, Cerutti F, Chiaveri E, Chin M,
  Colonna N, Cort{\'e}s G, Cort{\'e}s-Giraldo M, Diakaki M, Domingo-Pardo C,
  Duran I, Dressler R, Dzysiuk N, Eleftheriadis C, Ferrari A, Fraval K, Ganesan
  S, García A, Giubrone G, G{\'o}mez-Hornillos M, Gonçalves I,
  Gonz{\'a}lez-Romero E, Gunsing F, Gurusamy P, Hern{\'a}ndez-Prieto A, Jenkins
  D, Jericha E, Kadi Y, K{\"a}ppeler F, Karadimos D, Kivel N, Koehler P,
  Kokkoris M, Krti{\u{c}}ka M, Kroll J, Lampoudis C, Langer C, Leal-Cidoncha E,
  Lederer C, Leeb H, Leong L, Losito R, Mallick A, Manousos A, Marganiec J,
  Martínez T, Massimi C, Mastinu P, Mastromarco M, Meaze M, Mendoza E, Mengoni
  A, Milazzo P, Mingrone F, Mirea M, Mondalaers W, Paradela C, Pavlik A,
  Perkowski J, Plompen A, Praena J, Quesada J, Rauscher T, Reifarth R, Riego A,
  Robles M, Roman F, Rubbia C, Sabat{\'e}-Gilarte M, Sarmento R, Saxena A,
  Schillebeeckx P, Schmidt S, Schumann D, Tagliente G, Tain J, Tarrio D,
  Tassan-Got L, Tsinganis A, Valenta S, Vannini G, Variale V, Vaz P, Ventura A,
  Versaci R, Vermeulen M, Vlachoudis V, Vlastou R, Wallner A, Ware T, Weigand
  M, Wright T and {\v{Z}}ugec P 2013 {\em Nucl. Instr. Meth. A.\/} {\bf 732}
  190 -- 194

\bibitem{RaK88}
Ratynski W and K{\"a}ppeler F 1988 {\em Phys. Rev. C\/} {\bf 37} 595--604

\bibitem{LKM12}
Lederer C, K{\"a}ppeler F, Mosconi M, Nolte R, Heil M, Reifarth R, Schmidt S,
  Dillmann I, Giesen U, Mengoni A and Wallner A 2012 {\em Phys. Rev. C\/} {\bf
  85} 055809

\bibitem{FFK12}
Feinberg G, Friedman M, Krasa A, Shor A, Eisen Y, Berkovits D, Giorginis G,
  Hirsh T, Paul M, Plompen A and Tsuk E 2012 {\em Phys. Rev. C\/} {\bf 85}
  055810

\bibitem{HDJ05}
Heil M, Dababneh S, Juseviciute A, K{\"a}ppeler F, Plag R, Reifarth R and
  O'Brien S 2005 {\em Phys. Rev. C\/} {\bf 71} 025803

\bibitem{KNA87}
K{\"a}ppeler F, Naqvi A and Al-Ohali M 1987 {\em Phys. Rev. C\/} {\bf 35}
  936--941

\bibitem{Bee91}
Beer H 1991 {\em Ap. J.\/} {\bf 375} 823

\bibitem{RHF08}
Reifarth R, , Heil M, Forss{\'e}n C, Besserer U, Couture A, Dababneh S,
  D{\"o}rr L, G{\"o}rres J, Haight R, K{\"a}ppeler F, Mengoni A, O'Brien S,
  Patronis N, Plag R, Rundberg R, Wiescher M and Wilhelmy J 2008 {\em Phys.
  Rev. C\/} {\bf 77} 015804

\bibitem{MSG96}
Meissner J, Schatz H, G{\"o}rres J, Herndl H, Wiescher M, Beer H and
  K{\"a}ppeler F 1996 {\em Phys. Rev. C\/} {\bf 53} 459 -- 468

\bibitem{BRW94}
Beer H, Rupp G, Walter G, Voss F and K{\"a}ppeler F 1994 {\em Nucl. Instr.
  Meth. A\/} {\bf 337} 492 -- 503

\bibitem{PDA04}
Patronis N, Dababneh S, Assimakopoulos P, Gallino R, Heil M, K{\"a}ppeler F,
  Karamanis D, Koehler P, Mengoni A and Plag R 2004 {\em Phys. Rev. C\/} {\bf
  69} 025803

\bibitem{RAH03}
Reifarth R, Arlandini C, Heil M, K{\"a}ppeler F, Sedyshev P, Herman M, Rauscher
  T, Gallino R and Travaglio C 2003 {\em Ap. J.\/} {\bf 582} 1251 -- 1262

\bibitem{JaK95b}
Jaag S and K{\"a}ppeler F 1995 {\em Phys. Rev. C\/} {\bf 51} 3465 -- 3471

\bibitem{JaK96a}
Jaag S and K{\"a}ppeler F 1996 {\em Ap. J.\/} {\bf 464} 874--883

\bibitem{RHH03}
Reifarth R, Haight R, Heil M, Fowler M, K{\"a}ppeler F, Miller G, Rundberg R,
  Ullmann J and Wilhelmy J 2003 {\em Nucl. Phys.\/} {\bf A718} 478c

\bibitem{VDH07}
Vockenhuber C, Dillmann I, Heil M, K{\"a}ppeler F, Wallner A and Winckler N
  2007 {\em Phys. Rev. C\/} {\bf 75} 015804

\bibitem{URS09}
Uberseder E, Reifarth R, Schumann D, Dillmann I, Domingo~Pardo C, G{\"o}rres J,
  Heil M, K{\"a}ppeler F, Marganiec J, Neuhausen J, Pignatari M, Voss F, Walter
  S and Wiescher M 2009 {\em Phys. Rev. Lett.\/} {\bf 102} 151101

\bibitem{SND10}
Schumann D, Neuhausen J, Dillmann I, Domingo~Pardo C, K{\"a}ppeler F, Marganiec
  J, Voss F, Walter S, Heil M, Reifarth R, G{\"o}rres J, Uberseder E and
  Wiescher M 2010 {\em Nucl. Instr. Meth. A\/} {\bf 613} 347

\bibitem{PHK80}
Paul M, Henning W, Kutschera W, Stephenson E and Yntema J 1980 {\em Phys. Lett.
  B\/} {\bf 94} 303

\bibitem{DFK10}
Dillmann I, Faestermann T, Korschinek G, Lachner J, Maiti M, Poutivtsev M,
  Rugel G, Walter S, K{\"a}ppeler F, Erhard M, Junghans A, Nair C, Schwengner R
  and Wagner A 2010 {\em Nucl. Instr. Meth. B\/} {\bf 268} 1283

\bibitem{WCD08}
Wallner A, Coquard L, Dillmann I, Forstner O, Golser R, Heil M, K{\"a}ppeler F,
  Kutschera W, Mengoni A, Michlmayr L, Priller A, Steier P and Wiescher M 2008
  {\em J. Phys. G: Nucl. Part. Phys.\/} {\bf 35} 014018

\bibitem{WBD12}
Wallner A, Buczak K, Dillmann I, Feige J, K{\"a}ppeler F, Korschinek G, Lederer
  C, Mengoni A, Ott U, Paul M, Sch{\"a}tzel G, Steier P and Trautvetter H 2012
  {\em Publications of the Astronomical Society of Australia\/} {\bf 29} 115

\bibitem{DHK09}
Dillmann I, Heil M, K{\"a}ppeler F, Wallner A, Kutschera W, Priller A, Steier P
  and Paul M 2009 {\em Phys. Rev. C.\/} {\bf 79} 065805

\bibitem{CDW06}
Coquard L, Dillmann I, Wallner A, K{\"a}ppeler F and Kutschera W 2006 {\em
  Nuclei in the Cosmos-IX\/} http://pos.sissa.it/ ed Mengoni A and et~al
  (SISSA: Proceedings of Science) p contribution 274

\bibitem{RDF07}
Rugel G, Dillmann I, Faestermann T, Heil M, K{\"a}ppeler F, Knie K, Korschinek
  G, Kutschera W, Poutivtsev M and Wallner A 2007 {\em Nucl. Inst. Meth. B\/}
  {\bf 259} 683

\bibitem{DHK06a}
Dillmann I, Heil M, K{\"a}ppeler F, Faestermann T, Korschinek G, Knie K,
  Poutivtsev M, Rugel G, Wallner A and Rauscher T 2006 {\em Nuclei in the
  Cosmos-IX\/} http://pos.sissa.it ed Mengoni A and et~al (SISSA, Trieste:
  Proceedings of Science) p contribution 089

\bibitem{NFA09}
{Nakamura} T, {Fukuda} N, {Aoi} N, {Imai} N, {Ishihara} M, {Iwasaki} H,
  {Kobayashi} T, {Kubo} T, {Mengoni} A, {Motobayashi} T, {Notani} M, {Otsu} H,
  {Sakurai} H, {Shimoura} S, {Teranishi} T, {Watanabe} Y~X and {Yoneda} K 2009
  {\em Phys. Rev. C\/} {\bf 79} 035805

\bibitem{DAB03}
Datta~Pramanik U, Aumann T, Boretzky K, Carlson B, Cortina D, Elze T, Emling H,
  Geissel H, Gr{\"u}nschlo\ss{} A, Hellstr{\"o}m M, Ilievski S, Kratz J,
  Kulessa R, Leifels Y, Leistenschneider A, Lubkiewicz E, M{\"u}nzenberg G,
  Reiter P, Simon H, S{\"u}mmerer K, Wajda E and Walus W 2003 {\em Phys. Lett.
  B\/} {\bf 551} 63

\bibitem{BAA11}
Boretzky K, Agrawal B, Alkhazov G, Andreev G, Aumann T, Basu P, Bemmerer D,
  Bertini D, Bhattacharya P, Bhattacharya S, Blanco A, Caesar C, Chakraborty S,
  Chatterjee S, Cherciu M, Chulkov L, Ciobanu M, Cowan T, Datta~Pramanik U,
  Elekes Z, Endres J, Fetisov A, Fonte P, Galaviz D, Golotsov V, Haiduc M,
  Hehner J, Heil M, Heinz A, Hennig A, Ignatov A, Ickert G, Ivanov E, Kempe M,
  Kresan D, Krivshich A, Kumar~Das P, Leifels Y, Lopez L, Machado J, Maroussov
  V, Panja J, Potlog M, Rahaman A, Ray A, Reifarth R, R{\"o}der M, Rossi D, Roy
  J, Scheit H, Simon H, Sinha T, Sobiella M, Stach D, Stan E, Teubig P, Uvarov
  L, Vikhrov V, Volknandt M, Volkov S, Wagner A, W{\"u}stenfeld J, Yakorev D,
  Zhdanov A, Zilges A and Zuber K 2011 Neuland@r3b: A fully-active detector for
  time-of-flight and calorimetry of fast neutrons Tech. rep. GSI, Darmstadt gSI
  Sci. Rep., PHN-NUSTAR-NR-02, p. 174

\bibitem{SMV03}
Sonnabend K, Mohr P, Vogt K, Zilges A, Mengoni A, Rauscher T, Beer H,
  K{\"a}ppeler F and Gallino R 2003 {\em Ap. J.\/} {\bf 583} 506 -- 513

\bibitem{Wel08}
Weller H 2008 {\em Progress in Nuclear and Particle Physics\/} {\bf 62} 257 --
  303

\bibitem{EAB05}
Escher J, Ahle L, Bernstein L, Church J, Dietrich F, Forss{\'e}n C and Hoffman
  R 2005 {\em Nucl. Phys. A\/} {\bf 758} 86 -- 89

\bibitem{FDE07}
Forss\'{e}n C, Dietrich F~S, Escher J, Hoffman R~D and Kelley K 2007 {\em
  Physical Review C (Nuclear Physics)\/} {\bf 75} 055807

\bibitem{EBD12}
Escher J, Burke J, Dietrich F, Scielzo N, Thompson I and Younes W 2012 {\em
  Rev. Mod. Phys.\/} {\bf 84} 353

\bibitem{AAA01}
Abbondanno U, Andriamonje S, Andrzejewski J, Angelopoulos A, Assimakopoulos P,
  Bacri C~O, Badurek G, Beer H, Berthier B, Bondarenko I, Bos A, Bustreo N,
  Calvino F, Cano-Ott D, Capote R, Carlson P, Charpak G, Chauvin N, Cennini P,
  Chepel V, Colonna N, Cortes G, Corvi F~Damianoglou D, David S, Dimovasili E,
  Domingo C, Doroshenko A, Duran~Escribano I, Eleftheriadis C, Embid M, Ferrant
  L, Ferrari A, Ferreira-Marques R, Frais-Koelbl H, Furman W, Fursov B, Garzon
  J~A, Giomataris I, Gledenov Y, Gonzalez-Romero E, Goverdovski A, Gramegna F,
  Griesmayer E, Gunsing F, Haight R, Heil M, Hollander P, Ioannides K, Ioannu
  P, Isaev S, Jericha E, Kadi Y, K{\"a}ppeler F, Karadimos D, Karamanis D,
  Kayukov A, Kazakov L, Kelic A, Ketlerov V, Kitis G, Koehler P, Kopach Y,
  Kossionides E, Kroshkina I, Lacoste V, Lamboudis C, Leeb H, Lepretre A, Lopes
  M, Lozano M, Marrone S, Martinez-Val J, Mastinu P, Mengoni A, Meunier R,
  Mezentseva J, Milazzo P, Minguiez E, Mitrofanov V, Nicolis N, Nikolenko V,
  Oberhummer H, Pakou A, Pancin J, Papadopoulos K, Papaevangelou T, Paradela C,
  Paradellis T, Pavlik A, Pavlopoulos P, Perez-Parra A, Perriale L, Perlando J,
  Peskov V, Piksaikin V, Plag R, Plompen A, Poch A, Policarpo A, Popov A, Popov
  Y, Pretel C, Quesada J, Radermacher E, Rauscher T, Reifarth R, Rejmund F,
  Rubbia C, Rudolf G, Rullhusen P, Sakelliou L, Saldana F, Tagliente G, Tain J,
  Tapia C, Tassan-Got L, Terchychnyi R, Tsabaris C, Tsangas N, van Eijk C,
  Vannini G, Ventura A, Villamarin A, Vlachoudis V, Vlastou R, Voinov A, Voss
  F, Wendler H, Wiescher M, Wisshak K, Zeinalov S and Zhuravlev B 2001 Neutron
  tof facility (ps213) status report Tech. rep. CERN, Geneva, Switzerland
  report CERN/INTC 2001-021

\bibitem{CCV12}
Chiaveri E, Calviani M, Vlachoudis V, Weiss C and {the n\_TOF collaboration}
  2012 {\em Compound nuclear reactions and related topics\/} vol~21 (EPJ Web of
  Conferences)

\bibitem{GTB13}
{Guerrero} C, {Tsinganis} A, {Berthoumieux} E, {Barbagallo} M, {Belloni} F,
  {Gunsing} F, {Wei{\ss}} C, {Chiaveri} E, {Calviani} M, {Vlachoudis} V,
  {Altstadt} S, {Andriamonje} S, {Andrzejewski} J, {Audouin} L, {B{\'e}cares}
  V, {Be{\v c}v{\'a}{\v r}} F, {Billowes} J, {Boccone} V, {Bosnar} D, {Brugger}
  M, {Calvi{\~n}o} F, {Cano-Ott} D, {Carrapi{\c c}o} C, {Cerutti} F, {Chin} M,
  {Colonna} N, {Cort{\'e}s} G, {Cort{\'e}s-Giraldo} M~A, {Diakaki} M,
  {Domingo-Pardo} C, {Duran} I, {Dressler} R, {Dzysiuk} N, {Eleftheriadis} C,
  {Ferrari} A, {Fraval} K, {Ganesan} S, {Garc{\'{\i}}a} A~R, {Giubrone} G,
  {G{\"o}bel} K, {G{\'o}mez-Hornillos} M~B, {Gon{\c c}alves} I~F,
  {Gonz{\'a}lez-Romero} E, {Griesmayer} E, {Gurusamy} P, {Hern{\'a}ndez-Prieto}
  A, {Gurusamy} P, {Jenkins} D~G, {Jericha} E, {Kadi} Y, {K{\"a}ppeler} F,
  {Karadimos} D, {Kivel} N, {Koehler} P, {Kokkoris} M, {Krti{\v c}ka} M,
  {Kroll} J, {Lampoudis} C, {Langer} C, {Leal-Cidoncha} E, {Lederer} C, {Leeb}
  H, {Leong} L~S, {Losito} R, {Manousos} A, {Marganiec} J, {Mart{\'{\i}}nez} T,
  {Massimi} C, {Mastinu} P~F, {Mastromarco} M, {Meaze} M, {Mendoza} E,
  {Mengoni} A, {Milazzo} P~M, {Mingrone} F, {Mirea} M, {Mondalaers} W,
  {Papaevangelou} T, {Paradela} C, {Pavlik} A, {Perkowski} J, {Plompen} A,
  {Praena} J, {Quesada} J~M, {Rauscher} T, {Reifarth} R, {Riego} A, {Roman} F,
  {Rubbia} C, {Sabate-Gilarte} M, {Sarmento} R, {Saxena} A, {Schillebeeckx} P,
  {Schmidt} S, {Schumann} D, {Steinegger} P, {Tagliente} G, {Tain} J~L,
  {Tarr{\'{\i}}o} D, {Tassan-Got} L, {Valenta} S, {Vannini} G, {Variale} V,
  {Vaz} P, {Ventura} A, {Versaci} R, {Vermeulen} M~J, {Vlastou} R, {Wallner} A,
  {Ware} T, {Weigand} M, {Wright} T and {{\v Z}ugec} P 2013 {\em Eur. Phys. J.
  A\/} {\bf 49} 27

\bibitem{LBR90}
Lisowski P, Bowman C, Russell G and Wender S 1990 {\em Nucl. Sci. Eng.\/} {\bf
  106} 208

\bibitem{JPC04}
J-PARC 2008 J-parc annual report 2008 Tech. rep.
  http://j-parc.jp/documents/annual\_report/a\_report\_2008.pdf

\bibitem{Led13}
Lederer C and {the n\_TOF collaboration} 2013 {\em Nuclear Data for Science and
  Technology\/} (New York: BNL)

\bibitem{TMF11a}
{Tagliente} G, {Milazzo} P~M, {Fujii} K, {Abbondanno} U, {Aerts} G,
  {{\'A}lvarez} H, {Alvarez-Velarde} F, {Andriamonje} S, {Andrzejewski} J,
  {Audouin} L, {Badurek} G, {Baumann} P, {Be{\v c}v{\'a}{\v r}} F, {Belloni} F,
  {Berthoumieux} E, {Bisterzo} S, {Calvi{\~n}o} F, {Calviani} M, {Cano-Ott} D,
  {Capote} R, {Carrapi{\c c}o} C, {Cennini} P, {Chepel} V, {Chiaveri} E,
  {Colonna} N, {Cortes} G, {Couture} A, {Cox} J, {Dahlfors} M, {David} S,
  {Dillmann} I, {Domingo-Pardo} C, {Dridi} W, {Duran} I, {Eleftheriadis} C,
  {Embid-Segura} M, {Ferrari} A, {Ferreira-Marques} R, {Furman} W, {Gallino} R,
  {Goncalves} I, {Gonzalez-Romero} E, {Gramegna} F, {Guerrero} C, {Gunsing} F,
  {Haas} B, {Haight} R, {Heil} M, {Herrera-Martinez} A, {Jericha} E,
  {K{\"a}ppeler} F, {Kadi} Y, {Karadimos} D, {Karamanis} D, {Kerveno} M,
  {Kossionides} E, {Krti{\v c}ka} M, {Lamboudis} C, {Leeb} H, {Lindote} A,
  {Lopes} I, {Lozano} M, {Lukic} S, {Marganiec} J, {Marrone} S,
  {Mart{\'{\i}}nez} T, {Massimi} C, {Mastinu} P, {Mengoni} A, {Moreau} C,
  {Mosconi} M, {Neves} F, {Oberhummer} H, {O'Brien} S, {Pancin} J,
  {Papachristodoulou} C, {Papadopoulos} C, {Paradela} C, {Patronis} N, {Pavlik}
  A, {Pavlopoulos} P, {Perrot} L, {Pigni} M~T, {Plag} R, {Plompen} A, {Plukis}
  A, {Poch} A, {Praena} J, {Pretel} C, {Quesada} J, {Rauscher} T, {Reifarth} R,
  {Rosetti} M, {Rubbia} C, {Rudolf} G, {Rullhusen} P, {Salgado} J, {Santos} C,
  {Sarchiapone} L, {Savvidis} I, {Stephan} C, {Tain} J~L, {Tassan-Got} L,
  {Tavora} L, {Terlizzi} R, {Vannini} G, {Vaz} P, {Ventura} A, {Villamarin} D,
  {Vincente} M~C, {Vlachoudis} V, {Vlastou} R, {Voss} F, {Walter} S, {Wiescher}
  M and {Wisshak} K 2011 {\em Phys. Rev. C\/} {\bf 84} 015801

\bibitem{TMF11b}
{Tagliente} G, {Milazzo} P~M, {Fujii} K, {Abbondanno} U, {Aerts} G,
  {{\'A}lvarez} H, {Alvarez-Velarde} F, {Andriamonje} S, {Andrzejewski} J,
  {Audouin} L, {Badurek} G, {Baumann} P, {Be{\v c}v{\'a}{\v r}} F, {Belloni} F,
  {Berthoumieux} E, {Calvi{\~n}o} F, {Calviani} M, {Cano-Ott} D, {Capote} R,
  {Carrapi{\c c}o} C, {Cennini} P, {Chepel} V, {Chiaveri} E, {Colonna} N,
  {Cortes} G, {Couture} A, {Dahlfors} M, {David} S, {Dillmann} I,
  {Domingo-Pardo} C, {Dridi} W, {Duran} I, {Eleftheriadis} C, {Embid-Segura} M,
  {Ferrari} A, {Ferreira-Marques} R, {Furman} W, {Goncalves} I,
  {Gonzalez-Romero} E, {Gramegna} F, {Guerrero} C, {Gunsing} F, {Haas} B,
  {Haight} R, {Heil} M, {Herrera-Martinez} A, {Jericha} E, {K{\"a}ppeler} F,
  {Kadi} Y, {Karadimos} D, {Karamanis} D, {Kerveno} M, {Kossionides} E,
  {Krti{\v c}ka} M, {Lamboudis} C, {Leeb} H, {Lindote} A, {Lopes} I, {Lukic} S,
  {Marganiec} J, {Marrone} S, {Mart{\'{\i}}nez} T, {Massimi} C, {Mastinu} P,
  {Mengoni} A, {Moreau} C, {Mosconi} M, {Neves} F, {Oberhummer} H, {O'Brien} S,
  {Pancin} J, {Papachristodoulou} C, {Papadopoulos} C, {Paradela} C, {Patronis}
  N, {Pavlik} A, {Pavlopoulos} P, {Perrot} L, {Pigni} M~T, {Plag} R, {Plompen}
  A, {Plukis} A, {Poch} A, {Praena} J, {Pretel} C, {Quesada} J, {Reifarth} R,
  {Rosetti} M, {Rubbia} C, {Rudolf} G, {Rullhusen} P, {Salgado} J, {Santos} C,
  {Sarchiapone} L, {Savvidis} I, {Stephan} C, {Tain} J~L, {Tassan-Got} L,
  {Tavora} L, {Terlizzi} R, {Vannini} G, {Vaz} P, {Ventura} A, {Villamarin} D,
  {Vincente} M~C, {Vlachoudis} V, {Vlastou} R, {Voss} F, {Walter} S, {Wiescher}
  M and {Wisshak} K 2011 {\em Phys. Rev. C\/} {\bf 84} 055802

\bibitem{TMF13}
{Tagliente} G, {Milazzo} P~M, {Fujii} K, {Abbondanno} U, {Aerts} G,
  {{\'A}lvarez} H, {Alvarez-Velarde} F, {Andriamonje} S, {Andrzejewski} J,
  {Audouin} L, {Badurek} G, {Baumann} P, {Be{\v c}v{\'a}{\v r}} F, {Belloni} F,
  {Berthoumieux} E, {Calvi{\~n}o} F, {Calviani} M, {Cano-Ott} D, {Capote} R,
  {Carrapi{\c c}o} C, {Cennini} P, {Chepel} V, {Chiaveri} E, {Colonna} N,
  {Cortes} G, {Couture} A, {Dahlfors} M, {David} S, {Dillmann} I,
  {Domingo-Pardo} C, {Dridi} W, {Duran} I, {Eleftheriadis} C, {Embid-Segura} M,
  {Ferrari} A, {Ferreira-Marques} R, {Furman} W, {Goncalves} I,
  {Gonzalez-Romero} E, {Gramegna} F, {Guerrero} C, {Gunsing} F, {Haas} B,
  {Haight} R, {Heil} M, {Herrera-Martinez} A, {Jericha} E, {K{\"a}ppeler} F,
  {Kadi} Y, {Karadimos} D, {Karamanis} D, {Kerveno} M, {Kossionides} E,
  {Krti{\v c}ka} M, {Lamboudis} C, {Leeb} H, {Lindote} A, {Lopes} I, {Lukic} S,
  {Marganiec} J, {Marrone} S, {Mart{\'{\i}}nez} T, {Massimi} C, {Mastinu} P,
  {Mengoni} A, {Moreau} C, {Mosconi} M, {Neves} F, {Oberhummer} H, {O'Brien} S,
  {Papachristodoulou} C, {Papadopoulos} C, {Paradela} C, {Patronis} N, {Pavlik}
  A, {Pavlopoulos} P, {Perrot} L, {Pigni} M~T, {Plag} R, {Plompen} A, {Plukis}
  A, {Poch} A, {Praena} J, {Pretel} C, {Quesada} J, {Reifarth} R, {Rosetti} M,
  {Rubbia} C, {Rudolf} G, {Rullhusen} P, {Salgado} J, {Santos} C, {Sarchiapone}
  L, {Savvidis} I, {Stephan} C, {Tain} J~L, {Tassan-Got} L, {Tavora} L,
  {Terlizzi} R, {Vannini} G, {Vaz} P, {Ventura} A, {Villamarin} D, {Vincente}
  M~C, {Vlachoudis} V, {Vlastou} R, {Voss} F, {Walter} S, {Wiescher} M and
  {Wisshak} K 2013 {\em Phys. Rev. C\/} {\bf 87} 014622

\bibitem{TAA07}
{Terlizzi} R, {Abbondanno} U, {Aerts} G, {{\'A}lvarez} H, {Alvarez-Velarde} F,
  {Andriamonje} S, {Andrzejewski} J, {Assimakopoulos} P, {Audouin} L, {Badurek}
  G, {Baumann} P, {Be{\v c}v{\'a}{\v r}} F, {Berthoumieux} E, {Calviani} M,
  {Calvi{\~n}o} F, {Cano-Ott} D, {Capote} R, {Albornoz} A~C~D, {Cennini} P,
  {Chepel} V, {Chiaveri} E, {Colonna} N, {Cortes} G, {Couture} A, {Cox} J,
  {Dahlfors} M, {David} S, {Dillmann} I, {Dolfini} R, {Domingo-Pardo} C,
  {Dridi} W, {Duran} I, {Eleftheriadis} C, {Embid-Segura} M, {Ferrant} L,
  {Ferrari} A, {Ferreira-Marques} R, {Fitzpatrick} L, {Frais-Koelbl} H, {Fujii}
  K, {Furman} W, {Gallino} R, {Goncalves} I, {Gonzalez-Romero} E, {Goverdovski}
  A, {Gramegna} F, {Griesmayer} E, {Guerrero} C, {Gunsing} F, {Haas} B,
  {Haight} R, {Heil} M, {Herrera-Martinez} A, {Igashira} M, {Isaev} S,
  {Jericha} E, {Kadi} Y, {K{\"a}ppeler} F, {Karamanis} D, {Karadimos} D,
  {Kerveno} M, {Ketlerov} V, {Koehler} P, {Konovalov} V, {Kossionides} E,
  {Krti{\v c}ka} M, {Lamboudis} C, {Leeb} H, {Lindote} A, {Lopes} I, {Lozano}
  M, {Lukic} S, {Marganiec} J, {Marques} L, {Marrone} S, {Massimi} C, {Mastinu}
  P, {Mengoni} A, {Milazzo} P~M, {Moreau} C, {Mosconi} M, {Neves} F,
  {Oberhummer} H, {O'Brien} S, {Pancin} J, {Papachristodoulou} C,
  {Papadopoulos} C, {Paradela} C, {Patronis} N, {Pavlik} A, {Pavlopoulos} P,
  {Perrot} L, {Pignatari} M, {Plag} R, {Plompen} A, {Plukis} A, {Poch} A,
  {Pretel} C, {Quesada} J, {Rauscher} T, {Reifarth} R, {Rosetti} M, {Rubbia} C,
  {Rudolf} G, {Rullhusen} P, {Salgado} J, {Sarchiapone} L, {Savvidis} I,
  {Stephan} C, {Tagliente} G, {Tain} J~L, {Tassan-Got} L, {Tavora} L, {Vannini}
  G, {Vaz} P, {Ventura} A, {Villamarin} D, {Vincente} M~C, {Vlachoudis} V,
  {Vlastou} R, {Voss} F, {Walter} S, {Wendler} H, {Wiescher} M and {Wisshak} K
  2007 {\em Phys. Rev. C\/} {\bf 75} 035807

\bibitem{LCD11}
Lederer C, Colonna N, Domingo-Pardo C, Gunsing F, K{\"a}ppeler F, Massimi C,
  Mengoni A, Wallner A and {the n\_TOF Collaboration} 2011 {\em Phys. Rev. C\/}
  {\bf 83} 034608

\bibitem{DAA07b}
{Domingo-Pardo} C, {Abbondanno} U, {Aerts} G, {{\'A}lvarez} H,
  {Alvarez-Velarde} F, {Andriamonje} S, {Andrzejewski} J, {Assimakopoulos} P,
  {Audouin} L, {Badurek} G, {Baumann} P, {Be{\v c}v{\'a}{\v r}} F,
  {Berthoumieux} E, {Bisterzo} S, {Calvi{\~n}o} F, {Calviani} M, {Cano-Ott} D,
  {Capote} R, {Carrapi{\c c}o} C, {Cennini} P, {Chepel} V, {Chiaveri} E,
  {Colonna} N, {Cortes} G, {Couture} A, {Cox} J, {Dahlfors} M, {David} S,
  {Dillman} I, {Dolfini} R, {Dridi} W, {Duran} I, {Eleftheriadis} C,
  {Embid-Segura} M, {Ferrant} L, {Ferrari} A, {Ferreira-Marques} R,
  {Fitzpatrick} L, {Frais-Koelbl} H, {Fujii} K, {Furman} W, {Gallino} R,
  {Goncalves} I, {Gonzalez-Romero} E, {Goverdovski} A, {Gramegna} F,
  {Griesmayer} E, {Guerrero} C, {Gunsing} F, {Haas} B, {Haight} R, {Heil} M,
  {Herrera-Martinez} A, {Igashira} M, {Isaev} M, {Jericha} E, {K{\"a}ppeler} F,
  {Kadi} Y, {Karadimos} D, {Karamanis} D, {Kerveno} M, {Ketlerov} V, {Koehler}
  P, {Konovalov} V, {Kossionides} E, {Krti{\v c}ka} M, {Lamboudis} C, {Leeb} H,
  {Lindote} A, {Lopes} I, {Lozano} M, {Lukic} S, {Marganiec} J, {Marrone} S,
  {Massimi} C, {Mastinu} P, {Mengoni} A, {Milazzo} P~M, {Moreau} C, {Mosconi}
  M, {Neves} F, {Oberhummer} H, {Oshima} M, {O'Brien} S, {Pancin} J,
  {Papachristodoulou} C, {Papadopoulos} C, {Paradela} C, {Patronis} N, {Pavlik}
  A, {Pavlopoulos} P, {Perrot} L, {Plag} R, {Plompen} A, {Plukis} A, {Poch} A,
  {Pretel} C, {Quesada} J, {Rauscher} T, {Reifarth} R, {Rosetti} M, {Rubbia} C,
  {Rudolf} G, {Rullhusen} P, {Salgado} J, {Sarchiapone} L, {Savvidis} I,
  {Stephan} C, {Tagliente} G, {Tain} J~L, {Tassan-Got} L, {Tavora} L,
  {Terlizzi} R, {Vannini} G, {Vaz} P, {Ventura} A, {Villamarin} D, {Vincente}
  M~C, {Vlachoudis} V, {Vlastou} R, {Voss} F, {Walter} S, {Wendler} H,
  {Wiescher} M and {Wisshak} K 2007 {\em Phys. Rev. C\/} {\bf 76} 045805

\bibitem{DAA06a}
{Domingo-Pardo} C, {Abbondanno} U, {Aerts} G, {{\'A}lvarez-Pol} H,
  {Alvarez-Velarde} F, {Andriamonje} S, {Andrzejewski} J, {Assimakopoulos} P,
  {Audouin} L, {Badurek} G, {Baumann} P, {Be{\v c}v{\'a}{\v r}} F,
  {Berthoumieux} E, {Calvi{\~n}o} F, {Cano-Ott} D, {Capote} R, {Albornoz}
  A~C~D, {Cennini} P, {Chepel} V, {Chiaveri} E, {Colonna} N, {Cortes} G,
  {Couture} A, {Cox} J, {Dahlfors} M, {David} S, {Dillman} I, {Dolfini} R,
  {Dridi} W, {Duran} I, {Eleftheriadis} C, {Embid-Segura} M, {Ferrant} L,
  {Ferrari} A, {Ferreira-Marques} R, {Fitzpatrick} L, {Frais-Koelbl} H, {Fujii}
  K, {Furman} W, {Gallino} R, {Goncalves} I, {Gonzalez-Romero} E, {Goverdovski}
  A, {Gramegna} F, {Griesmayer} E, {Guerrero} C, {Gunsing} F, {Haas} B,
  {Haight} R, {Heil} M, {Herrera-Martinez} A, {Igashira} M, {Isaev} S,
  {Jericha} E, {Kadi} Y, {K{\"a}ppeler} F, {Karamanis} D, {Karadimos} D,
  {Kerveno} M, {Ketlerov} V, {Koehler} P, {Konovalov} V, {Kossionides} E,
  {Krti{\v c}ka} M, {Lamboudis} C, {Leeb} H, {Lindote} A, {Lopes} I, {Lozano}
  M, {Lukic} S, {Marganiec} J, {Marques} L, {Marrone} S, {Mastinu} P, {Mengoni}
  A, {Milazzo} P~M, {Moreau} C, {Mosconi} M, {Neves} F, {Oberhummer} H,
  {Oshima} M, {O'Brien} S, {Pancin} J, {Papachristodoulou} C, {Papadopoulos} C,
  {Paradela} C, {Patronis} N, {Pavlik} A, {Pavlopoulos} P, {Perrot} L, {Plag}
  R, {Plompen} A, {Plukis} A, {Poch} A, {Pretel} C, {Quesada} J, {Rauscher} T,
  {Reifarth} R, {Rosetti} M, {Rubbia} C, {Rudolf} G, {Rullhusen} P, {Salgado}
  J, {Sarchiapone} L, {Savvidis} I, {Stephan} C, {Tagliente} G, {Tain} J~L,
  {Tassan-Got} L, {Tavora} L, {Terlizzi} R, {Vannini} G, {Vaz} P, {Ventura} A,
  {Villamarin} D, {Vincente} M~C, {Vlachoudis} V, {Vlastou} R, {Voss} F,
  {Walter} S, {Wendler} H, {Wiescher} M and {Wisshak} K 2006 {\em Phys. Rev.
  C\/} {\bf 74} 025807

\bibitem{MKB12}
Massimi C, Koehler P, Bisterzo S and {the n\_TOF collaboration} 2012 {\em Phys.
  Rev. C\/} {\bf 85} 044615

\bibitem{MFM10}
Mosconi M, Fujii K, Mengoni A, Heil M, K{\"a}ppeler F and n\_TOF Collaboration)
  T 2010 {\em Phys. Rev. C\/} {\bf 82} 015802

\bibitem{MHK10}
Mosconi M, Heil M, K{\"a}ppeler F and Mengoni A 2010 {\em Phys. Rev. C\/} {\bf
  82} 015803

\bibitem{FMM10}
Fujii K, Mosconi M, Mengoni A and et a 2010 {\em Phys. Rev. C\/} {\bf 82}
  015804

\bibitem{LMA13}
{Lederer} C, {Massimi} C, {Altstadt} S, {Andrzejewski} J, {Audouin} L,
  {Barbagallo} M, {B{\'e}cares} V, {Be{\v c}v{\'a}{\v r}} F, {Belloni} F,
  {Berthoumieux} E, {Billowes} J, {Boccone} V, {Bosnar} D, {Brugger} M,
  {Calviani} M, {Calvi{\~n}o} F, {Cano-Ott} D, {Carrapi{\c c}o} C, {Cerutti} F,
  {Chiaveri} E, {Chin} M, {Colonna} N, {Cort{\'e}s} G, {Cort{\'e}s-Giraldo}
  M~A, {Diakaki} M, {Domingo-Pardo} C, {Duran} I, {Dressler} R, {Dzysiuk} N,
  {Eleftheriadis} C, {Ferrari} A, {Fraval} K, {Ganesan} S, {Garc{\'{\i}}a} A~R,
  {Giubrone} G, {G{\'o}mez-Hornillos} M~B, {Gon{\c c}alves} I~F,
  {Gonz{\'a}lez-Romero} E, {Griesmayer} E, {Guerrero} C, {Gunsing} F,
  {Gurusamy} P, {Jenkins} D~G, {Jericha} E, {Kadi} Y, {K{\"a}ppeler} F,
  {Karadimos} D, {Kivel} N, {Koehler} P, {Kokkoris} M, {Korschinek} G, {Krti{\v
  c}ka} M, {Kroll} J, {Langer} C, {Leeb} H, {Leong} L~S, {Losito} R, {Manousos}
  A, {Marganiec} J, {Mart{\'{\i}}nez} T, {Mastinu} P~F, {Mastromarco} M,
  {Meaze} M, {Mendoza} E, {Mengoni} A, {Milazzo} P~M, {Mingrone} F, {Mirea} M,
  {Mondelaers} W, {Paradela} C, {Pavlik} A, {Perkowski} J, {Pignatari} M,
  {Plompen} A, {Praena} J, {Quesada} J~M, {Rauscher} T, {Reifarth} R, {Riego}
  A, {Roman} F, {Rubbia} C, {Sarmento} R, {Schillebeeckx} P, {Schmidt} S,
  {Schumann} D, {Tagliente} G, {Tain} J~L, {Tarr{\'{\i}}o} D, {Tassan-Got} L,
  {Tsinganis} A, {Valenta} S, {Vannini} G, {Variale} V, {Vaz} P, {Ventura} A,
  {Versaci} R, {Vermeulen} M~J, {Vlachoudis} V, {Vlastou} R, {Wallner} A,
  {Ware} T, {Weigand} M, {Wei{\ss}} C, {Wright} T~J and {{\v Z}ugec} P 2013
  {\em Phys. Rev. Lett.\/} {\bf 110} 022501

\bibitem{AAA04c}
Abbondanno U, Aerts G, Alvarez H, Andriamonje S, Angelopoulos A, Assimakopoulos
  P, Bacri C~O, Badurek G, Baumann P, Becvar F, Beer H, Benlliure J, Berthier
  B, Berthoumieux E, Bertuzzi J, Blanc D, Bonzano R, Borcea C, Bourquin P,
  Bustreo N, Buttkus J, Calvino F, Cano-Ott D, Capote R, Cappi R, Carlier J,
  Carlson P, Cennini E, Cennini P, Chapuis D~Chepel V, Chiaveri E, Chohan V,
  Coceva C, Colonna N, Come J, Cortes G, Cortina D, Couture A, Cox J, Dababneh
  S, Daems G, Dahlfors M, Dangendorf V, David S, Dobers C, Dolfini R,
  Domingo-Pardo C, Duran-Escribano I, Durieu L, Eleftheriadis C, Embid-Segura
  M, Ferrant L, Ferrari A, Ferreira-Lorenco L, Ferreira-Marques R, Flamant C,
  Frais-Koelbl H, Furman W, Gaidon M, Gascon J, Gasser D, Giomataris Y,
  Giovannozzi M, Goncalves I, Gonzalez-Romero E, Goulas I, Goverdovski A,
  Gramegna F, Griesmayer E, Gunsing F, Haight R, Heil M, Herrera-Martinez A,
  Ioannides K, Janeva N, Jericha E, Kadi Y, K{\"a}ppeler F, Karamanis D, Kelic
  A, Ketlerov V, Kitis G, Koehler P, Konovalov V, Kossionides E, Kowalik G,
  Kuhnl-Kinel J, Lacoste V, Laxroix J, Leeb H, Lindote A, Lopes I, Lozano M,
  Lukic S, Magnin R, Mahner E, Marie F, Markov S, Marrone S, Martinez-Val J,
  Mastinu P, Mengoni A, Messerli R~Metral G, Milazzo P, Minguez E,
  Molina-Coballes A, Monteiro J, Moreau C, Neves F, Niquevert B, Nolte R,
  Oberhummer H, O'Brien S, Pancin J, Paradela-Dobarro C, Pavlik A, Pavlopoulos
  P, Perez-Parra A, , Perlado J, Perrot L, Peskov V, Plag R, Plompen A, Plukis
  A, Poch A, Poehler M, Policarpo A, Pretel C, Quesada J, Radermacher E, Raich
  U, Raman S, Rapp W, Rauscher T, Reifarth R, Rejmund F, Riunaud J, Rollinger
  G, Rosetti M, Rubbia C, Rudolf G, Rullhusen P, Saldana F, Salgado J, Savvidis
  E, Silari M, Soares J, Stephan C~Tagliente G, Tain J, Tapia C, Tassan-Got L,
  Tavora L, Terlizzi R, Terrani M, Tsangas N, Van~Baaren W, Vannini G, Vaz P,
  Ventura A, Villamarin-Fernandez D, Vicente-Vicente M, Vlachoudis V, Vlastou
  R, Voss F, Weierganz M, Wendler H, Wiescher M, Wisshak K, Zanini L and
  Zanolli M 2004 {\em Phys. Rev. Lett.\/} {\bf 93} 161103

\bibitem{MAA06}
Marrone S, Abbondanno U, Aerts G, Alvarez-Velarde F, Alvarez-Pol H, Andriamonje
  S, Andrzejewski J, Badurek G, Baumann P, Becvar F, Benlliure J, Berthomieux
  E, Calvino F, Cano-Ott D, Capote R, Cennini P, Chepel V, Chiaveri E, Colonna
  N, Cortes G, Cortina D, Couture A, Cox J, Dababneh S, Dahlfors M, David S,
  Dolfini R, Domingo-Pardo C, Duran-Escribano I, Embid-Segura M, Ferrant L,
  Ferrari A, Ferreira-Marques R, Frais-Koelbl H, Fujii K, Furman W~I, Gallino
  R, Goncalves I~F, Gonzalez-Romero E, Goverdovski A, Gramegna F, Griesmayer E,
  Gunsing F, Haas B, Haight R, Heil M, Herrera-Martinez A, Isaev S, Jericha E,
  Kappeler F, Kadi Y, Karadimos D, Kerveno M, Ketlerov V, Koehler P~E,
  Konovalov V, Kritcka M, Lamboudis C, Leeb H, Lindote A, Lopes M~I, Lozano M,
  Lukic S, Marganiec J, Martinez-Val J, Mastinu P~F, Mengoni A, Milazzo P~M,
  Molina-Coballes A, Moreau C, Mosconi M, Neves F, Oberhummer H, O'Brien S,
  Pancin J, Papaevangelou T, Paradela C, Pavlik A, Pavlopoulos P, Perlado J~M,
  Perrot L, Pignatari M, Pigni M~T, Plag R, Plompen A, Plukis A, Poch A,
  Policarpo A, Pretel C, Quesada J~M, Raman S, Rapp W, Rauscher T, Reifarth R,
  Rosetti M, Rubbia C, Rudolf G, Rullhusen P, Salgado J, Soares J~C, Stephan C,
  Tagliente G, Tain J~L, Tassan-Got L, Tavora L~M~N, Terlizzi R, Vannini G, Vaz
  P, Ventura A, Villamarin-Fernandez D, Vincente-Vincente M, Vlachoudis V, Voss
  F, Wendler H, Wiescher M, Wisshak K and n~TOF~Collaborat 2006 {\em Phys. Rev.
  C\/} {\bf 73} 034604

\bibitem{KGO93}
Koehler P, Graff S, O'Brian H, Gledenov Y and Popov Y 1993 {\em Phys. Rev. C\/}
  {\bf 47} 2107 -- 2112

\bibitem{KKV97}
Koehler P, Kavanagh R, Vogelaar R, Gledenov Y and Popov Y 1997 {\em Phys. Rev.
  C\/} {\bf 56} 1138--1143

\bibitem{WBC12}
Weigand M, Bredeweg T, Couture A, Jandel M, K{\"a}ppeler F, Lederer C,
  Korschinek G, Krticka M, O’Donnell J, Reifarth R, Ullmann J and Wallner A
  2010 {\em Nuclei in the Cosmos-XII\/} http://pos.sissa.it/ ed Lugaro M and
  Lattanzio J (SISSA, Trieste: Proceedings of Science) contr. 184

\bibitem{KRU12}
Koehler P~E, Reifarth R, Ullmann J~L, Bredeweg T~A, O'Donnell J~M, Rundberg
  R~S, Vieira D~J and Wouters J~M 2012 {\em Physical Review Letters\/} {\bf
  108}(14) 142502

\bibitem{RAA06}
Reifarth R, Agvaanluvsan U, Alpizar-Vicente A, Bredeweg T, Esch E~I, Greife U,
  Haight R, Hatarik R, Herwig F, O’Donnell J, Rundberg R, Schwantes J,
  Ullmann J, Vieira D and Wouters J 2006 {\em New Astronomy Reviews\/} {\bf 50}
  644–647

\bibitem{KFG11}
Kimura A, Furutaka K, Goko S, Harada H, Kin T, Kitatani F, Koizumi M, Nakamura
  S, Ohta M, Toh Y, Fujii T, Fukutani S, Hori J, Takamiya K, Igashira M,
  Katabuchi T, Mizumoto M, Kamiyama T, Kino K and Kuynagi Y 2011 {\em Journal
  of the Korean Physical Society\/} {\bf 59} 1828

\bibitem{orela}
 {\em http://www.phy.ornl.gov/orela/orela.html/\/}

\bibitem{MHW79}
Macklin R, Halperin J and Winters R 1979 {\em Nucl. Instr. Meth. A\/} {\bf 164}
  213

\bibitem{MHW75}
Macklin R, Halperin J and Winters R 1975 {\em Phys. Rev. C.\/} {\bf 11} 1270

\bibitem{MHW77}
Macklin R, Halperin J and Winters R 1977 {\em Ap. J.\/} {\bf 217} 222

\bibitem{MaY87b}
Macklin R and Young P 1987 {\em Nucl. Sci. Eng.\/} {\bf 97} 239 -- 244

\bibitem{Mac88}
Macklin R 1988 {\em Nucl. Sci. Eng.\/} {\bf 99} 133

\bibitem{Mac82b}
Macklin R 1982 {\em Nucl. Sci. Eng.\/} {\bf 81} 520

\bibitem{Mac83}
Macklin R 1983 {\em Nucl. Sci. Eng.\/} {\bf 85} 350

\bibitem{Mac85a}
Macklin R 1985 {\em Nucl. Sci. Eng.\/} {\bf 89} 79

\bibitem{Mac85b}
Macklin R 1985 {\em Astrophys. Space Sci.\/} {\bf 115} 71

\bibitem{KSW97}
Koehler P, Spencer R, Winters R, Guber K, Harvey J, Hill N and Smith M 1997
  {\em Nucl. Phys.\/} {\bf A621} 258c -- 261c

\bibitem{KSG98}
Koehler P, Spencer R, Guber K, Winters R, Raman S, Harvey J, Hill N, Blackmon
  J, Bardayan D, Larson D, Lewis T, Pierce D and Smith M 1998 {\em Phys. Rev.
  C\/} {\bf 57} R1558 -- R1561

\bibitem{KWG01a}
Koehler P, Winters R, Guber K, Rauscher T, Harvey J, Raman S, Spencer R,
  Blackmon J, Larson D, Bardayan D and Lewis T 2001 {\em Phys. Rev. C\/} {\bf
  63} 049901

\bibitem{RKK03}
Rapp W, Koehler P, K{\"a}ppeler F and Raman S 2003 {\em Phys. Rev. C\/} {\bf
  68} 015802

\bibitem{KGR04}
Koehler P~E, Gledenov Y~M, Rauscher T and Fr{\"o}hlich C 2004 {\em Phys. Rev.
  C\/} {\bf 69} 15803

\bibitem{KoG13}
Koehler P~E and Guber K~H 2013 {\em Phys. Rev. C\/} {\bf 88} 035802

\bibitem{BeS78}
Bensussan A and Salome J 1978 {\em Nucl. Instr. Meth.\/} {\bf 155} 11

\bibitem{BDG09}
Block R, Danon Y, Gunsing F and Haight R 2009 {\em Handbook of Nuclear
  Engineering, Vol. I\/} (Berlin: Springer) iSBN: 9780387981307

\bibitem{GKD03}
Guber K, Koehler P, Derrien H, Valentine T, Leal L and Sayer R 2003 {\em Phys.
  Rev. C\/} {\bf 67} 062802

\bibitem{MBB05}
Mutti P, Beer H, Brusegan A, Corvi F and Gallino R 2005 {\em Nuclear Data for
  Science and Technology\/} AIP Conference Series 769 ed Haight R, Chadwick M,
  Kawano T and Talou P (New York: AIP) pp 1327 -- 1330

\bibitem{GSK97a}
Guber K, Spencer R, Koehler P and Winters R 1997 {\em Phys. Rev. Lett.\/} {\bf
  78} 2704 -- 2707

\bibitem{BCM97}
Beer H, Corvi F and Mutti P 1997 {\em Ap. J.\/} {\bf 474} 843 -- 861

\bibitem{WWG02}
Wagemans C, Wagemans J, Goeminne G, Serot O, Loiselet M and Gaelens M 202 {\em
  Phys. Rev. C\/} {\bf 65} 034614

\bibitem{DWG07}
De~Smet L, Wagemans C, Goeminne G, Heyse J and Van~Gils J 2007 {\em Phys. Rev.
  C\/} {\bf 75} 034617

\bibitem{BBE13}
Beyer R, Birgersson E, Elekesa Z, Ferrari A, Grosse E, Hannaske R, Junghans A,
  K{\"o}gler T, Massarczyk R, Matica A, Nolte R, Schwengner R and Wagner A 2013
  {\em Nucl. Inst. Meth. A\/} {\bf 723} 151

\bibitem{Ber55}
Bergman A and et~al 1955 {\em Peaceful Uses of Atomic Energy, Vol. 4\/}
  (Geneva: United Nations) p 135

\bibitem{AAA02}
Ab{\'a}nades A, Aleixandre J, Andriamonje S and {the TARC Collaboration} 2002
  {\em Nucl. Instrum. Methods Phys. Res.\/} {\bf A 478} 577--730

\bibitem{RHO05}
Rochman D, Haight R, O'Donnell J, Michaudon A, Wender S, Vieira D, Bond E,
  Bredeweg T, Kronenberg A, Wilhelmy J, Ethvignot T, Granier T, Petit M and
  Danon Y 2005 {\em Nucl. Instr. Meth. A\/} {\bf 550} 397 ISSN 0168-9002
  \urlprefix\url{http://www.sciencedirect.com/science/article/pii/S0168900205012283}

\bibitem{YKK93}
Yamanaka A, Kimura I, Kanazawa S, Kobayashi K, Yamamoto S, Nakagome Y, Fujita Y
  and Tamai T 1993 {\em Journal of Nuclear Science and Technology\/} {\bf 30}
  863--869 (\textit{Preprint}
  \eprint{http://www.tandfonline.com/doi/pdf/10.1080/18811248.1993.9734560})
  \urlprefix\url{http://www.tandfonline.com/doi/abs/10.1080/18811248.1993.9734560}

\bibitem{MSB85}
Maguire H, Stopa C, Block R, Harris D, Slovacek R, Dabbs J, Dougan R, Hoff R
  and Lougheed R {1985} {\em {Nucl. Sci. Eng.}\/} {\bf {89}} {293--304} ISSN
  {0029-5639}

\bibitem{KLY02}
Kobayashi K, Lee S, Yamamoto S, Cho H and Fujita Y 2002 {\em Journal of Nuclear
  Science and Technology\/} {\bf 39} 111--119 (\textit{Preprint}
  \eprint{http://www.tandfonline.com/doi/pdf/10.1080/18811248.2002.9715164})
  \urlprefix\url{http://www.tandfonline.com/doi/abs/10.1080/18811248.2002.9715164}

\bibitem{RHH04}
Reifarth R, Haight R, Heil M, K{\"a}ppeler F and Vieira D 2003 {\em Nucl.
  Instr. Meth. A\/} {\bf 524} 215 -- 226

\bibitem{WVT96a}
Wisshak K, Voss F, Theis C, K{\"a}ppeler F, Guber K, Kazakov L, Kornilov N and
  Reffo G 1996 {\em Phys. Rev. C\/} {\bf 54} 1451 -- 1462

\bibitem{WVK92}
Wisshak K, Voss F, K{\"a}ppeler F and Reffo G 1992 {\em Phys. Rev. C\/} {\bf
  45} 2470 -- 2486

\bibitem{RKV04}
Reifarth R, K{\"a}ppeler F, Voss F, Wisshak K, Gallino R, Pignatari M and
  Straniero O 2004 {\em Ap. J.\/} {\bf 614} 363 -- 370

\bibitem{VWG94b}
Voss F, Wisshak K, Guber K, K{\"a}ppeler F and Reffo G 1994 {\em Phys. Rev.
  C\/} {\bf 50} 2582 -- 2601

\bibitem{WGV93}
Wisshak K, Guber K, Voss F, K{\"a}ppeler F and Reffo G 1993 {\em Phys. Rev.
  C\/} {\bf 48} 1401 -- 1419

\bibitem{WVK95b}
Wisshak K, Voss F, K{\"a}ppeler F, Guber K, Kazakov L, Kornilov N, Uhl M and
  Reffo G 1995 {\em Phys. Rev. C\/} {\bf 52} 2762--2779

\bibitem{WVK06a}
Wisshak K, Voss F, K{\"a}ppeler F and Kazakov L 2006 {\em Phys. Rev. C\/} {\bf
  73} 015807

\bibitem{WVK02}
Wisshak K, Voss F, K{\"a}ppeler F and Kazakov L 2002 {\em Phys. Rev. C\/} {\bf
  66} 025801 (11)

\bibitem{WVA01}
Wisshak K, Voss F, Arlandini C, Be\u{c}v\'a\u{r} F, Straniero O, Gallino R,
  Heil M, K{\"a}ppeler F, Krti\u{c}ka M, Masera S, Reifarth R and Travaglio C
  2001 {\em Phys. Rev. Lett.\/} {\bf 87} 251102 (4)

\bibitem{WVa04}
Wisshak K, Voss F, Arlandini C, Heil M, K{\"a}ppeler F, Reifarth R,
  Be\u{c}v\'a\u{r} F and Krti\u{c}ka M 2004 {\em Phys. Rev. C\/} {\bf 69}
  055801 (12)

\bibitem{RHK09}
Reifarth R, Heil M, K{\"a}ppeler F and Plag R 2009 {\em Nucl. Instr. Meth. A\/}
  {\bf 608} 139

\bibitem{WiK81}
Wisshak K and K{\"a}ppeler F 1981 {\em Nucl. Sci. Eng.\/} {\bf 77} 58--70

\bibitem{NIM91}
Nagai Y, Igashira M, Mukai N, Ohsaki T, Uesawa F, Takeda K, Ando T, Kitazawa H,
  Kubono S and Fukuda T 1991 {\em Ap. J.\/} {\bf 381} 444--448

\bibitem{ONI94}
Ohsaki T, Nagai Y, Igashira M, Shima T, Seino S and Irie T 1994 {\em Ap. J.\/}
  {\bf 422} 912

\bibitem{SOK97}
Shima T, Okazaki F, Kikuchi T, Kobayashi T, Kii T, Baba T, Y N and Igashira M
  1997 {\em Nucl. Phys.\/} {\bf A621} 231c

\bibitem{INM95}
Igashira M, Nagai Y, Masuda K, Ohsaki T and Kitazawa H 1995 {\em Ap. J.\/} {\bf
  441} L89 -- L92

\bibitem{BPS79}
Bradley T, Parsa Z, Stelts M and Chrien R 1979 {\em Nuclear Cross Sections for
  Technology\/} ({\em NBS Special Publication\/} vol 594) ed Fowler J~L,
  Johnson C~H and Bowman C~D (Washington D.C.: National Bureau of Standards) p
  344

\bibitem{GKK08}
Gritzay O, Kolotyi V, Klimova N, Kalchenko O, Gnidak M and Corona P 2008 {\em
  Nuclear Data for Science and Technology\/} ed Bersillon O, Gunsing F, Bauge
  E, Jaqmin R and Leray S (Paris: EDP Sciences) p 543

\bibitem{Gri11}
Gritzay O 2011 {\em Neutron Sources Spectra for EXFOR\/} INDC(NDS)-0590 ed
  Simakov S and K{\"a}ppeler F (Vienna: IAEA) p~21

\bibitem{CoR07}
Couture A and Reifarth R 2007 {\em Atomic Data Nucl. Data Tables\/} {\bf 93}
  807--830

\bibitem{RaT00}
Rauscher T and Thielemann F~K 2000 {\em Atomic Data Nucl. Data Tables\/} {\bf
  75} 1

\bibitem{Rau12}
Rauscher T 2012 {\em Ap. J. Lett.\/} {\bf 755} L10

\bibitem{Rau12a}
Rauscher T 2012 code smaragd, version 0.9.0s unpublished

\bibitem{Gor05}
Goriely S 2005 Most statistical model code, version 2005 Tech. rep.
  http://www-astro.ulb.ac.be

\bibitem{KHD05}
Koning A, Hilaire S and Duijvestijn M 2005 {\em International Conference on
  Nuclear Data for Science and Technology\/} ed Haight R, Chadwick M, Kawano T
  and Talou P (New York: American Insitute of Physics) pp 1154--1157 {AIP
  Conference Series 769}

\bibitem{FBB13}
{Frenje} J~A, {Bionta} R, {Bond} E~J, {Caggiano} J~A, {Casey} D~T, {Cerjan} C,
  {Edwards} J, {Eckart} M, {Fittinghoff} D~N, {Friedrich} S, {Glebov} V~Y,
  {Glenzer} S, {Grim} G, {Haan} S, {Hatarik} R, {Hatchett} S, {Gatu Johnson} M,
  {Jones} O~S, {Kilkenny} J~D, {Knauer} J~P, {Landen} O, {Leeper} R, {Le Pape}
  S, {Lerche} R, {Li} C~K, {Mackinnon} A, {McNaney} J, {Merrill} F~E, {Moran}
  M, {Munro} D~H, {Murphy} T~J, {Petrasso} R~D, {Rygg} R, {Sangster} T~C,
  {S{\'e}guin} F~H, {Sepke} S, {Spears} B, {Springer} P, {Stoeckl} C and
  {Wilson} D~C 2013 {\em Nuclear Fusion\/} {\bf 53} 043014

\bibitem{TaY87}
Takahashi K and Yokoi K 1987 {\em Atomic Data Nucl. Data Tables\/} {\bf 36} 375

\bibitem{Gor99}
Goriely S 1999 {\em Astron. Astrophys.\/} {\bf 342} 881

\bibitem{KlK88}
Klay N and K{\"a}ppeler F 1988 {\em Phys. Rev. C\/} {\bf 38} 295 -- 306

\bibitem{JBB92}
Jung M, Bosch F, Beckert K, Eickhoff H, Folger H, Franzke B, Gruber A, Kienle
  P, Klepper O, Koenig W, Kozhuharov C, Mann R, Moshammer R, Nolden F, Schaaf
  U, Soff G, Sp{\"a}dke P, Steck M, St{\"o}hlker T and S{\"u}mmerer K 1992 {\em
  Phys. Rev. Lett.\/} {\bf 69} 2164 -- 2167

\bibitem{BFF96}
Bosch F, Faestermann T, Friese J, Heine F, Kienle P, Wefers E, Zeitelhack K,
  Beckert K, Franzke B, Klepper O, Kozhuharov C, Menzel G, Moshammer R, Nolden
  F, Reich H, Schlitt B, Steck M, St{\"o}hlker T, Winkler T and Takahashi K
  1996 {\em Phys. Rev. Lett.\/} {\bf 77} 5190 -- 5193

\bibitem{FIK08}
Faber M, Ivanov A, Kienle P, Kryshen E, Pitschmann M and Troitskaya N 2008 {\em
  Phys. Rev. C\/} {\bf 78} 061603(R)

\bibitem{FFN80}
{Fuller} G~M, {Fowler} W~A and {Newman} M~J 1980 {\em Ap. J. Suppl.\/} {\bf 42}
  447--473

\bibitem{LaM00}
{Langanke} K and {Mart{\'{\i}}nez-Pinedo} G 2000 {\em Nuclear Physics A\/} {\bf
  673} 481--508 (\textit{Preprint} \eprint{arXiv:nucl-th/0001018})

\bibitem{LaM03}
{Langanke} K and {Mart{\'{\i}}nez-Pinedo} G 2003 {\em Reviews of Modern
  Physics\/} {\bf 75} 819--862 (\textit{Preprint}
  \eprint{arXiv:nucl-th/0203071})

\bibitem{TGC87}
Taddeucci T, Goulding C, Carey T, Byrd R, Goodman C, Gaarde C, Larsen J, Horen
  D, Rapaport J and Sugarbaker E 1987 {\em Nucl. Phys.\/} {\bf A469} 125 -- 172

\bibitem{Fre05}
Frekers D 2005 {\em Nucl. Phys. A\/} {\bf 752} 580c

\bibitem{Chi13}
Chiaveri E and {for the n\_TOF collaboration} 2013 {\em Nuclear Data for
  Science and Technology 2013\/}

\bibitem{RCH09}
Reifarth R, Chau L~P, Heil M, K{\"a}ppeler F, Meusel O, Plag R, Ratzinger U,
  Schempp A and Volk K 2009 {\em PASA\/} {\bf 26} 255

\bibitem{RCM07}
Ratzinger U, Chau L, Meusel O, Schempp A, Volk K, Heil M, K{\"a}ppeler F and
  Stieglitz R 2007 {\em Proceedings of the 18th International Collaboration on
  Advanced Neutron Sources, ICANS XVIII\/}
  http://www.ihep.ac.cn/english/conference/icans/proceedings/indexed/copyr/31.pdf
  ed Wei J, Wang S, Huang W and Zhao J (Beijing, China: Institute of High
  Energy Physics, Chinese Academy of Science) p 210

\bibitem{CMR06}
Chau L, Meusel O, Ratzinger U, Schempp A, Volk K and Heil M 2006 {\em EPAC
  Edinburgh 2006\/} p 1690 available online at
  http://accelconf.web.cern.ch/AccelConf/e06/PAPERS/TUPLS082.PDF

\bibitem{MCM06}
Meusel O, Chau L, Mueller I, Ratzinger U, Schempp A, Volk K, Zhang C and Minaev
  S 2006 {\em Linac 2006, Knoxville, USA\/}
  (http://accelconf.web.cern.ch/AccelConf/106/PAPERS/MO051.PDF) p 159

\bibitem{FRIB12}
Wei J, Arenius D, Bernard E, Bultman N, Casagrande F, Chouhan S, Compton C,
  Davidson K, Facco A, Ganni V, Gibson P, Glasmacher T, Harle L, Holland K,
  Johnson M, Jones S, Leitner D, Leitner M, Machicoane G, Marti F, Morris D,
  Nolen J, Ozelis J, Peng S, Popielarski J, Popielarski L, Pozdeyev E, Russo T,
  Saito K, Webber R, Weisend J, Williams M, Yamazaki Y, Zeller A, Zhang Y and
  Zhao Q 2012 {\em Proceedings of HIAT 2012, Chicago, IL USA\/} vol MOB01
  (North-Holland, Amsterdam: CERN) p
  http://accelconf.web.cern.ch/accelconf/HIAT2012/papers/mob01.pdf

\bibitem{Tan98}
{Tanihata} I 1998 {\em Journal of Physics G Nuclear Physics\/} {\bf 24}
  1311--1323

\bibitem{Fai07}
Gutbrod H~H 2006 Facility for antiproton and ion research fair baseline
  technical report, isbn-3-9811298-0-6 Tech. rep.
  http://www.fair-center.eu/for-users/publications/fair-publications.html

\bibitem{NMB06}
Nagler A, Mardor I, Berkovits D, Dunkel K, Pekeler M, Piel C, vom Stein P and
  Vogel H 2006 {\em 2006 Linear Accelerator Conference\/} ed Busso M, Gallino R
  and Raiteri C (CERN: Joint Accelerator Conferences Website, JACoW) p 168
  http://accelconf.web.cern.ch/accelconf/l06/PAPERS/MOP054.PDF

\bibitem{FPA09}
Feinberg G, Paul M, Arenshtam A, Berkovits D, Kijel D, Nagler I and Silverman I
  2009 {\em Nucl. Phys. A\/} {\bf 827} 590c

\bibitem{MMP09}
Mastinu P, Mart{\'i}n~Hern{\'a}ndez G and Praena J 2009 {\em Nucl. Instr. Meth.
  A\/} {\bf 601} 333--338

\bibitem{MPM12}
Mastinu P, Praena J, Mart{\'i}n~Hern{\'a}ndez G, Dzysiuk N, Prete G, Capote R,
  Pignatari M and Ventura A 2012 {\em Physics Procedia\/} {\bf 26} 261

\bibitem{CGM04}
Culbertson C, Green S, Mason A, Picton D, Baugh G, Hugtenburg R, Yina Z, Scott
  M and Nelson J 2004 {\em Applied Radiation and Isotopes\/} {\bf 61} 733 –
  738

\bibitem{DGT13}
Domingo-Pardo C and Guerrero C~Taín J December 2013 {\em n\_TOF Collaboration
  Meeting\/} (Bologna)

\bibitem{NFG11}
Nakamura S, Furutaka K, Goko S, Harada H, Kimura A, Kin T, Kitatani F, Koizumi
  M, Ohta M, Oshima M, Toh Y, Hori J, Fujii T, Fukutani S, Takamiya K, Igashira
  M, Katabuchi T, Mizumoto M, Kamiyama T, Kino K and Kiyanagi Y 2011 {\em
  Journal of the Korean Physical Society\/} {\bf 59} 1773 -- 1776

\bibitem{WLK12}
Woods P, Lederer C and K{\"a}ppeler F 2012 Destruction of the cosmic gamma-ray
  emitter 26al by neutron-induced reactions Tech. rep.
  \urlprefix\url{http://cds.cern.ch/record/1410662?ln=en}

\bibitem{SKK93}
Schatz H, K{\"a}ppeler F, Koehler P, Wiescher M and Trautvetter H~P 1993 {\em
  Ap. J.\/} {\bf 413} 750 -- 755

\bibitem{PTA10}
Paradela C, Tassan-Got L, Audouin L, Berthier B, Duran I, Ferrant L, Isaev S,
  Le~Naour C, Stephan C, Tarrio D, Trubert D and the~n\_TOF collaboration 2010
  {\em Phys. Rev. C\/} {\bf 82} 034601

\bibitem{RSK00}
Reifarth R, Schwarz K and K{\"a}ppeler F 2000 {\em Ap. J.\/} {\bf 528} 573 --
  581

\bibitem{ReL14}
Reifarth R and Litvinov Y~A 2014 {\em Phys. Rev. ST Accel. Beams\/} {\bf 17}(1)
  014701 \urlprefix\url{http://link.aps.org/doi/10.1103/PhysRevSTAB.17.014701}

\bibitem{HHS13}
{Hillenbrand} P~M, {Hagmann} S, {St{\"o}hlker} T, {Litvinov} Y, {Kozhuharov} C,
  {Spillmann} U, {Shabaev} V, {Stiebing} K, {Lestinsky} M, {Surzhykov} A,
  {Voitkiv} A, {Franzke} B, {Fischer} D, {Brandau} C, {Schippers} S, {Mueller}
  A, {Schneider} D, {Jakubassa} D, {Artiomov} A, {DeFilippo} E, {Ma} X,
  {D{\"o}rner} R and {Rothard} H 2013 {\em Physica Scripta Volume T\/} {\bf
  156} 014087

\bibitem{HBB11}
{von Hahn} R, {Berg} F, {Blaum} K, {Crespo Lopez-Urrutia} J~R, {Fellenberger}
  F, {Froese} M, {Grieser} M, {Krantz} C, {K{\"u}hnel} K~U, {Lange} M, {Menk}
  S, {Laux} F, {Orlov} D~A, {Repnow} R, {Schr{\"o}ter} C~D, {Shornikov} A,
  {Sieber} T, {Ullrich} J, {Wolf} A, {Rappaport} M and {Zajfman} D 2011 {\em
  Nuclear Instruments and Methods in Physics Research B\/} {\bf 269} 2871--2874

\bibitem{BaK87}
Bao Z and K{\"a}ppeler F 1987 {\em Atomic Data Nucl. Data Tables\/} {\bf 36}
  411--451

\bibitem{GBP88}
Gallino R, Busso M, Picchio G, Raiteri C and Renzini A 1988 {\em Ap. J.\/} {\bf
  334} L45

\bibitem{HoI88}
Hollowell D and Iben I 1988 {\em Ap. J.\/} {\bf 333} L25

\bibitem{BVW92}
Beer H, Vo{\ss} F and Winters R 1992 {\em Ap. J. Suppl.\/} {\bf 80} 403

\bibitem{DHK05}
Dillmann I, Heil M, K{\"a}ppeler F, Plag R, Rauscher T and Thielemann F~K 2005
  {\em Capture Gamma-Ray Spectroscopy and Related Topics\/} AIP Conference
  Series 819 ed Woehr A and Aprahamian A (New York: AIP) p 123
  http://www.kadonis.org

\bibitem{CGS06}
Cristallo S, Gallino R, Straniero O, Piersanti L and Dom{\'i}nguez I 2006 {\em
  Mem. Soc. Astron. Italiana\/} {\bf 77} 774

\end{thebibliography}

\end{document}